\documentclass[12pt,a4paper,twoside,titlepage]{book}  
\usepackage[american,italian]{babel}     
\usepackage{frontespizio}
\usepackage{amsmath,amsthm,amssymb,latexsym}
\usepackage{amsfonts}           
\usepackage{geometry}
\usepackage{graphicx}         
\usepackage{fancyhdr}
\usepackage{tikz}
\def\RR{\mathbb{R}}

\def\NN{\mathbb{N}}

\def\CC{\mathbb{C}}

\newtheorem{definition}{Definition}

\newtheorem{theorem}{Theorem}[section]    
\newtheorem{lemma}{Lemma}[section]   
\newtheorem{pro}{Proposition}[chapter]
\newtheorem{cor}{Corollary}[section]    
\newtheorem{remark}{Remark}[chapter]   
\newenvironment { abstract }

\begin{document}

\begin{titlepage}
\begin{rmfamily}
  \begin{large}
  \begin{center}
  \vbox to3cm{\vfill
  \includegraphics[width=3cm]{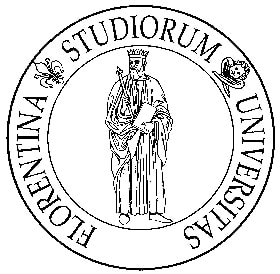}
    \vfill}
  \end{center}
  \begin{center}
  \vspace{0cm}
  \begin{LARGE}
    \uppercase{Universit\`a degli Studi di Firenze}
    \end{LARGE}\\
  \rule{9cm}{.4pt}\\
  \smallskip
  Facolt\`a di Scienze Matematiche, Fisiche e Naturali\\
  \medskip
  Dipartimento di Matematica "Ulisse Dini"\\
  \medskip
  Dottorato di Ricerca in Matematica\\
  \bigskip
  \vfill
  \begin{LARGE} Tesi di Dottorato\end{LARGE}\\
  \vfill
  \bigskip
    \begin{Huge}
      {\renewcommand{\\}{\cr}
     \halign{\hbox to\textwidth{\strut\hfil#\hfil}\cr
     {\textbf {A Study of a}}\cr
     
     {\textbf {Nonlinear Schr\"odinger Equation}}\cr
     
     {\textbf{for Optical Fibers}}\cr}}
     
    \end{Huge}
  \bigskip
  \bigskip
  \bigskip
  \bigskip
  \vfill
  \begin{tabular*}{\textwidth}{@{}l@{\extracolsep{\fill}}l@{\extracolsep{\fill}}l@{}}
    & Candidato:& \\
   & Domenico Felice &  \\
 & &\\
Tutor: & & Coordinatore del Dottorato:\\
Dr. Luigi Barletti& & Prof. Alberto Gandolfi\\
  \end{tabular*}

  \end{center}
 
  \begin{center}
  \vfill
  \vbox to-1cm{\vfil
  \rule{9cm}{.4pt}\\
  \medskip
  \uppercase{Ciclo XXIII, Settore disciplinare MAT/07 Fisica Matematica}\\
  \vfil}
  \end{center}
  \end{large}
\end{rmfamily}
\end{titlepage}
\newpage

\begin{flushright}
To the memory of my grandmother Adele.
\end{flushright}
\chapter*{Introduction}
Non linear fiber optics concerns with the non linear optical phenomena occurring inside optical fibers. Although the field of optical fiber communications traces its beginning to the $1960$ \cite{Franken}, the use of optical fibers for commercial purpose became feasible only after $1970$ when fiber losses were reduced to below $20 dB/Km$ \cite{Maurer}. Further progress in fabrication technology resulted by $1979$ in a loss of only $0.2 dB/Km$ in the $1.55\mu m$ wavelength region \cite{Miya}, a loss level limited mainly by the fundamental process of Rayleigh scattering.

The field of fiber optics has continued to grow during the decade of $1990s .$ This growth was boosted by several advances in lightwave technology, the most important being the advent of high-capacity fiber optic communication systems \cite{Agrawal}. In such systems, the transmitted signal is amplified periodically by using optical amplifiers to compensate for residual fiber loss. As a result, the non linear effects, which are present in the fiber, accumulate over long distances, and the effective interaction length can exceed thousands Kilometers. 

An optical fiber looks like a thin strand of glass and consists of a central core surrounded by a cladding whose refractive index is slightly lower than that of the core. The guiding properties of an optical fiber are characterized by a dimensionless parameter defined as $V=a(\omega/c)(n_1^2-n_2^2)^{1/2}$ where $a$ is the core radius, $\omega$ is the frequency of the light, and $n_1$ and $n_2$ are the refractive indices of the core and the cladding, respectively. The parameter $V$ determines the number of modes supported by the fiber. Optical fibers with $V<2.405$ support only a single mode and are called \textit{single-mode} fibers \cite{Agrawal}.

Our work is focused on the one-dimensional nonlinear Schr\"odinger equation for the propagation of light in single-mode fibers, in the presence of attenuation, dispersion, and non linear effects. We immediately point out that, by contrast to the usual theory of evolutionary PDEs, in our NLSE the evolution variable is the 'space' variable, namely the longitudinal coordinate of the fiber. The derivation of this NLSE from Maxwell's equations relies on some assumptions that are commonly met in today's fiber optic communication systems. The lowest-order nonlinear effects originate from the third-order susceptibility $\chi^{(3)},$ \cite{Agrawal,Boiti}, which is responsible for phenomena such as third-harmonic generation and nonlinear refraction. Among these, nonlinear refraction, a phenomenon referring to the intensity dependence of the refractive index, plays the most important role. We refer to Agrawal's book \cite{Agrawal} for a detailed derivation of the NLSE from Maxwell equations.

From the mathematical point of view, the treatment of Schr\"odinger equation (even the linear one) may be delicate since such equation possesses a mixture of the properties of parabolic and hyperbolic equation. Particularly useful tools are energy and Strichartz's estimates. In the NLSE for optical fibers, because of attenuation, there is not conservation of energy. In this thesis, we study problems of local nature (local existence of solutions, uniqueness) and problems of global nature (global existence) by the Kato's method \cite{Kato} based on a fixed point argument and Strichartz's estimates. Moreover, the tools used in \cite{Cazenave} to study global problems connected with finite-time blow up of solutions, are used in our work to show a closeness result between the NLSE for optical fibers and an integrable NLSE. To achieve this result we first define the Painlev\'e property for a partial differential equation; then, we employ the Painlev\'e analysis \cite{He} to investigate the integrability of a general non autonomous one-dimensional Schr\"odinger equation: explicit constraints on the parameters of the general non autonomous equation can be obtained by the Painlev\'e analysis, which provides an effective way to construct integrable non autonomous NLSE systems \cite{He}. Under these conditions, these integrable non autonomous NLSE, can be converted into the standard NLSE with cubic nonlinearity \cite{He}. The parameters of the NLSE for single-mode fibers, in the presence of attenuation, dispersion, and non linear effects, does not satisfy these conditions. To obtain an integrable equation we need to introduce one more term in the NLSE: the equation so-constructed is now integrable. Under special conditions on the added term we show that the two equations are close each other in the sense of the $L^2$- norm, i.e. two arbitrary solutions of the integrable equation and of the NLSE are close in the $L^2$- norm. The proof uses argumentations usually employed to prove finite-time blow up of solutions of the standard cubic NLSE \cite{Cazenave}. 

Conversion of integrable system into the standard cubic NLSE is a fantastic tool to obtain analytical solutions for general integrable non autonomous NLSE starting from the solutions of the standard NLSE. In particular, we shall apply such technique to the so-called soliton-like solutions and to the continuous wave solutions.

Group Velocity Dispersion in a nonlinear optical fiber plays a very crucial role in the shaping of the pulse propagating through the fiber. In a fiber having a constant anomalous dispersion, due to exact balancing between the effects of nonlinearity and linearity dispersion, the shape of a suitably prepared pulse does not change as it propagates through the nonlinear optical fiber. Such a pulse is called a \textit{soliton} \cite{Hasegawa,Agrawal}.

The continuous wave solutions are related with the analysis of noisy fibers. Through the fiber, an optical pulse is affected by the amplified spontaneous emission (ASE) noise generated by optical amplifiers. In systems operating in the linear regime, the characteristics of ASE noise do not change after propagation through the optical fiber. On the other hand, when nonlinear effects are not negligible, signal and noise interactions occur during propagation. As a result, at certain frequencies, the noise may be amplified or attenuated. In \cite{Forestieri} it is assumed that for the purpose of investigating the interaction between signal and noise, the signal can be considered a continuous wave (CW) i.e. an electromagnetic wave of constant amplitude, and the noise can be treated as a perturbation of the CW solution of the NLSE. Then, it is reasonable to study stability against small perturbations of modulus and phase of such solutions. From a mathematical point of view, such stability results are obtained in \cite{Zhidkov} for the standard NLSE in a fiber having constant normal dispersion. In this case, because of these solutions are time-homogeneous, the transformation obtained by Painlev\'e analysis, does not work properly over all \textit{real} time; thus, we prove a stability result depending on the width of an arbitrary interval of time for the standard NLSE. Such result can be easily converted into integrable non autonomous NLSE system. In the regime of anomalous dispersion there is not stability of such type. This phenomenon is referred as \textit{modulation instability} and it admits solutions whose shape does not change during the propagation (such as the \textit{soliton}-like solutions \cite{Agrawal}). In the case of \textit{soliton}-like solutions for the standard NLSE, because of phase and translation symmetries there exists an orbital stability, i.e. stability modulo such symmetries \cite{Weinstein}. This type of stability are obtained by Lyapunov method and it is strongly based on the conservation of the Hamiltonian and square integral functionals. In \cite{Grillakis} this approach is extended to solitary waves of general Hamiltonian systems that are invariant under a group of transformations. By contrast to the normal dispersion regime, in this case, transformation obtained by Painlev\'e analysis works properly for every \textit{real} time. Thus the well-known results for the standard NLSE can be easily converted into integrable non autonomous NLSE system. The idea for a future development is to convert the functionals which are conserved in the standard NLSE into functionals which are invariant under any group of transformations for an integrable non autonomous NLSE; thus, in view of \cite{Grillakis} obtain a type of stability modulo such transformations.

This thesis is organized as follows. In Chapter $1$ we describe the characteristic of the nonlinear optical fibers. Then we derive the NLSE for a single-mode optical fiber, under suitable physical approximations. In Chapter $2$ we adapt to our case well-know results about local and global existence and uniqueness of solutions of the Cauchy problem for the Schr\"odinger equations. In particular we are interested to solutions in the Hilbert spaces $H^1$ and $H^2.$ In Chapter $3$ we describe the Painlev\'e analysis, giving conditions under which a general non autonomous NLSE is integrable. Under these conditions we get a transformation that allows to convert integrable non autonomous NLSE into the standard cubic NLSE. Then we specialize the general context to the NLSE for optical fiber systems, and we show the closeness between such a NLSE and an integrable NLSE. Finally, in Chapter $4$ we deal with solitary-wave solutions of the NLSE. In particular, we present results about existence and stability of CW solutions for the Standard NLSE. Then we convert these results to the integrable NLSE by means of the tools obtained in Chapter $3.$ Moreover, we consider the CRLP perturbation method \cite{Forestieri} and obtain the perturbed phase within a class of functions which derivative has $L^2$-norm inside a definite bounded interval of $\RR^+.$ At last we show existence and orbital stability of soliton-like solutions of the standard NLSE. Then we convert these results to the integrable NLSE. 
\vspace{.3cm}

This thesis work has been carried within the joint research project between the 
Department of Mathematics 'Ulisse Dini' in Florence and the CNIT (Consorzio Nazionale 
Interuniversitario per le Telecomunicazioni) in Pisa, which has sponsored my PhD grant.

\tableofcontents

\chapter{Optical Fiber}
\section{Fiber characteristics}
An optical fiber (or optical fibre) is a flexible, transparent fiber made of a pure glass (silica) not much wider than a human hair. It functions as a waveguide to transmit light between the two ends of the fiber. Optical fibers are widely used in fiber-optic communications, which permit transmission over longer distances and at higher bandwidths (data rates) than other forms of communication. Fibers are used instead of metal wires because signals travel along them with less loss and are also immune to electromagnetic interference.

Optical fiber typically includes a transparent core surrounded by a transparent cladding material with a lower index of refraction. Light is kept in the core by \textit{total internal reflection}, which is an optical phenomenon that happens when a ray of light strikes a medium boundary at an angle larger than a particular critical angle with respect to the normal to the surface. If the refractive index is lower on the other side of the boundary and the incident angle is greater than the critical angle, no light can pass through and all of the light is reflected. This causes the fiber to act as a waveguide.

 The material of choice for low-loss fibers is pure silica glass synthesized by fusing $SiO_2$ molecules. The refractive-index difference between the core and the cladding is realized by the selective use of dopants during the fabrication process. For example $GeO_2$  increases the refractive index of pure silica and is suitable for the core, while materials such as boron and fluorine are used for the cladding because they decrease the refractive index of silica. 

Depending on refraction index $n_1$ of the core, optical fibers can be classified as follows
\begin{itemize}
\item \textit{step-index fibers} if $n_1$ is constant inside the core;
\item \textit{graded-index
fibers} in which the refractive index of the core decreases gradually from the center
to the core boundary.
\end{itemize}

 Two parameters that characterize an optical
fiber are the relative core cladding index difference
$$\Delta =\frac{n_1-n_2}{n_1}$$
where $n_2$ is the cladding refractive index and $n_2<n_1,$ and the so called $V$- parameter defined as
$$V=k_0a\sqrt{n_1^2-n_2^2},$$
where $k_0=\frac{2\pi}{ \lambda},$ $a$ is the core radius and $\lambda$ is the wavelength of light.



The V-parameter determines the number of modes supported by the fiber. Modes are the possible solutions of the Helmholtz equation for waves, which is obtained by combining Maxwell's equations and the boundary conditions, as we will see in the next. These modes define the way the wave travels through space, i.e. how the wave is distributed in space. Waves can have the same mode but have different frequencies. This is the case in single-mode fibers, where we can have waves with different frequencies, but of the same mode, which means that they are distributed in space in the same way, and that gives us a single ray of light. A step-index fiber supports
a single mode if $V<2.405.$ Optical fibers designed to satisfy this condition are called
single-mode fibers. The main difference between the single-mode and multi mode fibers
is the core size. The core radius $a$ is typically $25 \mu m$ for multi mode fibers, while for 
single-mode fibers is required to be $<5 \mu m.$ The numerical value of the
outer radius $b$ is less critical as long as it is large enough to confine the fiber modes
entirely. A standard value of $b = 62.5 \mu m$ is commonly used for both single-mode and
multi mode fibers. Since nonlinear effects are mostly studied using single-mode fibers,
the term optical fiber in what follows refers to single-mode fibers.

\section{Fiber losses}
An important fiber parameter provides a measure of power loss during transmission of
optical signals inside the fiber. If $P_0$ is the power launched at the input of a fiber of
length $L,$ the transmitted power $P_T$ is given by
$$
P_T=P_0\exp(-\alpha L)
$$
where the attenuation real positive constant $\alpha$ is a measure of total losses of all sources. It is customary to express $\alpha$ in units of $dB/km$ using the relation
$$\alpha_{dB}=-\frac{10}{L}\log \Bigg( \frac{P_T}{P_0}\Bigg)\approx 4,343 \alpha ,$$
where the previous equation was used to relate $\alpha_{dB}$ and $\alpha.$

Fiber losses depend on the wavelength of light. Silica fiber
exhibits a minimum loss of about $0.2 dB/km$ near $1.55 \mu m.$ Losses are considerably higher at shorter wavelengths, reaching a level of a few $dB/km$ in the visible region,i.e. $\lambda$ is in the range of $380 \div 760 nm.$
Note, however, that even a $10 dB/km$ loss corresponds to an attenuation constant of
only $\alpha \simeq 2 \times 10^{-5}cm^{-1} $, an incredibly low value compared to that of most other
materials.

\section{Chromatic Dispersion}
When an electromagnetic wave interacts with the bound electrons of a dielectric, the medium response, in general, depends on the optical frequency $\omega.$  This property, referred to as chromatic dispersion, manifests through the frequency dependence of the refractive index $n(\omega).$

Fiber dispersion plays a critical role in propagation of short optical pulses because different spectral components associated with the pulse broadening can be detrimental for optical communication system. Mathematically, the effects of fiber dispersion are accounted for by expanding the mode-propagation coefficient $\beta$ in a Taylor series about the frequency $\omega_0$ at which the pulse spectrum is centred:
$$
\beta(\omega)\Doteq n(\omega)\frac{\omega}{c}=\beta_0+\beta_1(\omega-\omega_0)+\frac 1 2 \beta_2(\omega-\omega_0)^2+\cdots,
$$
where $c$ is the speed of light and 
$$\beta_m=\Big(\frac{d^m\beta}{d\omega^m}\Big)_{\omega=\omega_0}\qquad (m=0,1,2,\cdots).$$

The parameters $\beta_1$ and $\beta_2$ are related to the refractive index $n$ and its derivative through the relations
$$
\beta_1\Doteq\frac{1}{v_g}\Doteq\frac{n_g}{c}=\frac 1 c \Big(n+\omega\frac{dn}{d\omega}\Big),\quad\beta_2=\frac 1 c \Big(2\frac{dn}{d\omega}+\omega \frac{d^2n}{d\omega^2}\Big),
$$
where $n_g$ is the group index and $v_g$ is the group velocity. Physically speaking, the envelope of an optical pulse moves at the group velocity while the parameter $\beta_2$ represents dispersions of the group velocity and is responsible for pulse broadening, in other words the frequency dependence of the group velocity leads to pulse broadening simply because different spectral components of the pulse do not arrive simultaneously at the fiber output. Consider an optical fiber of length $L,$ a specific spectral component at the frequency $\omega$ would arrive at the output end of the fiber after a time delay $T=L/v_g.$ If $\Delta \omega$ is the spectral width of the pulse, the extent of pulse broadening is governed by
$$
\Delta T=\frac{dT}{d\omega}\Delta \omega=\frac{d}{d\omega}\Big(\frac{L}{v_g}\Big)\Delta\omega=L\frac{d}{d\omega}\Big(\frac{d \beta}{d\omega}\Big)\Delta\omega=\frac{d^2\beta}{d\omega^2}\Delta\omega\equiv L\beta_2\Delta\omega.
$$
This phenomenon is known as the group-velocity dispersion (GVD), and $\beta_2$ is the GVD parameter. It follows from this equation that the physical dimension of $\beta_2$ is $L^{-1}T^2.$ Nonlinear effects in optical fiber can manifest qualitatively different behaviors depending on the sign of the GVD parameter. In standard silica fibers, $\beta_2$ vanishes at a wavelength of about $1.27 \mu m$ and becomes negative for longer wavelength. This wavelength is referred to as the zero-dispersion wavelength and is denoted as $\lambda_D.$ For wavelength such that $\lambda<\lambda_D,$ the fiber is said to exhibit normal dispersion as $\beta_2>0.$ In the normal-dispersion regime, high frequency (blue-shifted) components of optical pulse travel slower than low-frequency (red-shifted) components of the same pulse. By contrast, the opposite occurs in the anomalous-dispersion regime in which $\beta_2$ is negative. Silica fibers exhibit anomalous dispersion when the light wavelength exceeds the zero-dispersion wavelength. The anomalous dispersion regime is of considerable interest for the study of nonlinear effects because it is in this regime that optical fibers supports solitons through a balance between the dispersive and nonlinear effects.

\section{Polarization-Mode Dispersion}
Even a single-mode fiber is not truly single mode because it can support two degenerate modes that are polarized in two orthogonal directions. In a Cartesian reference system $x,y,z,$ we assume that the wave vector is referred along the $z-$ axis and so the polarization occurs in the plane $x,y.$ Under ideal conditions (perfect cylinder symmetry and stress-free fiber), a mode excited with its polarization in the $x$ direction would not couple to the mode with the orthogonal $y$-polarization state. In real fibers, small departures from cylinder symmetry because of random variations in core shape and stress-induced anisotropy result in a mixing of the two polarization states by breaking the mode degeneracy. Mathematically, the mode-propagation constant $\beta$ becomes slightly different for the modes polarized in the $x$ and $y$ direction. This property is referred to as modal birefringence. 

The axis along which the mode index is smaller is called the fast axis because the group velocity is larger for light propagating in that direction. For the same reason, the axis with the larger mode index is called the slow axis.
 As a result of this mode-propagation constant difference, light launched into the fiber with a fixed state of polarization changes its polarization in a random fashion. If an input pulse excites both polarization components, the two components travel along the fiber at different speeds because of their different group velocities. The pulse becomes broader at the output and because group velocities change randomly in response to random changes in fiber birefringence. This phenomenon is referred to as polarization-mode dispersion (PMD).

The extent of pulse broadening can be estimated from the variance of the time delay $\Delta T$ occurring between the to polarization components during propagation of an optical pulse,  because of random changes in birefringence occurring along the fiber. These changes tend to equalize the propagation times for the two polarization components. In fact, PMD is characterized by the root-mean-square value of $\Delta T$ obtained after averaging over random perturbations. The variance of $\Delta T$ is found to depend on $\sqrt{L},$ where $L$ denotes the fiber length. Because of its $\sqrt{L}$ dependence, PMD-induced pulse broadening is relatively small compared with the GVD effects.

For some applications it is desirable that fibers transmit light without changing its state of polarization. Such fibers are called \textit{polarization maintaining} or \textit{polarization preserving} fibers \cite{Agrawal}. The use of polarization maintaining fibers requires identification of the slow and fast axes before an optical signal can be launched into the fiber. Structural changes are often made to the fiber for this purpose. For example a way is that, cladding is flattened in such a way that the flat surface is parallel to the slow axis of the fiber. Such a fiber is called ``D fiber'' after the shape of the cladding and makes axes identification relatively easy. When the polarization direction of the linearly polarized light coincides with the slow or fast axis, the state of polarization remains unchanged during propagation. 

\section{Pulse propagation in fibers}
Like all electromagnetic phenomena, the propagation of optical fields in fibers is governed by Maxwell's equations. In the International System of Units, these equations are \cite{Agrawal}

$$
\nabla\times \mathbf{E}=-\frac{\partial \mathbf{B}}{\partial t},
$$
$$
\nabla\times \mathbf{H}=\mathbf{J}+\frac{\partial \mathbf{D}}{\partial t},
$$
$$
\nabla \cdotp \mathbf{D}=\rho_f,
$$
$$
\nabla \cdotp \mathbf{B}=0,
$$
where $\mathbf{E}$ and $\mathbf{H}$ are electric and magnetic field, respectively, and $\mathbf{D}$ and $\mathbf{B}$ are corresponding electric and magnetic flux densities. The current density vector $\mathbf{J}$ and the charge density $\rho_f$ represent the sources for the electromagnetic field. The flux densities $\mathbf{D}$ and $\mathbf{B}$ arise in response to the electric and magnetic fields propagating inside the medium and are related to them through the constitutive equations given by \cite{Agrawal}
$$
\mathbf{D}=\epsilon_0\mathbf{E}+\mathbf{P}
$$
$$
\mathbf{B}=\mu_0\mathbf{H}+\mathbf{M}
$$
where $\epsilon_0$ is the vacuum permittivity and $\mu_0$ is the vacuum permeability, and $\mathbf{P}$ and $\mathbf{M}$ are the induced electric and magnetic polarizations. 

Optical fiber are a non magnetic medium: this is expressed by the relation $\mathbf{M}=0;$ moreover optical fiber is a medium without free charges, and so it occurs that $\mathbf{J}=0$ and $\rho_f=0.$ 

Maxwell's equations can be used to obtain the wave equation that describes light propagation in optical fibers. By taking the curl of the first equation of the previous system and using the second one and the constitutive equations one can eliminate $\mathbf{B},$ $\mathbf{H}$ and $\mathbf{D}$ in favour of $\mathbf{E}$ and $\mathbf{P}$ and obtain
\begin{equation}
\nabla \times(\nabla \times \mathbf{E})=-\frac{1}{c^2}\frac{\partial^2 \mathbf{E}}{\partial t^2}-\mu_0\frac{\partial^2 \mathbf{P}}{\partial t^2},
\end{equation}
where $c$ is the speed of the light in the vacuum and the relation $\varepsilon_0\mu_0=1/c^2$ was used. To the equation $(1.1)$ it needs to add the condition
\begin{equation}
\nabla\cdot( \epsilon_0\mathbf{E}+\mathbf{P})=0
\end{equation}
and the boundary conditions that the passage to the surface that separates different dielectric media the tangential component of $\mathbf{E}$ and the normal component of $\mathbf{D}$ are continuous.
The problem will be closed by expressing the polarization $\mathbf{P}$ in terms of the electric field $\mathbf{E}.$

If the medium is isotropic and the response of the medium is instantaneous to the electric field we have that
$$
\mathbf{P}=\epsilon_0\chi\mathbf{E},
$$
where $\chi$ is a positive real constant called electric susceptibility. Now we recall the relation for a vector in $\RR ^3$
$$
\nabla\times( \nabla\times \mathbf{E})=\nabla(\nabla\cdot \mathbf{E})-\Delta\mathbf{E}.
$$
So we can write eq. $(1.1)$ as
\begin{eqnarray}
&&\Delta\mathbf{E}-\nabla(\nabla\cdot \mathbf{E})-\frac{1}{c^2}   \frac{\partial^2\mathbf{E}}{\partial t^2}-\mu_0\frac{\partial^2\mathbf{P}}{\partial t^2}=0\\
&&\Delta\mathbf{E}-\nabla(\nabla\cdot \mathbf{E})-\frac{1}{c^2}   \frac{\partial^2\mathbf{E}}{\partial t^2}-\chi\frac{1}{c^2}\frac{\partial^2\mathbf{E}}{\partial t^2}=0\nonumber
\end{eqnarray}
and so we obtain that
\begin{eqnarray*}
&&\Delta\mathbf{E}-\frac{n^2}{c^2}\frac{\partial^2\mathbf{E}}{\partial t^2}=0\\
\\
&&\nabla\cdot\mathbf{E}=0
\end{eqnarray*}
where $n=\sqrt{1+\chi}$ is the refractive index of the medium.

\vspace{.5cm}

Generally the response of the medium to the electric field is not instantaneous. In an isotropic medium this delay si described by
\begin{equation}
\mathbf{P}(\mathbf{r},t)=\epsilon_0\int_{-\infty}^{+\infty}\chi(t-t^\prime)\mathbf{E}(\mathbf{r},t^\prime)dt^\prime,
\end{equation}
where due to causality, the electric susceptibility $\chi(t-t^\prime)$ is equal to zero for $t-t^\prime <0.$ It also rapidly decrease as $|t-t^\prime|\rightarrow \infty.$ By defining the Fourier transform and its inverse for a field $\mathbf{F}(t)$ as
$$
\widehat{\mathbf{F}}(\omega)=\int_{-\infty}^{+\infty}\mathbf{F}(t)e^{i\omega t}dt,\qquad \mathbf{F}(t)=\frac{1}{2\pi}\int_{-\infty}^{+\infty}\widehat{\mathbf{F}}(\omega)e^{-i\omega t}d\omega,
$$
we have from $(1.4)$ that
\begin{equation}
\widehat{\mathbf{P}}(\omega)=\epsilon_0\widehat{\chi}(\omega)\widehat{\mathbf{E}}.
\end{equation}
Then it follows from $(1.2)-(1.3)$ that
\begin{equation}
\Delta\widehat{\mathbf{E}}(\omega)+\frac{\omega^2}{c^2}(1+\widehat{\chi})\widehat{\mathbf{E}}(\omega)=0
\end{equation}
\begin{equation}
\nabla\cdot\widehat{\mathbf{E}}(\omega)=0.
\end{equation}
The refractive index $n(\omega)=\sqrt{1+\widehat{\chi}(\omega)}$ in this case depends on the frequency. Moreover it is necessarily a complex number. In fact, we observe that, because of causality, the Fourier transform
$$
\widehat{\chi}(\omega)=\int_{-\infty}^{+\infty}\chi(t)e^{i\omega t}dt,
$$
is a well-defined analytic function over the upper half complex plane such that $\widehat{\chi}(\omega)=0$ for $\omega\rightarrow\infty.$ Then by applying the Cauchy theorem over a boundary $C$ described in the figure below we obtain that
$$
\int_C
\frac{\widehat{\chi}(\omega)}{\omega^\prime-\omega}d\omega^\prime=0.
$$
No we refer to $C_R$ as the outer semi-circumference and to $C_\epsilon$ as the inner one. By sending $R$ to infinity, the contribute of this integral along $C_R$ is zero and so we have that
$$
\int_{|\omega^\prime-\omega|\geq \epsilon}\frac{\widehat{\chi}(\omega)}{\omega^\prime-\omega}d\omega^\prime+\int_{C_\epsilon}\frac{\widehat{\chi}(\omega)}{\omega^\prime-\omega}d\omega^\prime=0.
$$
By sending $\epsilon$ to zero we have that
\begin{equation}
P\int_{-\infty}^{+\infty}\frac{\widehat{\chi}(\omega)}{\omega^\prime-\omega}d\omega^\prime-i\pi\widehat{\chi}(\omega)=0
\end{equation}
where it is introduced the principal part
$$
P\int_{-\infty}^{+\infty}\frac{\widehat{\chi}(\omega)}{\omega^\prime-\omega}d\omega^\prime=\lim_{\epsilon\rightarrow 0}\int_{|\omega^\prime-\omega|\geq \epsilon}\frac{\widehat{\chi}(\omega)}{\omega^\prime-\omega}d\omega^\prime.
$$
Let us separate $\widehat{\chi}(\omega)$ into its real and imaginary part:
\begin{equation}
\widehat{\chi}(\omega)=\widehat{\chi}^\prime(\omega)+i\widehat{\chi}^{\prime\prime}(\omega)
\end{equation}
By substituting eq $(1.9)$ into eq. $(1.8)$ we obtain the Kramers-Kronig relations 
\begin{equation}
\widehat{\chi}^\prime(\omega)=\frac{1}{\pi}P\int_{-\infty}^{+\infty}\frac{\widehat{\chi}^{\prime\prime}(\omega)}{\omega^\prime-\omega}d\omega^\prime
\end{equation}
\begin{equation}
\widehat{\chi}^{\prime\prime}(\omega)=-\frac{1}{\pi}P\int_{-\infty}^{+\infty}\frac{\widehat{\chi}^{\prime}(\omega)}{\omega^\prime-\omega}d\omega^\prime
\end{equation}
They show the reason why the refractive index must be a \textit{complex} number.
\begin{center}

\begin{tikzpicture}

\draw[->] (-3.5,0) -- (4,0) node[right] {$Re\,w$} coordinate(Re w axis);
\draw[->] (0,-0.5) -- (0,3.8) node[above] {$Im\,w$} coordinate(Im w axis);
 \draw[dashed] (-4,0) -- (-3.5,0);
 \draw[dashed] (0,-1) -- (0,0);

 \draw (1.6,0.2) node[color=black]{\scriptsize{$w'$}};
 \draw (1.6,-0.01) node[color=black]{$\cdot$};

 \draw[] (-3,0) arc (180:0:3) -- cycle;

 \draw[->] (0:3) arc (0:45:3);%
 \draw[->] (0:1) arc (180:90:0.6);%
 \draw[] (0:2.2) arc (0:90:0.6);%
 \draw[->,>=stealth] (-3,0) --(-1,0);%


\end{tikzpicture}\\
\scriptsize{\textit{The boundary $C$.}}
\end{center}

\subsection{General Relation between $\mathbf{P}$ and $\mathbf{E}$}
The polarization $\mathbf{P}$ is usually a complicated nonlinear function of $\mathbf{E}.$ In the nonlinear case, when $\mathbf{E}$ is sufficiently weak, the polarization can be expand into a power series of $\mathbf{E}:$
$$
\mathbf{P}=\chi^{(1)}\cdot\mathbf{E}+\chi^{(2)}\colon \mathbf{E}\mathbf{E}+\chi^{(3)}\vdots \; \mathbf{E}\mathbf{E}\mathbf{E}+\cdots
$$
where $\chi^{(j)}$ is the $j$th order susceptibility. In general, $\chi^{j}$ is a tensor of rank $j+1.$ Thus, the $j$th component of $\mathbf{P}$ is given by
\begin{eqnarray*}
\frac{1}{\epsilon_0}P_j&=&\int_{-\infty}^{+\infty}\chi^{(1)}_{jk}(t-t^\prime)E_k(t^\prime)dt^\prime\\
&+&\int\int_{-\infty}^{+\infty}\chi^{(2)}_{jkl}(t-t_1,t-t_2)E_k(t_1)E_l(t_2)dt_1 dt_2\\
&+&\int\int\int_{-\infty}^{+\infty}\chi^{(3)}_{jklm}(t-t_1,t-t_2,t-t_3)E_k(t_1)E_l(t_2)E_m(t_3)dt_1 dt_2 dt_3\\
&+&\cdots,
\end{eqnarray*}
where we used the Einstein summation convention.

Without loss of generality we can assume that the coefficients are symmetric into the arguments and into the indices. Moreover if we assume that $\mathbf{P}$ depends linearly on the electric field and the response of the  isotropic and homogeneous medium is not instantaneous we have \cite{Newell} that 
\begin{eqnarray*}
\frac{1}{\epsilon_0}\mathbf{P}&=&\int_{-\infty}^{+\infty}\chi^{(1)}(t-t^\prime)\mathbf{E}(t^\prime)dt^\prime\\
&+&\int\int_{-\infty}^{+\infty}\chi^{(2)}(t-t_1,t-t_2)\mathbf{E}(t_1)\times\mathbf{E}(t_1) dt_1 dt_2\\
&+&\int\int\int_{-\infty}^{+\infty}\chi^{(3)}(t-t_1,t-t_2,t-t_3)(\mathbf{E}(t_1)\cdot\mathbf{E}(t_2))\mathbf{E}(t_3)dt_1 dt_2 dt_3
\end{eqnarray*}
where $\chi$  is a scalar and not a tensor and the nonlinear dependence of $\mathbf{E}$ is
built in the only way we can construct a vector bilinear and trilinear in
$\mathbf{E}.$
Moreover if we assume that the medium is centrosymmetric that is  when $\mathbf{E}$ is replaced by $-\mathbf{E},$ $\mathbf{P}$ is replaced by $-\mathbf{P},$ we see that the second order in $\mathbf{E}$ is zero and so the leading nonlinearity is cubic.

 An optical fiber has cylindrical symmetry, and so it is a centrosymmetric medium. Then in this case the leading nonlinearity is cubic. Let us define

\begin{equation}
 \mathbf{P}_L(\mathbf{r},t)=\epsilon_0\int_{-\infty}^{+\infty}\chi^{(1)}(t-t^\prime)\mathbf{E}(\mathbf{r},t^\prime)dt^\prime
\end{equation}
 and 
\begin{equation}
 \mathbf{P}_{NL}(\mathbf{r},t)=\epsilon_0\int_{-\infty}^{+\infty}\chi^{(3)}(t-t_1,t-t_2,t-t_3)( \mathbf{E}(\mathbf{r},t_1)\cdot\mathbf{E}(\mathbf{r},t_2))\mathbf{E}(\mathbf{r},t_3)dt_1 dt_2 dt_3
\end{equation}
 
 In view of these definitions eq. $(1.1)$ becomes

 \begin{equation}
 -\nabla\times (\nabla\times \mathbf{E})-\frac{1}{c^2}\frac{\partial^2 \mathbf{E}}{\partial t^2}=\mu_0\frac{\partial^2\mathbf{P}_L}{\partial t^2}+\mu_0\frac{\partial^2\mathbf{P}_{NL}}{\partial t^2}
 \end{equation}
 with
 \begin{equation}
 \nabla\cdot(\epsilon_0\mathbf{E}+\mathbf{P})=0.
 \end{equation}
 
 \subsection{Optical Fiber as a waveguide }
 We first consider the case when $\mathbf{P}_{NL}=0.$ In this way eq. $(1.14)$ becomes linear. We have to consider separately the equation $(1.14)$ in the core and the cladding. Specifically, we have to first find a solution in the inner cylinder of radius a, which is limited and propagates in the $z$ direction and then a solution in the outer cylinder of radius a, which vanishes at large $r,$ then we have to connect the two solutions on the surface which encloses the core, i.e. $r = a,$ so that they meet the conditions of continuity of the tangential component of $\mathbf{E}$ and normal component of $\epsilon_0\mathbf{E}+\mathbf{P}.$
 
  In the frequency domain, making use of Fourier transform with respect the time $t$ we have that
 
 $$
 \widehat{\frac{\partial^2 \mathbf{E}}{\partial t^2}}=-\omega^2\widehat{\mathbf{E}}\qquad\widehat{\frac{\partial^2 \mathbf{P}}{\partial t^2}}=-\omega^2\widehat{\mathbf{P}_L}=-\omega^2\widehat{\chi}^{(1)}\widehat{\mathbf{E}}
 $$
 because of $(1.12).$ Moreover we have that
 $$
 \widehat{\mathbf{D}}(\mathbf
 {x},\omega)=\epsilon_0\widehat{\mathbf{E}}(\mathbf
 {x},\omega)+\widehat{\mathbf{P}}(\mathbf
 {x},\omega)=\left\{\begin{array}{ll}
\epsilon_0 n_1^2(\omega)\widehat{\mathbf{E}}(\mathbf
{x},\omega)&\mbox{for}\quad r=\sqrt{x^2+y^2}<a\\
\epsilon_0 n_2^2(\omega)\widehat{\mathbf{E}}(\mathbf
{x},\omega)&\mbox{for}\quad r=\sqrt{x^2+y^2}>a
 \end{array}\right.
 $$
 where we set $n_i^2(\omega)=1+\widehat{\chi}^{(1)}(\omega).$
 
 We have seen that the refractive index is a complex number; the imaginary part takes to account the absorption of the fiber. In what follows we neglect this effect. We will include the loss fiber in the next. So we can approximate $n(\omega)$ as a \textit{real} number. Another approximation that is valid for a single-mode fiber is that
 $$
 0<\frac{n_1(\omega)-n_2(\omega)}{n_1(\omega)}\ll 1
 $$
 which allows to consider $\widehat{\mathbf{D}} $ continuous in $\widehat{\mathbf{E}}$ and so we can consider to be continue across the core-cladding interface also the normal component of $\widehat{\mathbf{E}}.$ This approximation may appear incoherent with the physical problem, because it is the refractive index difference that makes possible the fiber in a waveguide. But we will catch up this by taking the cylindrical symmetry of the solution and the single-mode propagation on $z$ axis.
 
 Therefore eq .$(1.14)$ becomes
 \begin{equation}
 \Delta \widehat{\mathbf{E}}(\mathbf
  {x},\omega)+\frac{n_1^2(\omega)\omega^2}{c^2}\widehat{\mathbf{E}}(\mathbf
   {x},\omega)=0\qquad \mbox{for} \quad r<a
 \end{equation}
  \begin{equation}
  \Delta \widehat{\mathbf{E}}(\mathbf
   {x},\omega)+\frac{n_2^2(\omega)\omega^2}{c^2}\widehat{\mathbf{E}}(\mathbf
    {x},\omega)=0\qquad \mbox{for} \quad r>a
  \end{equation}
  with
 \begin{equation}
  \nabla\cdot\widehat{\mathbf{E}}(\mathbf
      {x},\omega)=0.
 \end{equation}
 
 For each components $\widehat{E}_x,\widehat{E}_y,\widehat{E}_z$ we seek a dependence of type
 \begin{equation}
 \psi(r,\theta)\exp\Big[i\frac{\beta(\omega)\omega}{c}z\Big],
 \end{equation}
 which describes a wave packet with cylindrical symmetry, that propagates along $z$ direction with wave number 
 $$
 \kappa(\omega)=\frac{\beta(\omega)\omega}{c}.
 $$
 Now we assume that
 \begin{equation}
 n_1(\omega)>\beta(\omega)>n_2(\omega).
 \end{equation}
 This assumption makes sense in view to obtain solutions for which optical fiber works as a waveguide. Now it is useful to write eqs $(1.16),(1.17)$ into cylindrical coordinates $(r,\theta,z):$
 \begin{equation}
 \Big(\frac{1}{r}\frac{\partial}{\partial r}r\frac{\partial}{\partial r}+\frac{1}{r^2}\frac{\partial^2}{\partial \theta^2}\Big)\psi(r,\theta)+h^2\psi(r,\theta)=0,\qquad 0\leq r<a
 \end{equation}
\begin{equation}
 \Big(\frac{1}{r}\frac{\partial}{\partial r}r\frac{\partial}{\partial r}+\frac{1}{r^2}\frac{\partial^2}{\partial \theta^2}\Big)\psi(r,\theta)-q^2\psi(r,\theta)=0,\qquad  r>a,
 \end{equation}
 where 
 \begin{equation}
 h=\frac{\omega}{c}\sqrt{n_1^2-\beta^2}
 \end{equation}
\begin{equation}
 q=\frac{\omega}{c}\sqrt{\beta^2-n_2^2}
 \end{equation}
depend on $\omega.$ Now we separate the variables 
\begin{equation}
\psi(r,\theta)=\varphi(r)e^{\pm il\theta}
\end{equation}
where $l=0,1,2,\cdots$ and we obtain that the function $\varphi(r)$ satisfies the Bessel equations
\begin{equation}
r^2\frac{d^2\varphi(r)}{dr^2}+r\frac{d\varphi(r)}{dr}+(h^2r^2-l^2)\varphi(r)=0\qquad 0\leq r<a
\end{equation}
\begin{equation}
r^2\frac{d^2\varphi(r)}{dr^2}+r\frac{d\varphi(r)}{dr}-(q^2r^2+l^2)\varphi(r)=0\qquad r>a.
\end{equation}
It is well-known that the general solution of the first equation is
\begin{equation}
\varphi (r)=c_1J_l(hr)+c_2Y_l(hr),
\end{equation}
where $J_l$ and $Y_l$ are the Bessel functions of first and second kind, respectively. While the general solution of the eq. $(1.27)$ is
\begin{equation}
\varphi (r)=c_3I_l(qr)+c_4K_l(qr)
\end{equation}
where $I_l$ and $K_l$ are the modified Bessel functions of first and second kind respectively. Now we list some properties of these special functions.

For $x\ll 1$ we have that \cite{Gatteschi}
\begin{equation}
J_l(x)\approx \frac{1}{l!}\big(\frac{x}{2}\big)^l,\qquad l=0,1,2,3,\ldots
\end{equation} 
\begin{equation}
Y_0(x)\approx \frac{2}{2\pi}\big(\ln\frac{x}{2}+0.5772\ldots)
\end{equation}
\begin{equation}
Y_l(x)\approx-\frac{(l-1)!}{\pi}\big(\frac{2}{x}\big)^l \qquad l=1,2,3,\ldots
\end{equation}
\begin{equation}
K_0(x)\approx\log\frac{2}{x}-0.5772\ldots
\end{equation}
\begin{equation}
K_l(x)\approx\frac{(l-1)!}{2}\big(\frac{2}{x}\big)^l \qquad l=1,2,3,\ldots.
\end{equation}
While for $x\gg 1$ and $l=0,1,2,3,\ldots$ we have that \cite{Gatteschi}
\begin{equation}
I_l(x)\approx\Big(\frac{1}{2\pi x}\Big)^{1/2}e^x
\end{equation}
\begin{equation}
K_l(x)\approx\Big(\frac{\pi}{2 x}\Big)^{1/2}e^{-x}.
\end{equation}
Since we are interested to solutions which are limited for $r=0$ and which vanish for $r\rightarrow \infty,$ it needs to have \cite{Gatteschi} $c_2=c_3=0.$ 

Since eqs $(1.16),(1.17)$ are not coupled into Cartesian components of the field $\widehat{\mathbf{E}},$ each component can be chosen independently, provided that condition $(1.18)$ is satisfied. Therefore we can assume that $\widehat{\mathbf{E}}$ is polarized in the $y$ axis direction and so we have that the components of $\mathbf{E}$ along the $x$ and $y$ axes are
\begin{equation}
\widehat{E}_x=0
\end{equation} 
\begin{equation}
\widehat{E}_y=\left\{\begin{array}{ll}
A_lJ_l(hr)e^{il\theta +i\kappa z}& \mbox{for}\quad 0\leq r\leq a\\
B_lK_l(qr)e^{il\theta +i\kappa z}&\mbox{for}\quad  r\geq a
\end{array}\right.
\end{equation}
where
$A_l$ and $B_l$ are arbitrary constants. While the component $\widehat{E}_z$ is given by the requirement that the divergence of $\widehat{E}$ is zero. Then because of $(1.18)$ and $(1.19)$ we have that
$$
\widehat{E}_z=\frac{i}{\kappa}\frac{\partial \widehat{E}_y}{\partial y}
$$
and being
$$
\frac{\partial \widehat{E}_y}{\partial y}=\frac{\partial \widehat{E}_y}{\partial r}\sin\theta +\frac{\partial \widehat{E}_y}{\partial \theta}\frac{\cos \theta}{r}
$$
we obtain that
\begin{equation}
\widehat{E}_z=\left\{\begin{array}{ll}
iA_l\frac{h}{\kappa}(J^\prime_l\sin\theta+i\frac{l}{hr}J_l\cos\theta)e^{il\theta +i\kappa z}&\mbox{for}\quad 0\leq r\leq a\\
\\
iA_l\frac{q}{\kappa}(K^\prime_l\sin\theta+i\frac{l}{hr}K_l\cos\theta)e^{il\theta +i\kappa z}&\mbox{for}\quad r\geq a
\end{array}\right.
\end{equation}
where $J^\prime_l$ and $K^\prime_l$ are the first derivative with respect the argument. Now we use the continuity of components $\widehat{E}_y$ and $\widehat{E}_z$ across the core-cladding interface; so we obtain that
\begin{equation}
\frac{A_l}{B_l}=\frac{K_l(qa)}{J_l(ha)}
\end{equation}
and
\begin{equation}
hK_l(qa)J^\prime_l(ha)=qJ_l(ha)K^\prime_l(qa).
\label{interfacciacorecladding}
\end{equation}
Fixed $\omega$ and $l$ Eq. $(\ref{interfacciacorecladding})$ is a transcendental equation in $\beta.$ To study this equation,  we list some other properties of special functions, that is
\begin{equation}
J_l^\prime(x)=\frac{1}{2}(J_{l-1}(x)-J_{l+1}(x))=J_{l-1}(x)-\frac{l}{x}J_l(x)
\end{equation}
\begin{equation}
\frac{2l}{x}J_l(x)=J_{l-1}(x)+J_{l+1}(x)
\end{equation}
and
\begin{equation}
-K^\prime_l(x)=\frac{1}{2}(K_{l-1}(x)+K_{l+1}(x))=K_{l-1}(x)+\frac{l}{x}K_l(x)
\end{equation}
\begin{equation}
-\frac{2l}{x}K_l(x)=K_{l-1}(x)-K_{l+1}(x).
\end{equation}
To study Eq. $(\ref{interfacciacorecladding})$ we fix $x=ha=a\frac{\omega}{c}\sqrt{n_1^2-\beta^2}$ and $V=a\frac{\omega}{c}\sqrt{n_1^2-n_2^2}=\sqrt{(ha)^2+(qa)^2}.$ Therefore eq 
$(1.41)$ becomes
\begin{equation}
f_l(x)=g_l(x)
\label{equazionecorecladding}
\end{equation}
where
\begin{equation}
f_l(x)=x\frac{J_l^\prime(x)}{J_l(x)}+l
\end{equation}
\begin{equation}
g_l(x)=\sqrt{V^2-x^2}\frac{K_l^\prime(\sqrt{V^2-x^2})}{K_l(\sqrt{V^2-x^2})}+l
\end{equation}
So we seek solution of $(\ref{equazionecorecladding})$ such that $0\leq x\leq V.$ 

Let us consider first the function $f_l(x).$ We know that the function $J_l(x)$ has, for $x\geq 0,$ a countable infinite simple zeros $j_{l,r}$ where the index $r=0,1,2,3,\ldots$ is chosen in order to obtain an increasing series of zeros. In the regular interval $j_{l,r}\leq x\leq j_{l,r+1}$ the function $f_l(x)$ is decreasing. In fact
\begin{eqnarray*}
f^\prime_l(x)&=& x\frac{J_l^{\prime\prime}(x)J_l(x)-(J_l^\prime(x))^2}{J_l(x)^2}+\frac{J_l^\prime(x)}{J_l(x)}\\
&=&x\frac{J_l(x)^{\prime\prime}}{J_l(x)}-x\frac{(J_l^\prime(x))^2}{J_l(x)}+\frac{J_l^\prime(x)}{J_l(x)}\\
&=&\frac{1}{xJ_l(x)}[x^2J_l^{\prime\prime}(x)+xJ_l^{\prime}(x)]-x\bigg(\frac{J_l^\prime(x)}{J_l(x)}\bigg)^2
\end{eqnarray*}
Now because of $J_l$ solves the Bessel equation
$$
J_l^{\prime\prime}+\frac{1}{x}J_l^\prime(x)+\bigg(1-\frac{l^2}{x^2}\bigg)J_l(x)=0
$$
we obtain that
\begin{equation}
xf^\prime_l=-x^2+l^2-\bigg(x\frac{J_l^\prime(x)}{J_l(x)}\bigg)^2.
\end{equation}
From $(1.42)$ it follows that
\begin{eqnarray*}
x\frac{J_l^\prime(x)}{J_l(x)}&=&x\frac{J_{l-1}(x)}{J_l(x)}-x\frac{l}{x}\frac{J_{l}(x)}{J_l(x)}\\
&=&-l+x\frac{J_{l-1}(x)}{J_l(x)}
\end{eqnarray*}
Then we obtain that
\begin{eqnarray*}
xf^\prime_l&=&-x^2+l^2-\bigg(-l+x\frac{J_{l-1}(x)}{J_l(x)}\bigg)^2\\
&=&-x^2+l^2-l^2-x^2\bigg(\frac{J_{l-1}(x)}{J_l(x)}\bigg)^2+2 l x\frac{J_{l-1}(x)}{J_l(x)}\\
&=&-x^2+x^2\frac{J_{l-1}}{J_l^2}\bigg[\frac{2l}{x}J_l-J_{l-1}\bigg]\\
&=& -x^2+x^2\frac{J_{l-1}}{J_l^2}\bigg[J_{l-1}+J_{l+1}-J_{l-1}\bigg] \qquad\mbox{by}\quad(1.43)\\
&=&x^2\bigg(\frac{J_{l-1}J_{l+1}}{J_l^2}-1\bigg).
\end{eqnarray*}
From $(1.43)$ , for $x=j_{l,r},$ zero of $J_l(x)$ it follows that $J_{l-1}=-J_{l+1}$ and so we have that
$$
(J_{l-1}J_{l+1})(j_{l,r})=-J_{l+1}^2(j_{l,r}).
$$
Then
$$\lim_{x\rightarrow j_{l,r}^{\pm}}f^\prime_l=-\infty$$
Finally, if we prove that $\Delta_l\Doteq J_{l-1}J_{l+1}-J_l^2$ is negative we obtain that the function $f_l$ is decreasing. We note that for $x=j_{l,r},$ $\Delta_l=-J_{l+1}^2<0.$ Moreover if we prove that the possible maximum of the function $\Delta_l$ are negative we get our goal:
\begin{eqnarray*}
\Delta_l^\prime &=& J_{l-1}^\prime J_{l+1}+J_{l+1}^\prime J_{l-1}-2J_lJ^\prime_l\\
&=& \frac{1}{2}[J_{l-2}-J_l]J_{l+1}+\frac{1}{2}[J_l-J_{l+2}]J_{l-1}-J_l[J_{l-1}-J_{l+1}]\qquad \mbox{by}\quad (1.42)\\
&=& \frac{1}{2}[J_{l-2}+J_l]J_{l+1}-\frac{1}{2}[J_l+J_{l+2}]J_{l-1}\\
&=&\frac{1}{2}\Big[\frac{2(l-1)}{x}J_{l-1}\Big]J_{l+1}-\frac{1}{2}\Big[\frac{2(l+1)}{x}J_{l+1}\Big]J_{l-1}\qquad \mbox{by}\quad(1.43)\\
&=&\frac{l}{x}J_{l-1}J_{l+1}-\frac{1}{x}J_{l-1}J_{l+1}-\frac{l}{x}J_{l-1}J_{l+1}-\frac{1}{x}J_{l-1}J_{l+1}\\
&=&-\frac{2}{x}J_{l-1}J_{l+1}
\end{eqnarray*}

Now we observe that the zeros of the function $J_{l-1}$ are shifted to the left with respect to the zeros of $J_l$ so that we have
$$
0<j_{l-1,1}<j_{l,1}<j_{l-1,2}<j_{l,2}<\cdots<j_{l-1,r}<j_{l,r}<j_{l-1,r+1}<j_{l,r+1}.
$$
Because of $J_{-1}=J_1$ and thanks to $(1.43)$ we have that
\begin{eqnarray}
f_0(x)&=&-x\frac{J_1(x)}{J_0(x)}\\
f_l(x)&=&x\frac{J_{l-1}(x)}{J_l(x)},\qquad l=0,1,2,\cdots
\end{eqnarray}
Therefore we can define the behavior of the function $f_l(x):$ in the intervals of regularity $j_{l,r}<x<j_{l,r+1}$ $f_l(x)$ decreases from $+\infty$ to $-\infty.$

For $l\geq 1,$ $f_l(x)$ has simple zeros in $j_{l-1,r};$ so it is positive for $j_{l,r}<x<j_{l-1,r+1}$ and negative for $j_{l-1,r+1}<x<j_{l,r+1}.$ For $l=0,$ in the intervals $j_{0,r}<x<j_{0,r+1}$ the function $f_0(x)$ changes sign in $x=j_{1,r}.$

Now we consider the function $g_l(x)$ that we rewrite, thanks to Eqs $(1.45)$ and $K_{-1}=K_1,$ as
\begin{eqnarray}
g_0(x)&=&-\sqrt{V^2-x^2}\frac{K_1(\sqrt{V^2-x^2})}{K_0\sqrt{V^2-x^2}}\\
g_l(x)&=&-\sqrt{V^2-x^2}\frac{K_{l-1}(\sqrt{V^2-x^2})}{K_l\sqrt{V^2-x^2}},\quad l=1,2,3,\ldots
\end{eqnarray}
By repeating the same arguments used for the function $f_l(x),$ we prove that the function $g_l(x)$ is increasing in the interval $0\leq x\leq V.$ Moreover thanks to $(1.34)$ we have that, for $x\ll 1,$
$$
\frac{K_{l-1}(x)}{K_l(x)}\approx x \qquad l=0,1,2,3,\ldots
$$
and so approximately we have that $g_l(V)=0.$ So for the fiber mode $l=0$ equation $(\ref{equazionecorecladding})$ has $r$ solutions when
\begin{equation}
j_{1,r-1}<V<j_{1,r}, \quad r=1,2,\ldots,\quad j_{1,0}=0,
\end{equation}
while for each fiber mode $l\geq 1,$ it admits $r$ solutions when 
\begin{equation}
j_{l-1,r}<V<j_{l-1,r+1},\qquad r=1,2,\ldots
\end{equation}
In conclusion for
\begin{equation}
V=a\frac{\omega}{c}\sqrt{n_1^2-n_2^2}<j_{l,1}
\end{equation}
it propagates only fiber modes up to mode $l$ included. In particular for
\begin{equation}
V=a\frac{\omega}{c}\sqrt{n_1^2-n_2^2}<2.405,
\end{equation}
being $j_{0,1}\approx 2.405,$ the first zero of $J_0(x)$, there is one fiber mode of propagation for $l=0$ and so for values of $\omega$ such that it is valid eq $(1.57)$ the fiber is a single-mode fiber.

\subsection{Derivation of NLSE}
In this section we derive a basic equation that governs propagation of optical pulses in nonlinear dispersive fibers. The starting point is the wave equation $(1.14).$ By means of relations
$$
\mathbf{P}=\mathbf{P}_L+\mathbf{P}_{NL}
$$
and
$$
\nabla\times(\nabla\times\mathbf{E})=\nabla(\nabla\cdotp\mathbf{E})-\Delta\mathbf{E}=-\Delta\mathbf{E}
$$
it can be written in the form
\begin{equation*}
 \Delta \mathbf{E}-\frac{1}{c^2}\frac{\partial^2 \mathbf{E}}{\partial t^2}=\mu_0\frac{\partial^2\mathbf{P}_L}{\partial t^2}+\mu_0\frac{\partial^2\mathbf{P}_{NL}}{\partial t^2}.
\end{equation*}

As first simplifying assumption $\mathbf{P}_{NL}$ is treated as a small perturbation to $\mathbf{P}_L;$ this is justified because nonlinear changes in the refractive index are $<10^{-6}$ \cite{Agrawal}. Second, the optical field is assumed to maintain its polarization along the fiber length; this is not really the case, unless polarization-maintaining fibers are used. Third, the optical field is assumed to be quasi-monochromatic, i.e., the pulse spectrum, centered at $\omega_0,$ is assumed to have a spectral width $\Delta\omega$ such that $\Delta\omega/\omega_0 \ll 1.$ The value of $\omega_0$ is given to be  $\sim 10^{15}s^{-1}$ \cite{Agrawal}, so the last assumption is valid for pulses as short as $0.1 ps,$ for example. To study eq $(1.14)$ we pursue the  approach  described in \cite{Agrawal}. Let us write the electric field as
\begin{equation}
\mathbf{E}(\mathbf{r},t)=\frac 1 2 \hat{y}[E(\mathbf{r},t)\exp(-i\omega_0 t)+c.c.]
\end{equation}
where $\hat{y}$ is the polarization unit vector, $E(\mathbf{r},t)$ is a slowly varying complex function of time (relatively to the optical period). In a similar way, we write the polarization as 
\begin{equation}
\mathbf{P}_L(\mathbf{r},t)=\frac 1 2 \hat{y}[P_L(\mathbf{r},t)\exp(-i\omega_0 t)+c.c.]
\end{equation}
\begin{equation}
\mathbf{P}_{NL}(\mathbf{r},t)=\frac 1 2 \hat{y}[P_{NL}(\mathbf{r},t)\exp(-i\omega_0 t)+c.c.].
\end{equation}
The linear component $P_L$ can be obtained by substituting eq $(1.59)$ in the expression  $
 \mathbf{P}_L(\mathbf{r},t)=\epsilon_0\int_{-\infty}^{+\infty}\chi^{(1)}(t-t^\prime)\mathbf{E}(\mathbf{r},t^\prime)dt^\prime
 $  and making use of eq $(1.58)$ we have that 
$$
 \frac 1 2 \hat{y}[P_L(\mathbf{r},t)\exp(-i\omega_0 t)+c.c.]=\frac 1 2 \hat{y} \int_{-\infty}^{+\infty}\chi^{(1)}(t-t^\prime)[E(\mathbf{r},t^\prime)\exp(-i\omega_0 t^\prime)+c.c.]dt^\prime
 $$
and so
$$
P_L(\mathbf{r},t)=\epsilon_0 \int_{-\infty}^{+\infty}\chi^{(1)}(t-t^\prime) E(\mathbf{r},t)\exp[i\omega_0 (t-t^\prime)]dt^\prime.
$$

To evaluate $P_{NL}(\mathbf{r},t)$ we make another considerable simplification that occurs if the nonlinear response is assumed to be instantaneous, so that the time dependence of $\chi^{(3)}$ is given by the product of three delta function of the form $\delta(t-t_i).$ From this eq. $(1.13)$ reduces to
\begin{equation}
\mathbf{P}_{NL}(\mathbf{r},t)=\epsilon_0\chi^{(3)}( \mathbf{E}(\mathbf{r},t)\cdot\mathbf{E}(\mathbf{r},t))\mathbf{E}(\mathbf{r},t)
\end{equation}
and thus $ \mathbf{P}_{NL}(\mathbf{r},t)=\epsilon_0\chi^{(3)}| \mathbf{E}(\mathbf{r},t)|^2\mathbf{E}(\mathbf{r},t).$ 

When eq. $(1.58)$ is substituted in eq $(1.61),$ $\mathbf{P_{NL}}(\mathbf{r},t)$ is found  to have a term oscillating at $\omega_0$ and another term oscillating at the third-harmonic frequency $3\omega_0.$ The letter term is generally negligible in optical fiber.  Because of this fact and $
\mathbf{P}_{NL}(\mathbf{r},t)=\frac 1 2 \hat{y}[P_{NL}(\mathbf{r},t)\exp(-i\omega_0 t)+c.c.]
$ it is found that \cite{Agrawal}
$$
P_{NL}(\mathbf{r},t)\approx \epsilon_0\epsilon_{NL}E(\mathbf{r},t)
$$
where the nonlinear dielectric constant is defined as $\epsilon_{NL}\Doteq\frac 3 4 \chi^{(3)}|E(\mathbf{r},t)|^2.$  Because of the slowly varying envelope approximation and the perturbative nature of $P_{NL},$ $\epsilon_{NL}$ is treated as a constant during the derivation of the propagation equation. Therefore we can work in the Fourier domain and so the wave equation for the Fourier transform of the slowly varying amplitude $E(\mathbf{r},t)$  it is found to be 
\begin{equation}
\Delta\widehat{E}+\epsilon(\omega)k_0^2\widehat{E}=0,
\end{equation}
where $k_0=\omega/c$ , 
$\epsilon(\omega)=1+\widehat{\chi}_1(\omega)+\epsilon_{NL}$ and $$\widehat{E}(\mathbf{r},\omega-\omega_0)=\int_{-\infty}^{+\infty}E(\mathbf{r},t)\exp[i(\omega-\omega_0)t]dt .$$
The dielectric constant $\epsilon(\omega)$ takes into account both linear and nonlinear effects of refractive index and absorption. In our perturbative approach it is approximately assumed as
$$
\epsilon\approx n^2 +2n\epsilon_1
$$
where $\epsilon_1$ is a small perturbation given by \cite{Agrawal}
\begin{equation}
\epsilon_1=\tilde{n}|E|^2+\frac{i\alpha}{2k_0},
\end{equation}
where $\tilde{n}$  is a measure of the fiber nonlinearity and $\alpha$ takes into account the fiber absorption.

Equation $(1.62)$ can be solved by using the method of separation variables.  If we assume a solution of the form
$$
\widehat{E}(\mathbf{r},\omega-\omega_0)=F(x,y)\widehat{A}(z,\omega-\omega_0)\exp(i\beta_0z),
$$ 
where $\widehat{A}(z,\omega)$ is slowly varying function of $z$ and $\beta_0$ is the wave number to be determined later, Eq. $(1.62)$ leads 
$$
\frac{\partial^2 F}{\partial x^2}\widehat{A}+\frac{\partial^2 F}{\partial y^2}\widehat{A}+F\frac{\partial^2 \widehat{A}}{\partial z^2}+F 2i\beta_0\frac{\partial^2 \widehat{A}}{\partial z^2}-F\beta_0\widehat{A}+\epsilon(\omega)k_0^2F\widehat{A}+\tilde{\beta}F\widehat{A}-\tilde{\beta}F\widehat{A}=0
$$
where we introduced the coefficient $\tilde{\beta}.$ Because of $\widehat{A}$ is assumed to be a slowly varying function of $z,$ i.e. $\partial^2\widehat{A}/\partial z^2\ll\partial \widehat{A}/\partial z,$ we can neglect the second derivative $\partial^2\widehat{A}/\partial z^2.$ Then we obtain that
$F$ and $\widehat{A}$ solve the following equations:
\begin{equation}
\frac{\partial^2 F}{\partial x^2}+\frac{\partial^2 F}{\partial y^2}+[\epsilon(\omega)k_0^2-\tilde{\beta}^2]F=0,
\end{equation}
\begin{equation}
2i\beta_0\frac{\partial\widehat{A}}{\partial z}+(\tilde{\beta}^2-\beta_0^2)\widehat{A}=0.
\end{equation}
 The coefficient $\tilde{\beta}$ is determined by solving the eigenvalue equation $(1.64).$  This equation can be solved by using the first order perturbation theory \cite{Agrawal}; we first replace $\epsilon $ with the refractive index $n^2$ and obtain the modal distribution $F(x,y)$ as we claimed in the section $1.5.2.$ Then we include the effect of $\epsilon_1$ in Eq. $(1.64.)$ In the first order perturbation theory, $\epsilon_1$ does not affect the modal distribution $F(x,y).$ However the eigenvalue $\tilde{\beta}$ it is found \cite{Agrawal} to be
\begin{equation}
\tilde{\beta}(\omega)=\beta(\omega)+\Delta \beta,
\end{equation}
and $\Delta \beta$ it is found  \cite{Agrawal} to be
$$
\Delta \beta=\frac{k_0\int_{-\infty}^{+\infty}\epsilon_1|F(x,y)|^2 dx dy}{\int_{-\infty}^{+\infty}|F(x,y)|^2 dx dy}
$$
In view af all this the electric field $\mathbf{E}(\mathbf{r},t)$ can be written as
\begin{equation}
\mathbf{E}(\mathbf{r},t)=\frac 1 2 \hat{y}\{F(x,y)A(z,t)\exp[i(\beta_0z-\omega_0t)]+c.c.\},
\end{equation}
where $A(z,t)$ is the slowly varying pulse envelope. The Fourier transform $\widehat{A}(z,\omega-\omega_0)$ of $A(z,t)$ satisfies eq. $(1.65),$ which, from eq $(1.66)$ and making use of the approximation $\tilde{\beta}^2-\beta_0^2=(\tilde{\beta}+\beta_0)(\tilde{\beta}-\beta_0)\approx 2\beta_0(\tilde{\beta}-\beta_0),$  becomes
\begin{equation}
\frac{\partial \widehat{A}}{\partial z}=i[\beta(\omega)+\Delta \beta -\beta_0]\widehat{A}.
\end{equation}
At this point, recalling the section $1.3,$ the function $\beta(\omega)$ can be expanded in a Taylor series about the carrier frequency $\omega_0$ as
$$
\beta(\omega)=\beta_0+\beta_1(\omega-\omega_0)+\frac 1 2 \beta_2(\omega-\omega_0)^2+\cdots,
$$
where 
$$\beta_m=\Big(\frac{d^m\beta}{d\omega^m}\Big)_{\omega=\omega_0}\qquad (m=0,1,2,\cdots).$$
The cubic and higher-order terms in this expansion are generally negligible if the spectral width $\Delta \omega\ll \omega_0,$ that is an approximation that our approach provide. For this reason eq. $(1.65)$ becomes
\begin{equation}
\frac{\partial \widehat{A}}{\partial z}=i[(\omega-\omega_0)\beta_1+\frac 1 2 (\omega-\omega_0)^2 \beta_2 +\Delta \beta]\widehat{A}
\end{equation}
Now, we can go back to the time domain by taking the inverse Fourier transform of eq. $(1.69),$ and we obtain the propagation equation for $A(z,t);$ during the Fourier transform operation, $\omega-\omega_0$ is replaced by the differential operator $i(\partial/\partial t).$ The resulting equation for $A(z,t)$ becomes
\begin{equation}
\frac{\partial A}{\partial z}=-\beta_1\frac{\partial A}{\partial t}-\frac{i\beta_2}{2}\frac{\partial^2 A}{\partial t^2}+i\Delta \beta A.
\end{equation}
The term with $\Delta \beta$ includes the effect of fiber loss and nonlinearity. From eq.$(1.63)$ we obtain that
$$
\Delta \beta=i\frac \alpha 2 +\frac{k_0\tilde{n}|A|^2\int_{-\infty}^{+\infty}|F(x,y)|^4 dx dy}{\int_{-\infty}^{+\infty}|F(x,y)|^2 dx dy};
$$
so the result is
\begin{equation}
\frac{\partial A}{\partial z}+\beta_1\frac{\partial A}{\partial t}+\frac{i\beta_2}{2}\frac{\partial^2 A}{\partial t^2}+\frac \alpha 2 A=i\gamma|A|^2 A,
\end{equation}
where $\gamma$ is a nonlinear parameter defined as 
$$
\gamma\Doteq\frac{\tilde{n}\omega_0}{c A_{\mbox{eff}}},
$$
where 
$$
A_{\mbox{eff}}\Doteq\frac{\int_{-\infty}^{+\infty}|F(x,y)|^2 dx dy}{\int_{-\infty}^{+\infty}|F(x,y)|^4 dx dy}
$$

\vspace{.7cm}

Eq. $(1.71)$ describes propagation of picosecond optical pulse in single-mode fibers. It is often referred as the nonlinear Schr$\ddot{o}$dinger (NLS) equation because it can be reduced to that form under certain conditions. It includes the effects of fiber losses through $\alpha,$ of chromatic dispersion through $\beta_1$ and $\beta_2,$ and of fiber nonlinearity through $\gamma.$ The physical significance of the parameters $\beta_1$ and $\beta_2$ has been discussed. Briefly, the pulse envelope moves at the group velocity $v_g\equiv 1/\beta_1$ while the effects of group-velocity dispersion (GVD) are governed by $\beta_2.$ The GVD parameter can be positive or negative depending on whether the wavelength $\lambda$ is below or above the zero-dispersion wavelength $\lambda_D$ of the fiber. In the anomalous-dispersion regime($\lambda>\lambda_D$), $\beta_2$ is negative, and the fiber can support optical solitons. In standard silica fibers, $\beta_2\sim 50 ps^2/Km$ in the vicible region but becomes close to $-20 ps^2/Km$ near wavelengths $\sim 1.5\mu m,$ the change in sign occurring in the vicinity of $1.3\mu m.$

We conclude this chapter by showing that equation $(1.71)$ is actually a Nonlinear Schrodinger Equation. In fact, in a time frame moving with the signal group velocity, where we introduce the change of variable $T=t-z/v_g,$ eq.$(1.71)$ becomes
$$
i\frac{\partial A}{\partial z}=-i\frac \alpha 2 A+\frac{\beta_2}{2}\frac{\partial^2 A}{\partial T^2}-\gamma|A|^2 A.
$$
Moreover, let us introduce the amplitude $U(z,T)$ as
$$
A(z,T)=\exp[-\alpha z/ 2]U(z,T)
$$
in this way we obtain that eq. $(1.71)$ becomes
\begin{equation}
i\frac{\partial U}{\partial z}(z,T)=\frac{\beta_2}{2}\frac{\partial^2 U}{\partial T^2}(z,T)-\gamma e^{-\alpha z}|U(z,T)|^2 U(z,T)
\end{equation}

\chapter{Nonlinear Schr\"odinger Equation}
\section{Introduction}
In chapter $1$ we obtained the non linear Schrodinger equation that governs propagation of optical pulses inside single-mode fiber:
\begin{equation}
i\frac{\partial A(z,t)}{\partial z}=-i\frac{\alpha}{2}A(z,t)+\frac{\beta_2}{2}\frac{\partial^2 A(z,t)}{\partial t^2}-\gamma |A(z,t)|^2A(z,t),
\label{ing}
\end{equation}
where $A$ is the slowly varying amplitude of the pulse envelope and $t$ is measured in a frame of reference moving with the pulse at the group velocity $v_g.$ The three terms on the right-hand side of Eq. $(\ref{ing})$ govern, respectively, the effects of fiber loss ($\alpha >0$), dispersion ($\beta_2\in \RR\backslash\{0\}$) and non linearity ($\gamma>0$) on pulses propagating inside optical fibers. Depending on the initial width $T_0$ and the peak power $P_0$ of the incident pulse, either dispersive or non linear effects may dominate along the fiber. It is useful to introduce two length scales, known as the dispersion length $L_D$ and the nonlinear length $L_{NL}.$ Depending on the relative magnitude of $L_D,\quad L_{NL},$ and the fiber length $L,$ pulses can evolve quite differently.

Let us consider a time scale normalized to the input width $T_0$ as
$$\tau=\frac{t}{T_0}.$$
In addition, we introduce a normalized amplitude $U$ as
$$A(z,\tau)=\sqrt{P_0}\exp(-\alpha z/2)(U(z,\tau)),$$
where $P_0$ is the peak power of the incident pulse. We now take into consideration a space scale normalized to the fiber length as
$$\zeta=\frac{z}{L},$$
where $L$ is the fiber length. Thus, it turns out that $U(\zeta,\tau)$ satisfies
\begin{equation}
i\frac{\partial U(\zeta,\tau)}{\partial \zeta}=\frac{L}{L_D}\frac{\mbox{sgn}\beta_2}{2}\frac{\partial^2  U(\zeta,\tau)}{\partial \tau^2}- e^{-\alpha L\zeta}\frac{L}{L_{NL}} |U(\zeta,\tau)|^2U(\zeta,\tau),
\label{ing1}
\end{equation}
where $\mbox{sgn}\beta_2=\pm 1$ depending on the sign of the coefficient $\beta_2$ and
$$
L_D=\frac{T_0^2}{|\beta_2|}\qquad L_{NL}=\frac{1}{\gamma P_0}.
$$
The dispersion length $L_D$ and the nonlinear length $L_{NL}$ provide the length scales at which dispersive or nonlinear effects become important for pulse evolution. Depending on the relative magnitudes of $L,L_D,L_{NL},$ we notice from Eq. $(\ref{ing1})$ that when $L$ is longer or comparable to both $L_D$ and $L_{NL},$ dispersion and nonlinearity act together as the pulse propagates along the fiber. We observe that Eq. $(\ref{ing1})$ is now dimensionless; thus, we refer to this particular propagation regime, being $L\sim L_D\sim L_{NL},$ as the following mathematical model 
 
\begin{equation}
\left \{\begin{array}{l}
iv_z(z,t)+ v_{tt}(z,t) +C_1e^{-C_2z}|v(z,t)|^2v(z,t)=0\\
v(0,t)=\varphi(t)
\end{array} \right.
\label{modello}
\end{equation}
where $C_1=\pm 1,\; C_2>0,\;z$  and $t$ are \textit{real} variables. The sign of $C_1$ depends on the sign of the  parameter $\beta_2$ which represents dispersion of the group velocity and refers to normal dispersion if $\beta_2>0$ or anomalous dispersion if $\beta_2<0.$ While the coefficient $C_2$ takes into account fiber loss. In what follows we show some well-known results about uniqueness and local and global existence of solutions of the problem $(\ref{modello}),$ referring to \cite{Cazenave},\cite{Taylor}. The physical sense of equation $(\ref{ing})$ suggests to study problem $(\ref{modello})$ in the Hilbert space $H^1(\RR);$ as matter of fact, in chapter $1$ we pointed out that the function $|v|^2$ in eq. $(\ref{ing}1)$ has dimension of physical power, that is $[|v|^2]=W;$ the dimension of power is energy divided by time. The SI unit of power is the watt ($W$), which is equal to one joule per second.  Therefore, being power the first temporal derivative of the energy, it is natural to consider the Hilbert space $H^1(\RR)$ with respect time variable $t.$

In addition, we present also well-known results about uniqueness and existence of solutions of problem $(\ref{modello})$ in the Hilbert space $H^2(\RR);$ this will allow us to consider a class of functions such that if $v$ is in it then the function $t\mapsto t^2v(z,t)$ belongs to $L^2(\RR)$ for every $z$ in a suitable interval of $\RR.$

\section{Sobolev space}
Throughout these notes, we consider $L^p$ spaces of complex-valued functions (unless the contrary is explicitly stated, all functions we will deal with are complex-valued). Being $\Omega$ an open set of $\RR,$ $L^p(\Omega)$ denotes the space of (class of) measurable functions $u:\Omega \rightarrow \CC$ such that $||u||_{L^p}<\infty $ with
$$
||u||_{L^p}=\left\{ \begin{array}{ll}
\Bigg(\int_{\Omega}|u(t)|^p dt\Bigg)^{1/p}& \mbox{if}\quad p\in[0,\infty[\\
\sup\mbox{ess}_{\Omega}|u(t)|& \mbox{if}\quad p=\infty.
\end{array}\right.
$$
The quantity $L^p(\Omega)$ denotes a Banach space and $L^2(\Omega)$ is a Hilbert space when equipped with the scalar product
$$
(u,v)_{L^2}=\int_{\Omega}u(t)\overline{v(t)}dt.
$$

\vspace{.7cm}
We refer to $q$ as the conjugate of $p$ such as $1/p + 1/q=1.$

\begin{pro}[\bf Holder inequality]
Let $p\in[0,\infty]$ and let $q$ its conjugate. If $u\in L^p(\Omega)$ and $v\in L^q(\Omega)$ then we have that
\begin{equation}
\int_{\Omega}|u(t)v(t)|dt\leq||u||_{L^p}||v||_{L^q}.
\end{equation}
\end{pro}

Let $I$ be an interval of $\RR,$ and $X$ a Banach space; for $p\in[0,\infty]$ we denote with $L^p(I,X)$ the space of (class of) measurable functions $f:I\rightarrow X$ such that the function $z\mapsto||f(z)||$ belongs to $L^p(I).$ For $f\in L^p(I,X),$ one defines
$$
||u||_{L^p(I,X)}=\left\{ \begin{array}{ll}
\Bigg(\int_{\Omega}||u(t)||^p dt\Bigg)^{1/p}& \mbox{if}\quad p\in[0,\infty[\\
\\
\sup\mbox{ess}_{\Omega}||u(t)||& \mbox{if}\quad p=\infty.
\end{array}\right.
$$

\vspace{1cm}
Let $\Omega$ be an open subset of $\RR.$ Let $u\in C^m(\Omega),\;m\geq 1.$  Let $\alpha\in \NN $ and $D^\alpha=\frac{\partial^\alpha}{\partial t^\alpha},$ by integration by parts we have that
\begin{equation}
\int_{\Omega}D^\alpha u\varphi=(-1)^\alpha\int_\Omega uD^\alpha\varphi,
\end{equation}
for all $\varphi \in C^m_c(\Omega),$ that is for all functions compactly supported with all their derivatives up to order $m$. We note that both integrals in $(2.4)$ are well-defined since $D^\alpha u\varphi \in C_c(\Omega)$ and $uD^\alpha\varphi \in C_c(\Omega).$

A function $u$ defined almost everywhere on $\Omega$ is said to be {\textit {locally integrable}} on $\Omega$ provided $u\in L^1(U)$ for every open $U\Subset\Omega,$ that is for every $U$ such that $\overline{U}\subset \Omega$ and $\overline{U}$ is compact. Referring to eq. $(2.4),$ the right-hand side makes sense as soon as $f\in L_{loc}^1(\Omega)$ and the left-hand side makes sense as soon as $D^\alpha u\in L_{loc}^1(\Omega).$ This motives the following definitions of {\textit weak} derivative.
\begin{definition}
Let $u\in L_{loc}^1(\Omega)$ and $\alpha \in \NN .$ We say that $D^\alpha u\in L_{loc}^1(\Omega)$ if there exists $u_\alpha\in L_{loc}^1(\Omega)$ such that
\begin{equation}
\int_{\Omega}u_\alpha\varphi=(-1)^\alpha\int_\Omega uD^\alpha\varphi,
\end{equation}
for all $\varphi \in C^\alpha_c(\Omega).$ Such a function $u_\alpha$ is then unique and we set $D^\alpha u=u_\alpha .$ If $u_\alpha \in L_{loc}^p(\Omega)$(respectively, $u\in L^p(\Omega)$) for some $1\leq p\leq \infty,$ we say that $D^\alpha u\in L_{loc}^p(\Omega)$ (respectively, $D^\alpha u\in L^p(\Omega)$).
\end{definition}

For $m\in \NN$ and $p\in[0,\infty]$ the Sobolev spaces are defined as follows.
\begin{equation}
W^{m,p}(\Omega)\Doteq \{u\in L^p(\Omega):\quad D^\alpha u\in L^p(\Omega)\;\mbox{for}\; \alpha \leq m \}
\end{equation}
For $u\in W^{m,p}(\Omega),$ we set
$$
||u||_{W^{m,p}}= \sum_{\alpha\leq m}||D^\alpha u||_{L^p},
$$
which defines a norm on $W^{m,p}(\Omega).$ We set $H^m(\omega)=W^{m,2}(\Omega),$ and we equip $H^m(\Omega)$ with the scalar product
$$
(u,v)_{H^m}=\sum_{\alpha\leq m}\int_\Omega(D^\alpha u,D^\alpha v),
$$
which defines on $H^m(\Omega)$ the norm
$$
||u||_{H^m}=\Big(\sum_{\alpha\leq m}||D^\alpha u||_{L^2}^2\Big)^{\frac{1}{2}},
$$
which is equivalent to the norm $||\cdot||_{W^{m,2}}.$

We have \cite{Adams}
\begin{pro}
$W^{m,p}(\Omega)$ is a Banach space and $H^m(\Omega) $ is a Hilbert space. If $p<\infty,$ then $W^{m,p}(\Omega)$ is separable, and if $1<p<\infty,$ then $W^{m,p}(\Omega)$ is reflexive.
\end{pro}

We now define the subspaces $W_0^{m,p}(\Omega)$ for $p<\infty.$ Formally, $W_0^{m,p}(\Omega)$ is the subspace of functions of $W^{m,p}(\Omega)$ that vanish on $\partial \Omega$ together with their derivatives up to order $m-1.$ 
\begin{definition}
Let $1\leq p<\infty$ and let $m\in \NN.$ We denote by $W_0^{m,p}(\Omega)$ the closure of $C^\infty_c(\Omega)$ in $W^{m,p}(\Omega),$ and we set $H^m_0(\Omega)=W_0^{m,2}(\Omega).$
\end{definition}

\vspace{1cm}
When $\Omega=\RR ,$ one can characterize the space $W^{m,p}(\RR)$ by means of the Fourier transform which is defined by
$$
\mathcal{F}u(\omega)=\hat{u}(\omega)=\int_{\RR}u(t)e^{-2\pi it\omega}dt
$$
when $u\in L^1(\RR).$ It is clear that $\mathcal{F}:L^1(\RR)\rightarrow L^\infty(\RR).$ Let us define the Schwartz space functions:
$$
\mathcal{S}(\RR)=\{u\in C^\infty(\RR):t^\beta D^\alpha u\in L^\infty(\RR)\quad \mbox{for all}\quad \alpha, \beta \geq 0\}
$$
where $\alpha,\beta$ are positive integer. The Fourier transform verify \cite{Taylor}
$$
\mathcal{F}:\mathcal{S}(\RR)\longrightarrow \mathcal{S}(\RR).
$$

We define the inverse of the Fourier transform $\mathcal{F}^*$ by
$$
\mathcal{F}^*(\hat{u})(t)=\int_{\RR}\hat{u}(\omega)e^{2\pi i\omega t}d\omega.
$$
It is clear that $\mathcal{F}^*$ satisfies
$$
(\mathcal{F}u,v)=(u,\mathcal{F}^*v),
$$
for $u,v\in \mathcal{S}(\RR).$
 From this and the inversion formula it follows that, for $u,v\in \mathcal{S}(\RR),$
$$
(\mathcal{F}u,\mathcal{F}v)=(u,v)=(\mathcal{F}^*u,\mathcal{F}^*v).
$$
Thus $\mathcal{F}$ and $\mathcal{F}^*$ extend uniquely from $\mathcal{S}(\RR)$ to isometries on $L^2(\RR)$ and are inverse to each other. Thus we have the Plancherel theorem \cite{Taylor}:
\begin{pro}
The Fourier transform
$$
\mathcal{F}:L^2(\RR)\longrightarrow L^2(\RR)
$$
is unitary, with inverse $\mathcal{F}^*.$
\end{pro}
Furthermore we have that \cite{Cazenave}
\begin{pro}
Let $u\in L^2(\RR)$ and $\alpha \in\NN .$ Then $D^\alpha u\in L^2(\RR)$ if and only if $|\cdot|^\alpha \widehat{u}\in L^2(\RR).$ Moreover, $\mathcal{F}(D^\alpha u)(\omega)=(2\pi i)^\alpha\omega^\alpha\widehat{u}(\omega),$ where $\omega\in \RR .$ In particular $||D^\alpha(u)||_{L^2}=(2\pi)^\alpha||\;|\cdot|\widehat{u}||_{L^2}.$
\end{pro}

\vspace{.7cm}

A tempered distribution is a continuous linear functional
$
w:\mathcal{S}(\RR)\rightarrow\CC.
$
The space $\mathcal{S}(\RR)$ has a topology, determined by the seminorms
$$
p_k(u)=\sum_{j=0}^k\sup_{t\in \RR}\langle t \rangle^k|D^ju(t)|,
$$
where $\langle t \rangle=(1+|t|^2)^{1/2}.$ The distance function
$$
d(u,v)=\sum_k^\infty 2^{-k}\frac{p_k(u-v)}{1+p_k(u-v)}
$$
makes $\mathcal{S}(\RR)$ a complete metric space; with such a topology it is a Frech\'et space. For a linear map $w:\mathcal{S}(\RR)\rightarrow\CC$ to be continuous, it is necessary and sufficient that, for some $k$ and $C,$
$$
w(u)\leq Cp_k(u)\qquad \mbox{for all}\qquad u\in \mathcal{S}(\RR).
$$

The set of all continuous linear functional on $\mathcal{S}(\RR)$ is denoted $\mathcal{S}^\prime(\RR)$ and is called the space of tempered distributions. In view of all this, we can produce the following definition
\begin{definition}
Given $m\in  \NN ,$
$$
H^m(\RR)=\{u\in\mathcal{S}^\prime(\RR):\quad (1+|\cdot|^2)^{\frac{m}{2}}\widehat{u}\in L^2(\RR) \}.
$$
\end{definition}

For these spaces we have the following embedding result \cite{Cazenave}
\begin{pro}
Let $m\geq 1$ then $H^m(\RR)\hookrightarrow L^\infty(\RR). $
\end{pro} 

\section{The Schrodinger Operator}
The operator $\Delta=D^2=\partial^2/\partial t^2$ acts on $L^2(\RR).$  It is well-known that $\Delta$ is self-adjoint on $L^2(\RR)$ and that $(\Delta u,u)\leq 0;$ this implies that the operator $A=i\Delta$ generates the group of isometries on $L^2(\RR),$ that we denote as $(\mathcal{T}(z))_{z\in \RR}.$ 

Given $\varphi\in L^2(\RR),$ it is well-known that the linear problem
$$
\left\{\begin{array}{ll}
u\in C(\RR,L^2(\RR))\cap C^1(\RR,(D(\Delta)^*)\\
iu_z+ u_{tt}=0 & \mbox{for all}\quad u\in (D(\Delta))^*\\
u(0)=\varphi
\end{array}\right.
$$
where $D(\Delta)=\{u\in H_0^1(\RR),u_{tt}\in L^2(\RR)\}$ and $(D(\Delta))^*$ is the dual space of $D(\Delta),$ posses only one solution given by $u(z)=\mathcal{T}(z)\varphi.$

The quantity $\mathcal{T}(z)$ can be expressed explicitly in Fourier domain. Indeed, given $\varphi \in \mathcal{S}(\RR),$ and given $u\in C^{\infty}(\RR,\mathcal{S}(\RR))$ to be defined by
$$
\widehat{u(z)}(\omega)=e^{-4\pi^2 i|\omega|^2z}\widehat{\varphi}(\omega)\quad \mbox{for all} \; z\;,\omega \in \RR.
$$
 We have $i\hat{u}_z-4\pi^2|\omega|^2\hat{u}=0$ in $\RR\times\RR ,$ and so $iu_z+u_{tt}=0$ in $\RR\times \RR. $ Since $u(0)=\varphi,$  we deduce that $u(z)=\mathcal{T}(z)\varphi.$ Thus, we obtain that
\begin{equation}
\mathcal{F}(\mathcal{T}(z)\varphi)(\omega)=e^{-4\pi^2i|\omega|^2z}\hat{\varphi}(\omega)
\label{gruppodischrodingerfourier}
\end{equation}
for all $\varphi \in \mathcal{S}(\RR),\quad z\in \RR,\quad \omega \in \RR.$

Given $z\neq 0,$ we define the function $K_z$ as
$$
K_z=\Big(\frac{1}{4\pi i t}\Big)^{1/2}e^{\frac{it^2}{4z}}\qquad \mbox{for}\quad t\in \RR.
$$
Since $\widehat{K_z}=e^{-i4\pi^2\omega^2 t},$ it easily follows  from $(\ref{gruppodischrodingerfourier})$ that, for all $\varphi \in \mathcal{S}(\RR),$\\ $\mathcal{T}(z)\varphi=K_z\star \varphi;$ i.e.
$$
\mathcal{T}(z)\varphi(t)=(4\pi i z)^{-1/2}\int_{\RR}e^{\frac{i|t-s|^2}{4z}}\varphi (s)ds.
$$

\vspace{1cm}
 Concerning the non homogeneous problem, we now set the problem in the Hilbert space $H^1(\RR).$ Let  $\varphi\in H^1(\RR)$ and  $g\in C(H^1(\RR),H^{-1}(\RR))$( where $H^{-1}$ is the dual space of $H^{1}$), we call a function $v(\cdot,t)\in C([0,Z],H^1)\cap C^1([0,Z],H^{-1})$ a \textit{generalized solution} (or $H^1$-solution) of the problem
\begin{equation}
\left \{\begin{array}{l}
iv_z+ v_{tt} +g(v)=0\\
v(0)=\varphi
\end{array} \right.
\label{problemadicauchy}
\end{equation}
if $v(0,t)=\varphi$ in the sense of the space $H^{1}$ and $v(z,t)$ solves the equation $iv_z+ v_{tt} +g(v)=0$ for every $z\in [0,Z],$ in the sense of the space $H^{-1}.$

Furthermore, because of \textit{Duhamel's formula} we can express the solution as
\begin{equation}
v(z)=\mathcal{T}(z)\varphi +i\int_0^z\mathcal{T}(z-z^\prime)g(v(z^\prime)) dz^\prime,\qquad {\mbox{for all}}\quad z \in [0,Z].
\label{Duhamel}
\end{equation}

\vspace{.5cm}
Moreover if $g$ is bounded on bounded sets, we have that a function $v\in L^{\infty}([0,Z],H^1(\RR))$ is a solution (\textit{weak solution}) of $(\ref{problemadicauchy})$ if and only if it satisfies $(\ref{Duhamel}).$ The same occurs if $v\in C([0,Z],H^1(\RR))(\mbox{\textit{strong solution}}).$ Finally we introduce the notion of uniqueness in $H^1.$
\begin{definition}
Consider $g\in C(H^1(\RR),H^{-1}(\RR)).$ We say that there is uniqueness in $H^1$ for the problem $(2.8)$ if, given $\varphi \in H^1(\RR)$ and any interval $I\ni 0,$ it follows that any two $H^1$-solution coincide on $I.$
\end{definition}

\section{$H^1$ Solution}
Let $f$ be a function of continuous complex argument, i.e. $f\in C(\CC,\CC)$, such that 
\begin{equation}
f(0)=0,
\label{proprietà1}
\end{equation}
\begin{equation}
|f(u)-f(v)|\leq L(M) |u-v|
\label{proprietà2}
\end{equation}
 for all $u,v \in \CC$ such that $|u|,|v|\leq M,$ with
 \begin{equation}
 L(\cdot)\in C([0,\infty)).
 \label{proprietà3}
 \end{equation}
 Set 
 $$
 g(v)(t)=f(v(t))
 $$
 for all measurable $v:\RR\rightarrow \CC$ and almost all $t\in \RR.$
 \begin{theorem}
Let  $\varphi \in H^1(\RR).$ If $f\in C^1(\CC,\CC),$ then there exists a maximal interval $I\subset \RR$ on which is defined a unique function $v\in C(I,H^1(\RR))\cap C^1(I,H^{-1}(\RR))$ such that $iv_z+v_{tt}+g(v)=0$ for all $z\in I$ and $v(0)=\varphi.$
 \end{theorem}
 {\bf Proof}\qquad We only sketch the idea of the Proof, for more details we refer to \cite{Cazenave} and \cite{Kato}.
 
 The uniqueness follows immediately from [\cite{Cazenave}, Proposition $4.2.3$]. For the local existence we proceed as follows. Let us set $M=C||\varphi||_{H^1}, $ where $C$ is a positive \textit{real} number and
 $$
 E=\{v\in L^{\infty}((-Z,Z),H^1(\RR))\quad :\quad ||v||_{L^{\infty}((-Z,Z),H^1(\RR))}\leq M\},
 $$
 where $Z>0.$ The function space $E$ equipped with the distance
 $$
 d(u,v)=||u-v||_{L^{\infty}((-Z,Z),H^1(\RR))}
 $$
 is a Banach space. Given $\varphi\in H^1(\RR)$ and $v\in E$ we now set
 $$
 \mathcal{H}(v)(z)=\mathcal{T}(z)\varphi +i\int_0^z\mathcal{T}(z-z^\prime)g(v(z^\prime))dz^\prime.
 $$
  From Strichartz's estimates [\cite{Cazenave}, Theorem $2.3.3$] we have that
 $$
 \mathcal{H}(v)\in C([-Z,Z],H^1(\RR))
 $$
 and
 $$
 ||\mathcal{H}(v)||_{L^{\infty}((-Z,Z),H^1(\RR))}\leq\eta(M,Z)M,
 $$
 where $\eta$ is a continuous function depending on the initial value $\varphi $ and on $Z.$ Furthermore if we choose  $Z$ small enough (in this case $Z$ depends on $\varphi$ only by $||\varphi||_{H^1}$) we have that $\mathcal{H}(v)\in E$ and $d(\mathcal{H}(v),\mathcal{H}(u))\leq \frac{1}{2}d(v,u)$. In particular, $\mathcal{H}$ is a strict contraction on $E.$ The Banach fixed point theorem implies that $\mathcal{H}$ has a unique fixed point $v\in E.$ Moreover, $v\in C([-Z,Z],H^1(\RR)).$ The maximality of the interval of existence immediately follows from the uniqueness properties that there exists a solution $v\in C([-Z,Z],H^1(\RR)).$\quad $\Box$
 \vspace{.7cm}
 
Let us consider the problem $(\ref{modello}),$ and set

  $$f(v)=C_1e^{-C_2z}|v|^2v .$$  If $|u|,|v|\leq M,$ we have that
  \begin{eqnarray*}
  |f(v)-f(u)|&=& |C_1|e^{-C_2z}|(|v|^2v-|u|^2u)|\\
  &=&e^{-C_2z}|(|v|^2v-|v|^2u)+(|v|^2u-|u|^2u)|\\
  &\leq&e^{-C_2z}\big[|v|^2|v-u|+||v|^2-|u|^2||u|\big]\\
  &=&e^{-C_2z}\big[|v|^2|v-u|+|(|v|-|u|)(|v|+|u|)||u|\big]\\
  &\leq&e^{-C_2z} (M^2|v-u|+2M|v-u|M)\\
  &=&3M^2e^{-C_2z}|v-u|.
  \end{eqnarray*}

   Let $I=[0,\Lambda]$ with $\Lambda>0.$  The function $f(v)=C_1e^{-C_2z}|v|^2v$ belongs to $ C(\CC,\CC)$ for every $z\in I.$ Moreover it satisfies properties $(\ref{proprietà1})\div(\ref{proprietà3}),$ that is $ f(0)=0,$ and $|f(u)-f(v)|\leq L(M) |u-v|,$ with $|v|,|u|\leq M$ and with $L(M)=3e^{-C_2\Lambda}M^2.$ Thus, as a simple consequence of theorem $2.4.1$ we have 
   \begin{pro}
   Let $\varphi\in H^1(\RR)$ and $I=[0,\Lambda](\Lambda>0)$ be a bounded interval of $\RR.$ Then, if $f(v)=C_1e^{-C_2z}|v|^2v$ there exists a maximal interval $J\subset I$ on which the problem
   \begin{equation*}
     \left \{\begin{array}{l}
     iv_z+v_{tt} +g(v)=0\\
     v(0)=\varphi
     \end{array} \right.
     \end{equation*}
     with $g(v)(t)=f(v(t))$ a.a. $t\in \RR$ has a unique solution $v\in C(J,H^1(\RR))\cap C^1(J,H^{-1}(\RR)).$ 
   \end{pro}
        
\begin{remark}
The coefficient $C_2$ is a positive \textit{real} number; therefore, the Proposition holds for every $z$ running along the positive \textit{real} numbers. Indeed, since $C_2 z>0$ for every $z>0$ we obtain that $|f(v)-f(u)|\leq 3M^2|v-u|,$ and so the thesis of Proposition still holds true.

By contrast, if $z<0,$ in order to satisfy $(\ref{proprietà2}) $ it is needed that $z$ runs in a bounded interval. However, since the length of an optical fiber is positive and finite, we always refer to a bounded interval of \textit{real} positive numbers.
\end{remark}

 \begin{remark}
 If $C_2 =0$ we easily have that $|f(v)-f(u)|\leq 3 M^2|u-v|$ and so Theorem $2.4.1$ still holds true in this case. Therefore the standard Nonlinear cubic problem
 $$\left\{\begin{array}{l}
iv_z+v_{tt}+C_1|v|^2v=0\\
v(0)=\varphi
 \end{array} \right.
 $$
 has a maximal $H^1$- solution, both if $C_1$ is positive or negative.
 \end{remark}
 
 Let us consider again the equation
 \begin{equation}
 iv_{z}+v_{tt} +C_1e^{-C_2z}|v|^2v=0,\qquad v(0)=\varphi.
 \end{equation}
 From $(\ref{Duhamel})$ it follows that a $H^1$ solution of the problem can be written as
 $$
 v(z)=\mathcal{T}(z)\varphi+iC_1\int_0^z e^{C_2z^\prime}\mathcal{T}(z-z^\prime)[|v|^2v] dz^\prime
 $$
 In order to establish the absolute continuity of $H^1$ norm of such solution, we take into consideration the following estimate:
\begin{eqnarray*}
 |||v|^2v||_{H^1}^2 &\Doteq&\int_{\RR}|(|v|^2v)|^2 dt+\int_{\RR}|\frac{\partial}{\partial t}(|v|^2v)|^2 dt\\
 &=&\int_{\RR}|(|v|^2v)|^2 dt+\int_{\RR}|(2v_t|v|^2+v^2\bar{v}_t)|^2 dt\\
 &\leq& ||v||_{L^\infty}^2||v||_{L^2}^2+\Big(2\Big(\int_{\RR}|(v_t|v|^2)|^2 dt\Big)^{1/2} +\Big(\int_{\RR}|(v^2\bar{v}_t)|^2 dt\Big)^{1/2}\Big)^2\\
 &\leq& ||v||_{L^\infty}^2||v||_{L^2}^2+\Big(2||v||_{L^\infty}^2||v_t||_{L^2}+||v||^2_{L^\infty}||v_t||_{L^2}\Big)^2\\
 &=& ||v||_{L^\infty}^2||v||_{L^2}^2+9||v||_{L^\infty}^4||v||_{L^2}^2\\
 &\leq& K^2||v||^2_{H^1}
\end{eqnarray*}
where $K=\max\{\sup|v|,3\sup|v|^2\}$ and we used Holder inequality and Sobolev immersion $H^1(\RR)\hookrightarrow L^\infty(\RR).$

 Then by recalling that the operator $\mathcal{T}(z)$ is a isometry on the functions space $H^1(\RR),$ we have that
 $$
 ||v(z)||_{H^1}\leq ||\varphi ||_{H^1}+|C_1|K\int_0^z e^{-C_2 z^\prime}||v||_{H^1}dz^\prime.
$$

We now need  the following statement,
\begin{theorem}[Gronwall Lemma]
Let $I\ni 0$ be an interval of $\RR,$ and $u,v:I\rightarrow \RR$ two non negative functions. Let $c\geq 0$ be a constant such that
$$
v(z)\leq c+\int_0^zu(z^\prime)v(z^\prime) dz^\prime,\qquad \mbox{for all}\quad z\in I,\;z\geq 0
$$
then it is valid
\begin{equation}
v(z)\leq ce^{\int_0^zu(z^\prime)dz^\prime}
\end{equation}
\end{theorem}

\vspace{1cm}
So by applying theorem $2.4.2$ we have that
\begin{equation}
||v(z)||_{H^1}\leq ||\varphi ||_{H^1}\exp\Big[|C_1|K\int_0^z e^{-C_2z^\prime }dz^\prime\Big]
\label{stimagronwall}
\end{equation}

From $(\ref{stimagronwall})$ it follows that 
\begin{eqnarray}
\sup_{z\in [0,Z)}||v(z)||_{H^1}\leq ||\varphi ||_{H^1}\exp\Bigg[K\Big[\frac{1-e^{C_2 Z}}{C_2}\Big]\Bigg].
\end{eqnarray}

We are now ready to state
\begin{pro}
Let $\varphi \in H^1(\RR).$ Then, the solution of the problem 
\begin{equation*}
     \left \{\begin{array}{l}
     iv_z+v_{tt} +C_1e^{-C_2z}|v|^2v=0\\
     v(0)=\varphi
     \end{array} \right.
     \end{equation*}
     there exists for every $z\in \RR.$
\end{pro}
{\bf Proof} \quad Let us suppose that the solution of the problem  exists only into the maximal interval $[0,Z).$ Let us consider $\delta$ such that $0<\delta <Z;$  by the local existence, we have for the initial data $v_{\delta}(t,Z-\delta)=v(t,Z-\delta)$ a solution $v_{\delta}$ on the time interval $[Z-\delta,Z-\delta + \eta)$ for some $\eta >0,$ where $v$ is the solution on the maximal interval $[0,Z].$ Since we have uniform bounds (independent of $\delta$) on $||u||_{H^1}$ if $Z<\infty,$ it follows that $\eta$ is independent of $\delta.$ Thus, if we choose $\delta$ sufficiently small, we have
$$Z-\delta +\eta >Z.$$
However, for the uniqueness, $v_\delta$ and $v$ coincide on $\RR \times [Z-\delta,Z),$ and therefore $v_\delta$ extends $v$ beyond the maximal interval of existence. This is a contradiction.\quad $\Box$

\begin{remark}
If $C_2=0$ Proposition $2.7$ works properly; indeed, $(\ref{stimagronwall})$ becomes 
\begin{equation*}
||v(z)||_{H^1}\leq ||\varphi ||_{H^1}\exp\Big[\int_0^z |C_1|K dz^\prime\Big]
\end{equation*}
and so
\begin{eqnarray*}
\sup_{z\in [0,Z)}||v(z)||_{H^1}\leq ||\varphi ||_{H^1}\exp\Big[K Z\Big].
\end{eqnarray*}
Thus, if $Z<\infty$ we obtain that the $H^1$-norm is absolutely continuous on $z\in[0,Z].$
\end{remark}

\section{$H^2$ Solution}
In this section we show a result about uniqueness and existence of a $H^2$ solution of the initial value problem (IVP)

\begin{equation}
     \left \{\begin{array}{l}
     iv_z(z,t)+v_{tt}(z,t) +g(v(z,t))=0\\
     v(0)=\varphi(t)
     \end{array} \right.
     \label{H^2problem}
     \end{equation}
     with $\varphi\in H^2(\RR).$ From [Cazenave, Lemma $4.2.8$] it follows that if $I\ni 0$ is an interval, and $g\in C(H^2(\RR),L^2)$ is bounded on bounded sets, then $v\in C(I,H^2(\RR))\cap C^1(I,L^2(\RR))$ is a solution of the IVP $(\ref{H^2problem})$ if and only if $v$ satisfies the integral equation
     $$
     v(z)=\mathcal{T}(z)\varphi +i\int_0^z\mathcal{T}(z-z^\prime)g(v(z^\prime)) dz^\prime,\qquad {\mbox{for all}}\quad z \in I.
     $$
      
Let us assume that $g:H^2(\RR)\rightarrow L^2(\RR).$ Suppose that there exist $0\leq s<2$ and $2\leq r,\rho<\infty$ such that
\begin{equation}
g\in C(H^s(\RR),L^2(\RR))\qquad \mbox{is bounded on bounded sets}
\label{H^2proprietà1}
\end{equation}
and
\begin{equation}
||g(u)-g(v)||_{L^{\rho^\prime}}\leq L(M)||u-v||_{L^r}
\label{H^2proprietà2}
\end{equation}
for all $u,v \in H^2(\RR)$ such that $||u||_{H^s}, ||v||_{H^s}\leq M.$

\begin{theorem}
Let $g:H^2(\RR)\rightarrow L^2(\RR)$ satisfy  $(\ref{H^2proprietà1}),(\ref{H^2proprietà2}). $ For every $\varphi \in H^2(\RR), $ there exists an open interval $I$ of $\RR$ and a unique, maximal solution $v\in C(I,H^2(\RR))\cap C^1(I,L^2(\RR))$ of IVP $(\ref{H^2problem}).$ 
\end{theorem}
{\bf Proof}\qquad The uniqueness follows from [\cite{Cazenave}, Proposition $4.2.9$]. The idea to prove the local existence is the same as the one we showed in Theorem $2.4.1.$ Let consider
\begin{equation*}
\mathcal{H}(v)(z)=\mathcal{T}(z)\varphi +i\int_0^z\mathcal{T}(z-z^\prime)g(v(z^\prime))dz^\prime.
\end{equation*}
Let define a suitable bounded Banach space $E.$ In [\cite{Cazenave},Theorem $4.8.1$] it is found that $\mathcal{H}:E\rightarrow E$ is a strict contraction. Therefore, from the Banach fixed point theorem we obtain a solution of the problem $(2.8).$ It is remarkable that the interval constructed, on which there is local existence, depends on the initial value $\varphi $ by $||\varphi||_{H^2}.$ $\Box$
\vspace{.5cm}

Furthermore we have
\begin{pro}
Let $v(z)$ be a solution given by the theorem. If
$$\mbox{Re}(g(w),iw)_{L^2}=0\quad \mbox{for all}\; w\in H^2(\RR),
$$
 then $||v(z)||_{L^2}=||\varphi ||_{L^2}$ for all $z\in I.$
\\
Moreover if there exists $G\in C^1(H^2(\RR),\RR)$ such that $g=G^\prime,$ then 
$E((v(z))=E(\varphi)$ for all $z\in I,$ where
$$
E(v)\Doteq\frac 1 2 \int_{\RR}|v_t|^2 dt -\int_{\RR}G(v(t))dt\qquad \mbox{for all}\quad u\in H^2(\RR)
$$
\end{pro} 
{\bf Proof}\quad We only sketch the proof. For more details we refer to \cite{Cazenave}. 

Multiplying the equation $iv_z+v_{tt}+g(v)=0$ by $iv$ we obtain that
\begin{eqnarray*}
\frac{d}{dz}||v(z)||_{L^2}^2&=& 2\;\mbox{Re}(v_z,v)=2\;\mbox{Re}\big[(-v_{tt},v)+(g(v),iv)\big]=0.
\end{eqnarray*}

Moreover multiplying $iv_z+v_{tt}+g(v)=0$ by $v_z$ we have that
\begin{eqnarray*}
E(v(z))&=&\mbox{Re}(iv_z,v_z)=0,
\end{eqnarray*}
where $E(v(z))$ is obtained by $\mbox{Re}\big[(-v_{tt},v_t)+(g(v),v_t)\big]. \quad \Box$

\vspace{1cm}
Now we show that there exists a unique maximal solution of the problem $(\ref{modello},)$ that is
$$
iv_{z}+ v_{tt} +C_1e^{-C_2z}|v|^2v=0,\qquad v(0,t)=\varphi(t),
$$
where $C_1=\pm 1$ and $C_2>0.$  

Let us assume that $v\in H^2(\RR),$ we define $g(v)=C_1e^{-C_2 z}|v|^2v;$ by using Holder inequality and the immersion $H^2\hookrightarrow L^\infty$ we obtain
\begin{eqnarray*}
||g(v)||_{L^2}^2 &=& \int_{\RR} |g(v)|^2 dt=|C_1|^2 e^{-2C_2 z}\int_{\RR}||v|^2v|^2 dt\\
&\leq &  e^{-2C_2 z}||v||_{L^\infty}^4||v||^2_{L^2}\\
&\leq&||v||_{L^\infty}^4||v||^2_{L^2}
\end{eqnarray*}
for every $z>0.$ To show that $g(v)\in L^2(\RR)$ it is needed that $||v(z,\cdot)||_{L^\infty},||v(z,\cdot)||_{L^2}$ are finite for  $||v(z)||_{H^2}\leq K$ for all $z \in I,$ interval of $\RR^+,$ and $K$ positive constant. We know that $||v||_{L^2}\leq \infty,$ easily. Then the $L^\infty$ norm of $v$ can be controlled by means of the $H^2$ norm of $v$ by the following argument:
let us consider the fundamental theorem of calculus for a function $v(t)\in H^2(\RR),$ that is
$$
v(t)-v(s)=\int_s^tv^{\prime}(\tau)d\tau,
$$
where $t>s$ and $v^\prime$ is the first derivative with respect to the variable $t$, in the sense of distributions.
From this it follows that
$$|v(t)-v(s)|\leq \int_s^t|v^\prime|(\tau)d\tau $$
Then, by using Holder inequality we obtain that
$$
|v(t)-v(s)|\leq ||v^\prime||_{L^2}|t-s|^{\frac 1 2}
$$
Now we integrate between $t-r$ and $t+r$ with $r>0$ with respect the variable $s.$ Then, using the inverse triangular inequality $||v(t)|-|v(s)||\leq |v(t)-v(s)|$ we have that
$$
|v(t)|\leq|v(s)|+||v^\prime||_{L^2}|t-s|^{\frac 1 2}
$$
$$
2r|v(t)|\leq \int_{t-r}^{t+r}|v(s)|ds+\int_{t-r}^{t+r}||v^\prime||_{L^2}|s|^{1/2}ds
$$
because of $|t-s|\leq |s|$ for $s\in (t-r,t+r).$ Again, from Holder inequality it follows that
$$
2r|v(t)|\leq r^{1/2}||v||_{L^2}+\frac 2 3 r^{3/2}||v^\prime||_{L^2}\leq \frac 3 2 \Big(r^{1/2}||v||_{L^2}+r^{3/2}||v^\prime||_{L^2}\Big)
$$
Now let us consider the function $h(r)=\frac 3 2 \Big(r^{1/2}||v||_{L^2}+r^{3/2}||v^\prime||_{L^2}\Big);$ this function is continuous in $r\in \RR-{0}.$ Thus, let us calculate the derivative and compute the minimum:
$$
h^\prime(r)=\frac 3 2 \Big(-\frac{||v||_{L^2}}{2r\sqrt{r}}+\frac{||v^\prime||_{L^2}}{2\sqrt{r}}\Big)
$$
The minimum is reached for $\bar{r}=||v||_{L^2}/||v^\prime||_{L^2}$ and so $ h(\bar{r})=3\Big(||v||_{L^2}||v^\prime||_{L^2}\Big)^{1/2}.$ 

In this way we have that
$$
|v(t)|\leq 3K
$$
for $||v||_{H^2}\leq K.$

In view of this we have that $g(v)\in L^2(\RR)$ and it satisfies property $(\ref{H^2proprietà1}).$ 

From the last argument it follows that, if $||u||_{H^2},||v||_{H^2}\leq M$ then $|u|,|v|\leq 3 M.$ Now, since 
$$
||v|^2v-|u|^2u|\leq 3(3M)^2|v-u|
$$
it follows that
\begin{eqnarray*}
||g(v)-g(u)||_{L^2}&=& |C_1| e^{-C_2z}\Big(\int_{\RR} |(|v|^2v-|u|^2u)|^2 dt\Big)^{1/2}\\
&\leq& 27 M^2   e^{-C_2z}\Big(\int_{\RR} |v-u|^2 dt\Big)^{1/2}\\
&\leq& 27 M^2 ||v-u||_{L^2}
\end{eqnarray*}
and so we uncover that $g$ verifies property $(\ref{H^2proprietà2}),$ for every $z\in I,$  interval of $\RR^+.$ We resume all this into
\begin{pro}
 For every $\varphi \in H^2(\RR)$ there exists a maximal interval $I$ of \textit{real} positive numbers such that the initial value problem
\begin{equation}
iv_{z}+v_{tt} +C_1e^{-C_2z}|v|^2v=0,\qquad v(0)=\varphi,
\end{equation}
has a unique, maximal solution $v\in C(I,H^2(\RR))\cap C^1(I,L^2(\RR)).$
\end{pro}

\begin{remark}
Let us consider $g(w)=C_1e^{-C_2 z}|w|^2w;$
we note that 
\begin{eqnarray*}
\mbox{Re}(g(w),iw)_{L^2}=\mbox{Re}\int_{\RR}g(w)(-i\bar{w})dt&=&-C_1e^{-C_2 z}\mbox{Re}\int_{\RR}i|w|^4 dt=0
\end{eqnarray*}
Therefore, from proposition $(2.8)$ it follows that $||v(z)||_{L^2}=||\varphi ||_{L^2}$ for every $z$ in a suitable interval on which $v(z,t)$ is a solution of the problem $(\ref{H^2problem}).$ 
\end{remark}

\begin{remark}
If $g(u)=C_1|u|^2u,$ that is  $C_2=0,$  then all conditions of theorem $2.5.1$ are satisfied, and so  the problem
$$
iv_{z}+v_{tt} +C_1|v|^2v=0,\qquad v(0)=\varphi,
$$ 
has a maximal $H^2$-solution. Moreover conditions of proposition $2.8$ also are satisfied and so  we have the conservation of the following quantities:
$$
\int_{\RR}|u(z,t)|^2 dt, \qquad \frac 1 2 \int_{\RR}|u_t|^2 dt -\frac{C_1}{4}\int_{\RR}|u(z,t)|^4 dt .
$$
\end{remark}

\chapter{Integrable System}
\section{Introduction}
The Painlev\' e property for ordinary differential equations is defined as follows. The solutions of a system of ordinary differential equations are regarded as analytic functions of a complex variable. The movable singularities of the solution are the singularities of the solution whose location depends on the initial conditions. Thus, they  are movable. The system is said to possess the Painlev\'e property when all the movable singularities are single-valued.

One major difference between analytic functions of one complex variable and analytic functions of several complex variables is that, in general, the singularities of a function of several complex variables cannot be isolated. If $f=f(\xi_1,\ldots,\xi_n)$ is a meromorphic function of $n$ complex variables ($2n$ real variables), the singularities of $f$ occur along analytic manifolds of (real) dimension $2n-2.$ These manifolds are determined by conditions of the form $\phi(\xi_1,\ldots,\xi_n)=0,$ where $\phi$ is an analytic function of $(\xi_1,\ldots,\xi_n)$ in a neighborhood of the manifold. Therefore, we say that a partial differential equation has the Painlev\'e property when the solutions of the PDE are single-valued about the movable, singularity manifolds. For the sake of correctness, if the singularity manifold is determined by the condition $\phi(\xi_1,\ldots,\xi_n)=0,$ and $u(\xi_1,\ldots,\xi_n)$ is a solution of PDE, then we assume that
$$
u=u(\xi_1,\ldots,\xi_n)=\phi^\alpha\sum_{j=0}^\infty u_j\phi^j,
$$
where $\phi=\phi(\xi_1,\ldots,\xi_n)$ and $u=u(\xi_1,\ldots,\ldots \xi_n)$ are analytic functions of $(\xi_1,\ldots,\ldots \xi_n)$ in a neighborhood of the singularity manifold, and $\alpha$ is an integer. Substitution of the expression of the $u$ into the PDE determines the possible value of $\alpha$ and defines the recursion relations for $ u_j,\; j=0,1,2,\ldots .$

In \cite{Weiss} it is indicated that the Painlev\'e property may provide a unified description of integrable behavior in dynamical systems, while, at the same time, providing an efficient method for determining the integrability of particular systems.

In this chapter we will find a  NLSE that is Painlev\'e integrable. Afterwards, we will introduce some transformations that allow us to pass to the standard NLSE : $iQ_Z+Q_{TT}+\rho|Q|^2Q=0$ where $\rho$ can be both positive or negative real number. Finally we prove a result of 'closeness'; let $u$ be a solution of the equation
\begin{equation}
iu_z+\frac{\beta_2}{2}u_{tt}-\gamma e^{-\alpha z}|u|^2u=0
\end{equation}
presented in the first chapter; we prove that this solution is close in the $L^2$ norm to a solution of a integrable NLSE. 

\section{Integrability and compatibility conditions}
To perform the analysis of the Painlev\'e test for partial differential equations, i.e., the well known Weiss-Tabor-Carnevale (WTC) test, we consider a non autonomous generalized NLSE with real coefficients:
\begin{eqnarray}
&& iv_z(z,t)+f(z,t)v_{tt}(z,t)+g(z,t)|v(z,t)|^2v(z,t)+V(z,t)v(z,t)+\nonumber\\
&&ih(z,t)v(z,t)=0.
\end{eqnarray}  
Here $f(z,t)$ and $g(z,t)$ ($z,t$ real variables) are  the dispersion and the nonlinearity managements parameters, respectively. The function $V(z,t)$  denotes the external potential applied and $h(z,t)$ is the dissipation ($h>0$) or gain ($h<0$).
We follow \cite{He} to obtain some conditions which guarantee that Eq. $(3.2)$ pass the WTC test.

In order to perform conveniently the Painlev\'e analysis we first complexify Eq. $(3.2),$ which becomes 
\begin{eqnarray}
&&i\frac{\partial u(z,t)}{\partial z}+f(z,t)\frac{\partial^2 u(z,t)}{\partial t^2}+g(z,t)v(z,t)u(z,t)^2+V(z,t)u(z,t)\nonumber\\
&&+ih(z,t)u(z,t)=0
\end{eqnarray}

\begin{eqnarray}
&&-i\frac{\partial v(z,t)}{\partial z}+f(z,t)\frac{\partial^2 v(z,t)}{\partial t^2}+g(z,t)u(z,t)v(z,t)^2+V(z,t)v(z,t)\nonumber\\
&&-ih(z,t)u(z,t)=0
\end{eqnarray}
where $u(z,t)$ and $v(z,t)$ are treated as independent complex functions of variables $z,t$ and the functions $f(z,t),g(z,t),V(z,t)$ and $h(z,t)$ are analytic on the non characteristic singularity manifold $\varphi(z,t)=0.$ By using the following ansatz $\varphi(z,t)=t+\phi(z)$ \cite{He} the solutions of Eqs. $(3.3),(3.4)$ can be expanded on the non characteristic singularity manifold as
\begin{equation}
u(z,t)=[t+\phi(z)]^{-p}\sum_{j=0}^\infty u_j(z)[t+\phi(z)]^j,
\end{equation}

\begin{equation}
v(z,t)=[t+\phi(z)]^{-q}\sum_{j=0}^\infty v_j(z)[t+\phi(z)]^j,
\end{equation}
where $u_0\neq 0, v_0\neq 0.$

We also expand $f,g,V,h$ on the same singularity manifold as follows:
\begin{equation*}
f(z,t)=\sum_{i=0}^\infty f_i(z)[t+\phi(z)]^i,\quad g(z,t)=\sum_{i=0}^\infty g_i(z)[t+\phi(z)]^i
\end{equation*}

\begin{equation}
V(z,t)=\sum_{i=0}^\infty V_i(z)[t+\phi(z)]^i,\quad h(z,t)=\sum_{i=0}^\infty h_i(z)[t+\phi(z)]^i
\end{equation}

where $f_i(z)=\frac{1}{i!}\frac{\partial f(z,t)}{\partial z^i}\mid_{t=-\phi(z)},$ and similarly for $g(z,t),V(z,t)$ and $h(z,t).$

Substituting expressions $(3.5)-(3.7)$ into Eqs. $(3.3)$ and $(3.4)$ and collecting the same power of $\phi(z),$ one can obtain \cite{He} $(i)$ the values of $p$ and $q,$ and the equations concerning the first terms derived; and $(ii)$ the general recursion relations of $u_j$ and $v_j.$

Therefore, by the standard procedure for the leading-order analysis, we get $p=q=1$ and 
$$2f_0(z)+g_0(z)u_0(z)v_0(z)=0.$$
The recursion relations are 
\begin{equation}
A(j){u_j \choose v_j}=\left(\begin{array}{cc}
Q_j & g_0u_0^2\\
g_0v_0^2 & Q_j
\end{array}\right) {u_j \choose v_j}={F_j \choose G_j},
\end{equation}
where 
\begin{equation} 
Q_j=(j-1)(j-2)f_0+2g_0u_0v_0,
\end{equation}
\begin{equation*} 
F_j=-i[u_{j-2,z}+(j-2)u_{j-1}\phi_z]-g_ju_0^2v_0-\sum_{k=1}^j(j-k-1)(j-k-2)f_ku_{j-k}
\end{equation*}
\begin{equation*} -g_0v_0\sum_{m=1}^{j-1}u_{j-m}u_m-g_0\sum_{m=1}^{j-1}\sum_{k=0}^mv_{j-m}u_{m-k}u_k-\sum_{m=1}^{j-1}\sum_{k=0}^m\sum_{l=0}^k g_{j-m}v_{m-k}u_{k-l}u_l
\end{equation*}
\begin{equation} 
-\sum_{m=0}^{j-2}V_{j-m-2}u_m -i\sum_{m=0}^{j-2}h_{j-m-2}u_m,
\end{equation}
with $i$ the imaginary unit and $G_j$ has a similar expression which can be obtained from $F_j$ by first interchanging $u_j$ and $v_j$ and then taking its complex conjugate. In addition we use the notation that once an index is less than zero, the expression itself is zero. 

The above-mentioned recursion relations uniquely determine the unknown expansion coefficients uniquely unless the determinant of the matrix in Eq. $(3.8)$ is zero. Those values of $j$ at which the determinant is equal to zero are called the \textit{resonances}, and the conditions which ensure Eq. $(3.8)$ to have solutions at the resonances are named \textit{compatibility conditions}.

In \cite{He} it is found that from Eqs. $(3.8)-(3.10)$ resonances only occur at $j=-1,0,3,4.$ So the compatibility conditions of Eq. $(3.2)$ are given by \cite{He}

\begin{pro}
\begin{equation}
f(z,t)=f(z), \; g(z,t)=g(z),\; h(z,t)=h(z),\; V(z,t)=V_0(z)+V_1(z)t+V_2(z)t^2,
\end{equation}
where $V_0(z)$ and $V_1(z)$ are arbitrary and $V_2(z),f(z),g(z),h(z)$ satisfy the relation
\begin{eqnarray}
&&(4f^2gg_z-2ff_zg^2)h-4f^2g^2h^2-2f^2g^2h_z-g^2ff_{zz}+f^2gg_{zz}-2f^2g_z^2\nonumber\\
&&+f_z^2g^2+f_zgfg_z+4V_2f^3g^2=0.
\end{eqnarray}
\end{pro}
Condition $(3.11)$ suggests that the Painlev\'e integrable class of Eq.$(3.2)$ should have the form of
\begin{eqnarray}
&&iv_z(z,t)+f(z) v_{tt}(z,t)+g(z)|v(z,t)|^2v(z,t)\nonumber\\
&&+[V_0(z)+V_1(z)t+V_2(z)t^2]v(z,t)+ih(z)v(z,t)=0,
\end{eqnarray}
where $f(z),g(z),h(z)$ and $V_2(z)$ are related by $(3.12)$ and $V_0,V_1$ are arbitrary. We now set the transformation
$$u(z,t)=v(z,t)\exp\Big[-\int_0^z h(z^\prime)dz^\prime \Big],\qquad \mbox{for every}\quad z\in \RR .$$
Thus, Eq. $(3.13)$ becomes 
\begin{eqnarray*}
&&i\frac{\partial v(z,t)}{\partial z}-ih(z)v(z,t)+f(z)\frac{\partial^2 v(z,t)}{\partial t^2}+g(z)|v(z,t)|^2v(z,t)\nonumber\\
&&+[V_0(z)+V_1(z)t+V_2(z)t^2]v(z,t)+ih(z)v(z,t)=0,
\end{eqnarray*}
and so the loss/gain term can be eliminated formally. Thus,  in what follows we only consider the model with $h(z)\equiv 0,$ i.e.
\begin{eqnarray}
&&i\frac{\partial v(z,t)}{\partial z}+f(z)\frac{\partial^2 v(z,t)}{\partial t^2}+g(z)|v(z,t)|^2v(z,t)\nonumber\\
&&+[V_0(z)+V_1(z)t+V_2(z)t^2]v(z,t)=0.
\end{eqnarray}

Then, for $h\equiv 0$ the relation between $V_2,f,g$ becomes
\begin{equation}
-g^2ff_{zz}+f^2gg_{zz}-2f^2g_z^2+g^2f_z^2+gfg_zf_z+4V_2f^3g^2=0.
\end{equation}

\section{Transformation to the standard NLSE}
In this section we look for a transformation which convert Eq $(3.14)$ into the Standard NLSE
\begin{equation}
i\frac{\partial Q}{\partial Z}(Z,T)+\kappa\frac{\partial^2 Q}{\partial^2 T}(Z,T)+\chi|Q(Z,T)|^2Q(Z,T)=0
\label{standardNLSE}
\end{equation}
where $\kappa$ and $\chi$ are \textit{real} coefficients.  If $\kappa\chi >0,$ Eq. $(3.16)$ is called focusing, while if $\kappa\chi <0,$ Eq. $(3.16)$ is called defocusing.

We look for a transformation in the form \cite{He}
\begin{equation}
v(z,t)=Q(q(z),p(z,t))e^{ia(z,t)+c(z)},
\end{equation}
where $p(z,t),q(z),a(z,t)$ and $c(z)$ are real functions to be determined and $v(z,t),Q(Z,T)$ are the solutions of Eqs. $(3.14),(3.16),$ respectively.

In order to substitute Eq. $(3.17)$ into Eq. $(3.14)$ we take the first derivative of $v(z,t)$ with respect the variable $z.$ Thus, using Eq. $(3.17)$ and letting $p(z,t)=T,\; q(z)=Z,$ we have that
\begin{eqnarray*}
v_z(z,t)&=& \Big[q_z(z)Q_Z(Z,T)+p_z(z,t)Q_T(Z,T)\\
&&+ia_z(z,t)Q(Z,T)+c_z(z,t)Q(Z,T)\Big]e^{ia(z,t)+c(z)}.
\end{eqnarray*}
We now take the second derivative with respect the variable $t,$ and we obtain that
\begin{eqnarray*}
v_{tt}(z,t)&=& \Big[(p_{t}(z,t))^2\;Q_{TT}(Z,T)+p_{tt}(z,t)Q_T(Z,T)\\
&&+ip_t(z,t)a_t(z,t)Q_T(Z,T)+ia_{tt}(z,t)Q(Z,T)\Big]e^{ia(z,t)+c(z)}\\
&&+\Big[ia_t(z,t)p_t(z,t)Q_T(Z,T)-(a_t(z,t))^2\;Q(Z,T)\Big]e^{ia(z,t)+c(z)}
\end{eqnarray*}

We are now ready to substitute Eq. $(3.17)$ into Eq. $(3.14);$ so we obtain that
\begin{eqnarray*}
&&iq_zQ_Z+fp_t^2Q_{TT}+ge^{2c}|Q|^2Q+i[Q(c_z+fa_{tt})+Q_T(p_z+2 fp_ta_t)]\\
&&+fp_{tt}Q_T-(a_z+fa_t^2-V_0-V_1t-V_2t^2)Q=0
\end{eqnarray*}

Comparing this with Eq. $(3.16)$ we obtain that
\begin{eqnarray}
&& c_z+f a_{tt}=0,\label{sistema}\\\nonumber
&& p_z+2f p_ta_t=0,\\\nonumber
&& a_z+f a_t^2-V_0-V_1t-V_2t^2=0,\\\nonumber
&& p_{tt}=0.\nonumber
\end{eqnarray}
From the first two and the fourth equations above we get
\begin{equation}
a(z,t)=-\frac{c_z(z)}{2f(z)}+h_1(z)t+h_2(z)
\label{ainiziale}
\end{equation}
and
\begin{equation}
p(z,t)=t\;e^{2c(z)}-2\int f(z)h_1(z)\;e^{2c(z)}dz,
\label{ciniziale}
\end{equation}
where $h_1(z)$ and $h_2(z)$ are functions to be determined. Inserting $a(z,t)$ and $p(z,t)$ into the third equation in Eqs. $(\ref{sistema}),$ collecting the coefficients of all powers of $t,$ and further
setting them as zero, we get
\begin{eqnarray}
-c_{zz}f+c_zf_z+2f c_z^2-2 V_2 f^2=0,\label{sistema2}\\
-2c_2 h_1+h_1^\prime-V_1=0,\nonumber\\
fh_1^2+h_2^\prime-V_0=0.\nonumber
\end{eqnarray}
From the last two equations of this system $h_1(z)$ and $h_2(z)$ can be solved:
\begin{equation}
h_1(z)=h(z)\;e^{2c(z)},
\label{h1}
\end{equation}
\begin{equation}
h_2(z)=\int\big[V_0(z)-f(z)h^2(z)e^{4c(z)}dz\big]+A_2,
\label{h2}
\end{equation}
where $h(z)=\int V_1(z)e^{-2c(z)}dz +A_1.$ Here $A_1$ and $A_2$ are \textit{real} constants.

Finally, in comparison to the standard NLSE equation one has
\begin{equation}
\frac{ge^{2c}}{q_z}=\chi \qquad \frac{fe^{4c}}{q_z}=\kappa
\label{comparison}
\end{equation}

The above equations give 
\begin{equation}
c(z)=\frac{1}{2}\ln\frac{\kappa g(z)}{\chi f(z)}, \qquad q(z)=\frac{\kappa}{\chi^2}\int \frac{g^2(z)}{f(z)}dz+K_1,
\label{ceqgenerali}
\end{equation}
where $K_1$ is a constant of integration. Usually we choose $K_1$ such that $q(0)=0.$ Then it is easy to determine the remaining transformation parameters $a(z,t)$ and $p(z,t)$ as \cite{He}
\begin{equation}
a(z,t)=\frac{1}{4f(z)}\frac{d}{dz}\Bigg(\ln\frac{\chi f(z)}{\kappa g(z)}\Bigg)t^2+\frac{g(z)}{f(z)}r(z)t-\int \Bigg(\frac{g^2(z)}{f(z)}r^2(z)-V_0(z) \Bigg)dz+K_2
\label{agenerale}
\end{equation}

\begin{equation}
p(z,t)=\frac{\kappa g(z)}{\chi f(z)}t-\frac{2\kappa}{\chi}\int \frac{g^2(z)}{f(z)}r(z)dz.
\label{pgenerale}
\end{equation}

Here we defined $r(z)$ as
\begin{equation} 
r(z)=\int\frac{f(z)}{g(z)}V_1(z) dz +K_3
\label{r}
\end{equation}

\subsection{An explicit integrable model}
We now consider the equation
\begin{equation}
iv_z(z,t)+\frac{\beta_2}{2}v_{tt}(z,t)-\gamma\;e^{-\alpha z}|v(z,t)|^2v(z,t)=0
\label{ingegneri}
\end{equation}
which we pointed out in Chapter $1$ and that governs propagation of optical pulses inside single-mode fibers. We know that the quantity $|v|^2$ has dimension of power physics, i.e. $[|v|^2]=W,$ in the \textit{International System of Units}. Moreover, we found that 
$$[\alpha]=L^{-1},\quad [\beta_2]=T^2L^{-1},\quad [\gamma]=W^{-1}L^{-1},$$
where in the \textit{International System of Units} $L$ and $T$ are dimension symbol of length and time respectively.

In order to specialize the previous general context to Eq. $(\ref{ingegneri}),$ we set
$$
f(z)=\frac{\beta_2}{2},\qquad g(z)=-\gamma e^{-\alpha z}.
$$
Thus, from compatibility condition $(3.15)$ it follows that 
$$
\Big(\frac{\beta_2}{2}\Big)^2\Big(-\gamma e^{-\alpha z}\Big)\Big(-\gamma\alpha^2 e^{-\alpha z}\Big)-2\Big(\frac{\beta_2}{2}\Big)^2\Big(\gamma \alpha e^{-\alpha z}\Big)^2+4V_2\Big(\frac{\beta_2}{2}\Big)^3\Big(-\gamma e^{-\alpha z}\Big)^2=0,
$$
then we have that
$$
-\gamma^2\alpha^2+4V_2\gamma^2\frac{\beta_2}{2}=0,
$$
and so it easily follows that
$$
V_2=\frac{\alpha^2}{2\beta_2}.
$$
For arbitrariness of the functions $V_0(z)$ and $V_1(z)$ we set $V_0(z)\equiv 0,V_1(z)\equiv 0.$ Therefore, Eq. $(3.14)$ becomes
\begin{equation}
iv_z+\frac{\beta_2}{2}v_{tt}-\gamma e^{-\alpha z}|v|^2v+\frac{\alpha^2}{2\beta_2 }t^2v=0.
\label{integrabileNLSE}
\end{equation}
We immediately find that Eq. $(\ref{integrabileNLSE})$ is dimensionally balanced; indeed the dimension of the last term is $L^{-1}[v],$ in accordance to the other terms in the equation.

We are now ready to present the transformation that allows us to convert Eq. $(\ref{integrabileNLSE})$ into the standard NLSE $(\ref{standardNLSE}).$ From the first of Eqs. $(\ref{ceqgenerali})$ it follows that
\begin{eqnarray*}
c(z) &=&\frac 1 2 \ln\Big(\frac{\kappa(-\gamma e^{-\alpha z})}{\chi \frac{\beta_2}{2}}\Big)=\frac 1 2 \ln\Big(\frac{-2\gamma\kappa}{\chi \beta_2}e^{-\alpha z}\Big)\\ \nonumber
\end{eqnarray*}
and so
\begin{equation}
c(z)=\ln\sqrt{K}-\frac{\alpha }{2}z
\label{c}
\end{equation}
where we defined $K=-\frac{2\gamma\kappa}{\chi\beta_2}.$ We observe that $K$ must be positive; this condition is achieved if the sign of the products $\beta_2\gamma$ and $\kappa\chi$ are opposite. In order to have $K$ dimensionless, we set
$$
[\kappa]=T^2L^{-1},\qquad [\chi]=W^{-1}L^{-1}.
$$
We will see that these assumptions are in accordance to the transformation that we will find in the next. From Eq. $(\ref{c})$ and these assumptions it easily follows that $c(z)$ is dimensionless, being $z$ a \textit{length}($[z]=L$).

By continuing to evaluate the transformations, from the second of $(\ref{ceqgenerali})$ we obtain  that 
\begin{eqnarray}
q(z)&=&\frac{\kappa}{\chi^2}\int_0^z\frac{2\gamma^2e^{-2\alpha z^\prime}}{\beta_2}dz^\prime+K_1=-\frac{\kappa\gamma^2}{\alpha\beta_2\chi^2}e^{-2\alpha z}+\frac{\kappa\gamma^2}{\alpha\beta_2\chi^2}+K_1\nonumber\\ 
&=&K\frac{\gamma}{2\alpha\chi}e^{-2\alpha z}-K\frac{\gamma}{2\alpha\chi}=K\frac{\gamma}{2\alpha\chi}(e^{-2\alpha z}-1)
\label{q}
\end{eqnarray}
where $K=-\frac{2\gamma\kappa}{\chi\beta_2}$ and we have assumed $K_1=0$ in order to get $q(0)=0.$ We notice that $q$ is a \textit{length}($[q]=L$), as expected.

We recall that we  assumed $V_0\equiv V_1\equiv 0$ and so we have from Eq. $(\ref{r}),$ by setting $K_3=0,$ that $r(z)=0.$ Thus, from Eqs. $(\ref{agenerale})$ and $(\ref{pgenerale}),$ setting $K_2=0,$ a straightforwardly calculation leads to
\begin{eqnarray}
a(z,t)&=&\frac{1}{2\beta_2}\frac{d}{dz}\ln\Big(\frac{\chi\beta_2}{-2\gamma\kappa}e^{\alpha z}\Big)t^2=\frac{\alpha}{2\beta_2}t^2\label{a}
\end{eqnarray}
and
\begin{eqnarray} 
p(z,t)&=&-\frac{2\gamma\kappa}{\chi\beta_2}e^{-\alpha z}t=Ke^{-\alpha z}t.\label{p}
\end{eqnarray}
We immediately find that $a$ is dimensionless, while $p$ is a \textit{time} as we expected.

We now collect these results in the following
\begin{theorem}
Let us set $T=p(z,t)$ and $Z=q(z).$ Then a solution $v(z,t)$ of 
$$
iv_z+\frac{\beta_2}{2}v_{tt}-\gamma e^{-\alpha z}|v|^2v+\frac{\alpha^2}{2\beta_2 }t^2v=0
$$
can be converted into a solution of the standard NLSE provided that $K>0$ by the following transformation
\begin{equation}
v(z,t)=\sqrt{K}\exp\Bigg[i \frac{\alpha}{2\beta_2}t^2 -\frac{\alpha}{2}z\Bigg]Q(Z,T),
\label{transformation}
\end{equation}
where $Q(Z,T)$ is a solution of the standard NLSE $iQ_Z+\kappa Q_{TT}+\chi|Q|^2Q=0.$
\label{teoremaditrasformazione}
\end{theorem}
{\bf Proof}\qquad Let us substitute $(\ref{c}),(\ref{q}),(\ref{a}),(\ref{p})$ into $(3.17).$ Then, we obtain that
\begin{eqnarray}
v(z,t)&=&\exp\Big[i\frac{\alpha}{2\beta_2}t^2+\ln\sqrt{K}-\frac \alpha 2 z\Big]Q(Z,T)\\ \nonumber
&=&\sqrt{K}\exp\Big[i\frac{\alpha}{2\beta_2}t^2-\frac \alpha 2 z\Big]Q(Z,T)\qquad\qquad\Box\\ \nonumber
\end{eqnarray}

\begin{remark}
From Eq. $(\ref{transformation})$ it easily follows that physical dimension of the solution $v$ of Eq. $(\ref{integrabileNLSE})$ equals the one of the solution $Q$ of Eq. $(\ref{standardNLSE})$ is the same.
\end{remark}
\vspace{.5cm}

We should now claim that the transformations obtained above are invertible. Let us first list them in the following way:
\begin{equation}
\left\{\begin{array}{l}
T(z,t)=Ke^{-\alpha z}t\\
\\
Z(z)=\frac{K\gamma}{2\alpha \chi}(e^{-2\alpha z}-1)\\
\\
v(z,t)=\sqrt{K}e^{i\frac{\alpha}{2\beta_2}t^2-\frac{\alpha}{2}z}Q(Z(z),T(z,t))\\
\end{array}\right.
\label{trasformazioni}
\end{equation}
where $Q(Z,T)$ is a solution of the standard NLSE $(\ref{standardNLSE}),$ and $v(z,t)$ fulfils Eq. $(\ref{integrabileNLSE}).$ 

In order to have inversion, we write the Jacobian matrix of the $(\ref{trasformazioni}):$

$$
\left(\begin{array}{cc}
\frac{\partial Z(z)}{\partial z}&\frac{\partial Z(z)}{\partial t}\\
\\
\frac{\partial T(z,t)}{\partial z}&\frac{\partial T(z,t)}{\partial t}
\end{array}\right)
$$
that is
$$
\left(\begin{array}{cc}
-\frac{K\gamma}{\chi}e^{-2\alpha z}&0\\
\\
-\alpha Ke^{-\alpha z}t&Ke^{-\alpha z}
\end{array}\right)
$$
\vspace{.5cm}

Thus, we observe that the Jacobian (i.e. the determinant of the Jacobian matrix) is different to zero for every $z\in \RR $ and $t\in \RR .$ In order to have a correspondence between the sign of $z$ and $Z$ we restrict $z\in\RR^+$ and we assume that $\chi<0.$ In this way we obtain that $Z(z)\in[0,\frac{K\gamma}{2\alpha |\chi|}),$ for every $z\geq 0.$
\vspace{.5cm}

\begin{remark}
In what follows, unless the contrary is explicitly stated, we assume $\chi<0.$ Thus, we always refer to $Z$ as to belong in a bounded interval of $\RR^+.$ The coefficient $\kappa$ in Eq. $(\ref{standardNLSE})$ can be both positive or negative \textit{real} number.
\end{remark}

\subsection{Mathematical Models}
In the introduction of Chapter $2$ we pointed out the dimensionless Eq. $(\ref{ing1});$ then, we referred to it as Eq. $(\ref{modello})$ that is
\begin{equation}
iu_z(z,t)+ u_{tt}(z,t) +C_1e^{-C_2z}|u(z,t)|^2u(z,t)=0,
\label{modelloing}
\end{equation}
where $C_1=\pm 1$ and $C_2$ is a \textit{real} positive coefficient. The sign of $C_1$ depends on the sign of the parameter $\beta_2$ that represents dispersion of the group velocity and refers to anomalous dispersion if $\beta_2<0$ and normal dispersion if $\beta_2>0.$ We notice that Eq. $(\ref{modelloing})$ is dimensionless, that is $u(z,t)$ is a \textit{complex}-valued function of \textit{real} variables $z,t.$

In this case, the compatibility condition $(3.15)$ leads by means of a straightforwardly calculation to $V_2=\frac{C_2^2}{4}.$  Thus, Eq. $(\ref{integrabileNLSE})$ becomes
\begin{equation}
iv_z(z,t)+v_{tt}(z,t)+C_1\;e^{-C_2 z}|v(z,t)|^2v(z,t)+\frac{C_2^2}{4}t^2v(z,t)=0.
\label{modellointegrabile}
\end{equation}
We immediately notice that this equation is dimensionless.

We now take into consideration a scaling-transformation for Eq. $(\ref{standardNLSE})$ by setting
$$
Q=\sqrt{P_0}Q_{*},\quad Z=\widetilde{L}Z_*,\quad T=\widetilde{T}_0T_*,
$$
where $P_0$ is the peak power of the incident pulse, $\widetilde{L}= \frac{K\gamma}{2\alpha \chi}(e^{-2\alpha L}-1),$ with $L$ the fiber length, and $\widetilde{T}_0=K\;e^{-\alpha L}T_0,$ with $T_0$ the input pulse width.

Therefore, Eq. $(\ref{standardNLSE})$ becomes
\begin{equation}
i\frac{\partial Q_*}{\partial Z_*}+\mbox{sgn}(\kappa)\frac{\widetilde{L}}{\widetilde{L}_D}\frac{\partial^2 Q_*}{\partial T_*^2}-\frac{\widetilde{L}}{\widetilde{L}_{NL}}|Q_*|^2Q_*=0,
\label{scaledSNLSE}
\end{equation}
where $\mbox{sgn}(\kappa)=\pm 1$ depending on the sign of the parameter $\kappa,$ and we defined
$$
\widetilde{L}_D=\frac{\widetilde{T}}{|\kappa|},\qquad \widetilde{L}_{NL}=\frac{1}{|\chi|P_0}.
$$
We now assume the approximation that $\widetilde{L}\sim\widetilde{L}_D\sim\widetilde{L}_{NL};$ moreover, we consider the change of variable $\bar{Z}_*=\mbox{sgn}(\kappa)Z_*.$ Thus, we can refer to Eq. $(\ref{scaledSNLSE})$ as the following equation
\begin{equation}
iQ_Z(Z,T)+Q_{TT}(Z,T)+\rho|Q(Z,T)|^2Q(Z,T)=0,
\label{modelloSNLSE}
\end{equation}
where $\rho=\pm 1.$ Henceforth, unless the contrary is explicitly stated, we refer to dimensionless Eqs. $(\ref{modelloing}),\;(\ref{modellointegrabile}),\;(\ref{modelloSNLSE})$ as Non Linear Schrodinger Equation (NLSE), Transformed Non Linear Schrodinger Equation (TNLSE) and Standard Non Linear Schrodinger Equation (SNLSE), respectively.

Because of Eqs. $(\ref{comparison}),\;(\ref{ceqgenerali}),\;(\ref{agenerale}),\;(\ref{pgenerale}),$ the transformations that allow us to convert Eq. $(\ref{modellointegrabile})$ into Eq. $(\ref{modelloSNLSE})$ are the following

\begin{equation}
\left\{\begin{array}{l}
T(z,t)=\frac{C_1}{\rho}e^{-C_2 z}t\\
\\
Z(z)=\frac{C_1^2}{2\rho^2 C_2}(1-e^{-2C_2 z})\\
\\
v(z,t)=\sqrt{\frac{C_1}{\rho}}\;e^{i\frac{C_2}{4}t^2-\frac{C_2}{2}z}Q(Z(z),T(z,t))\\
\end{array}\right.
\label{trasformazionimodelloinutili}
\end{equation}
provided that the product $C_1\rho>0.$ Thus, we can relax the previous transformations to have
\begin{equation}
\left\{\begin{array}{l}
T(z,t)=e^{-C_2 z}t\\
\\
Z(z)=\frac{1}{2 C_2}(1-e^{-2C_2 z})\\
\\
v(z,t)=\;e^{i\frac{C_2}{4}t^2-\frac{C_2}{2}z}Q(Z(z),T(z,t))\\
\end{array}\right.
\label{trasformazionimodello}
\end{equation}

In this case, we have that $Z(z)\in[0,\frac{1}{2C_2}),$ for every $z\geq 0.$ 

\section{Approximation}
In order to establish a closeness between a solution of Eq. $(\ref{modellointegrabile})$ and a solution of Eq. $(\ref{modelloing}),$ in this section we need to consider a class of functions $v(z,t)$ such that the finiteness of the integral $\int_\RR  t^4|v(z,t)|^2 dt$ holds for every $z$ in a suitable interval of $\RR.$

\vspace{.7cm}

We have seen that, because of Duhamel's formula, if $v_0\in H^1(\RR)$ and $I\ni 0$ is an interval of $\RR,$ an arbitrary $H^1$ solution of the problem
$$
\left\{\begin{array}{l}
iv_z+v_{tt}+C_1e^{-C_2 z}|v|^2v+\frac{C_2^2}{4}t^2v=0  \\
v(0)=v_0
\end{array}\right.
$$
 satisfies the following integral equation
\begin{eqnarray}
v(z)&=&\mathcal{T}(z)v_0+iC_1\int_0^z e^{-C_2 z^\prime}\mathcal{T}(z-z^\prime)\big[|v(z^\prime,\cdot)|^2v(z^\prime,\cdot)\big]dz^\prime\nonumber\\
&+& i\frac{C_2^2}{4}\int_0^z\mathcal{T}(z-z^\prime)t^2v(z^\prime,\cdot),
\label{TNLSEintegrale}
\end{eqnarray}
for all $z\in I,$ where $\mathcal{T}(z)$ is given by
$$
\mathcal{T}(z)\varphi(t)=(4\pi i |z|)^{-1/2}\int_{\RR}e^{\frac{i|t-s|^2}{4z}}\varphi (s)ds.
$$

Similarly, if $u_0\in H^1(\RR)$ and $I\ni 0$ is an interval of $\RR,$ an arbitrary $H^1$ solution of the problem
$$
\left\{\begin{array}{l}
iu_z+u_{tt}+C_1e^{-C_2 z}|u|^2u=0  \\
u(0)=u_0
\end{array}\right.
$$
 satisfies the following integral equation
\begin{eqnarray}
u(z)&=&\mathcal{T}(z)u_0+iC_1\int_0^z e^{-C_2 z^\prime}\mathcal{T}(z-z^\prime)\big[|u(z^\prime,\cdot)|^2u(z^\prime,\cdot)\big] dz^\prime,
\label{NLSEintegrale}
\end{eqnarray}
for all $z\in I.$

In order to establish a distance between the solutions of Eq. $(\ref{TNLSEintegrale})$ and Eq. $(\ref{NLSEintegrale})$ we need to claim that the function $z\rightarrow t^2v(z,t)$ belongs to $ L^2(\RR)$ for every $z$ in a suitable interval of $\RR.$ By using the transformations $(\ref{trasformazionimodello}),$ we easily notice that this claim is equivalent to show that the function $Z\rightarrow T^2 Q(Z,T)$ belongs to $L^2(\RR)$ for every $Z$ in a suitable interval of $\RR.$
\begin{lemma}
Let $I=[0,L]\;(L>0)$ be an interval of $\RR,$  assume that $v(z,t)$ is an arbitrary  solution of equation $(\ref{modellointegrabile})$ on $I,$ and $Q(Z,T)$ an arbitrary solution of equation $(\ref{modelloSNLSE})$ on $\widetilde{I},$ i.e. the transformed of $I$ by $Z(z).$ Then the function $z\rightarrow t^2 v(z,t)\in L^2(\RR)$ for every $z\in I$ if and only if the function $Z\mapsto T^2Q(Z,T)\in L^2(\RR)$ for every $Z\in \widetilde{I}.$
\label{shift}
\end{lemma}
{\bf Proof}\qquad From $(\ref{trasformazionimodello})$ we have that
$$
v(z,t)=e^{i\frac{C_2}{4}t^2-\frac{C_2}{2}z}Q(Z,T)
$$
where
$$
T(z,t)=e^{-C_2z}t\qquad Z(z)=\frac{1}{2 C_2 }(1-e^{-2C_2 z}).
$$
Thanks to these transformations we have that if $I=[0,L],$ then
 $$\widetilde{I}=[0,\frac{1}{2 C_2}(1-e^{-2C_2 L})].$$
Then  obtain that
\begin{eqnarray*}
||t^2v(z,t)||_{L^2}^2&=&\int_{\RR}|t^2v(z,t)|^2 dt\\
&=&\int_{\RR}e^{-C_2z}\Big|\Big(e^{C_2 z}T\Big)^2 Q(Z,T)\Big|^2e^{C_2z}dT\\
&=& e^{4C_2 z}\int_{\RR}|T^2Q(Z,T)|^2 dT
\end{eqnarray*}
Therefore $||(\cdot)^2v(z,\cdot)||_{L^2}=e^{2C_2 z}||(\cdot)^2Q(Z(z),\cdot)||_{L^2}.$ Thus, it follows the thesis of the Lemma. $\Box$

\vspace{.7cm}

Because of Lemma $\ref{shift}$ we can shift the claim that function $z\mapsto t^2v(z,t)$ belongs to $L^2(\RR)$ for every $z$ in a suitable interval or $\RR,$ in the context of standard NLSE. In what follows we carry out some arguments of \cite{Cazenave} to show that the function $Z\mapsto T^2 Q(Z,T)$ belongs to $ L^2(\RR),$ for every $Z$ in a suitable interval of $\RR.$ Recalling that $Q(Z,T)$ is a solution of the NLSE, we start multiplying formally the standard NLSE $(\ref{modelloSNLSE})$ by $|T|^4\overline{Q},$ where $\overline{Q}$ is the complex conjugate of $Q:$
\begin{equation}
iT^4Q_Z\overline{Q}+ T^4Q_{TT}\overline{Q}+\rho T^4|Q|^4=0.
\label{SNLSEmoltiplicata}
\end{equation}

We now integrate over $\RR$ with respect to the variable $T:$
\begin{equation}
i\int_{\RR}T^4Q_Z\overline{Q}dT+ \int_{\RR}T^4Q_{TT}\overline{Q}dT+\rho \int_{\RR}T^4|Q|^4 dT=0.
\label{SNLSEintegrata}
\end{equation}
Let us integrate by part in the second term of the previous equation:
\begin{eqnarray*}
\int_{\RR}T^4Q_{TT}\overline{Q}dT&=&-\int_{\RR}Q_T\frac{d}{dT}(T^4\overline{Q})+\Big[(T^4\overline{Q})Q_T\Big]_{-\infty}^{+\infty}\\
&=&-\int_{\RR}Q_T\frac{d}{dT}(T^4\overline{Q})+0\\
&=&-\int_{\RR}T^4|Q_T|^2 dT -4\int_{\RR}T^3Q_T\overline{Q}dT,
\end{eqnarray*}
then we take the imaginary part of Eq. $(\ref{SNLSEintegrata})$ and we obtain that
$$
\frac 1 2 \frac{d}{dZ}\int_{\RR}|T^2 Q|^2 dT=4\mbox{Im}\int_{\RR}(T^2\overline{Q})(TQ_T) dT
$$
We require the finiteness of the integral $\int_{\RR}|T|^2|Q_T(Z,T)|^2 dT,$ for every $Z$ in a suitable interval, in order to use Cauchy-Schwartz inequality. So let us set $W(Z,T)=Q_T(Z,T);$ furthermore, we differentiate the standard NLSE $(\ref{modelloSNLSE})$ with respect to the variable $T$ and obtain:
$$
iQ_{ZT}+ Q_{TTT}+\rho (2|Q|^2 Q_T+Q^2\overline{Q}_T)=0
$$
finally,  we have that $W$ fulfils
\begin{equation}
iW_Z+ W_{TT}+\rho (2|Q|^2W+Q^2\overline{W})=0
\label{W}
\end{equation}
We are now ready to prove the following result
\begin{pro}
Let $I=[0,L]\;(L>0)$ be a bounded interval of $\RR;$  let $Q(Z,T)$ be a $H^2$  solution on $I$ of the standard NLSE $(\ref{modelloSNLSE}).$ Let $W=Q_T$ be a solution of Eq. $(\ref{W})$ with $W(0)=Q_T(0).$
\\
 If $|\cdot|W(0)\in L^2(\RR)$ then the function $Z\mapsto |T|W(Z,T)$ belongs to $L^\infty(I,L^2(\RR)).$
\end{pro}
{\bf Proof}\qquad We start multiplying formally Eq. $(\ref{W})$  by $T^2\overline{W}:$
$$
iT^2W_Z\overline{W}+ T^2W_{TT}\overline{W}+\rho (2T^2|Q|^2|W|^2+T^2Q^2\overline{W}^2)=0.
$$

We now integrate over $\RR $ with respect to the variable $T,$ then we take the imaginary part, and so we have that
\begin{eqnarray*}
&&\frac 1 2 \frac{d}{dZ}\int_{\RR}T^2|W(Z,T)|^2 dT=-\mbox{Im}\Bigg(\int_{\RR}T^2W_{TT}(Z,T)\overline{W}(Z,T)dT\\
&& +2\rho\int_{\RR}T^2|Q(Z,T)|^2|W(Z,T)|^2 dT +\rho \int_{\RR}T^2Q^2(Z,T) \overline{W}^2(Z,T) dT\Bigg).
\end{eqnarray*}
We notice that the second integral on the right-hand side of the previous equation is a \textit{real} quantity, so we have that
\begin{eqnarray*}
&&\frac 1 2 \frac{d}{dZ}\int_{\RR}T^2|W(Z,T)|^2 dT=-\mbox{Im}\Bigg(\int_{\RR}T^2W_{TT}(Z,T)\overline{W}(Z,T)dT\\
&&  +\rho \int_{\RR}T^2Q^2(Z,T) \overline{W}^2(Z,T) dT\Bigg).
\end{eqnarray*}
Taking an integration by part with respect the variable $T$ in the first integral on the right-hand side of the previous equation we obtain that
\begin{eqnarray*}
&&\frac 1 2 \frac{d}{dZ}\int_{\RR}T^2|W(Z,T)|^2 dT= \mbox{Im}\Bigg(\int_{\RR}\Big[|W_T(Z,T)|^2 T^2\\
&& +2W_T(Z,T) T\overline{W}(Z,T)\Big] dT-\int_{\RR}\rho T^2Q^2(Z,T) \overline{W}^2(Z,T) dT\Bigg)
\end{eqnarray*}
finally, since the first integral on the right-hand side of previous equation is a \textit{real} quantity we arrive at
\begin{eqnarray*}
&&\frac 1 2 \frac{d}{dZ}\int_{\RR}T^2|W(Z,T)|^2 dT=\mbox{Im}  \int_{\RR}\Big[ 2W_T(Z,T) T\overline{W}(Z,T)\\
&&-\rho T^2Q^2(Z,T) \overline{W}^2(Z,T)\Big] dT\\
&&=\mbox{Im}  \int_{\RR}T\overline{W}(Z,T)\Big[2 W_T(Z,T)-\rho T\overline{W}(Z,T)Q^2(Z,T)\Big] dT .
\end{eqnarray*}
Let us set $h(Z)=||TW(Z)||_{L^2}^2;$ we have from the previous relation that 
$$
h^\prime(Z)=2\mbox{Im}  \int_{\RR}T\overline{W}(Z,T)\Big[2 W_T(Z,T)-\rho T\overline{W}(Z,T)Q^2(Z,T)\Big] dT
$$
We observe that we obtained the previous relation only by a formal calculus. It is not a rigorous equation.

So by the fundamental theorem of calculus we have that
$$
h(Z)=h(0)+\int_0^Z 2\mbox{Im} \Big( \int_{\RR}T\overline{W}(Z^{\prime})\Big[2 W_T(Z^{\prime})-\rho T\overline{W}(Z^{\prime})Q^2(Z^{\prime})\Big] dT\Big) dZ^{\prime} .
$$
From Cauchy-Schwartz inequality we obtain
$$
h(Z)\leq h(0)+2 \int_0^Z ||TW(Z^{\prime})||_{L^2}||2 W_T(Z^{\prime})-\rho T\overline{W}(Z^{\prime})Q^2(Z^{\prime})||_{L^2}dZ^{\prime}
$$
and from triangular inequality we have that
$$
h(Z)\leq h(0)+2 \int_0^Z ||TW(Z^{\prime})||_{L^2}\Big[||2 W_{T}(Z^{\prime})||_{L^2}+||\rho T\overline{W}(Z^{\prime})Q^2(Z^{\prime})||_{L^2}\Big]dZ^\prime .
$$

Now by using Holder inequality and the Sobolev immersion $H^2(\RR)\hookrightarrow L^\infty(\RR)$ we notice that:
\begin{eqnarray*}
||\rho T\overline{W}Q^2||_{L^2}&=&|\rho|  \Big(\int_{\RR}|TW|^2 |Q^2|^2 dT\Big)^{1/2}\\
&\leq& |\rho| \Big(||Q||_{L^\infty}^4||TW||_{L^2}^2\Big)^{1/2}\\
&=&|\rho|\; ||Q||_{L^\infty}^2||TW||_{L^2}
\end{eqnarray*}
So we uncover this relation
$$
h(Z)\leq h(0)+2\int_0^Z\Big[2||TW(Z^{\prime})||_{L^2}|| W_{T}(Z^{\prime})||_{L^2}+|\rho|\;||TW(Z^{\prime})||_{L^2}^2||Q(Z^{\prime})||^2_{L^\infty}\Big]dZ^\prime
$$

\vspace{.5cm}

 By assumption we have that $Q(Z,T)\in C(I,H^2(\RR)).$  Since  $W(Z,T)=Q_T(Z,T),$ we have that $||W_T(Z)||_{L^2}$ is finite for every $Z\in I.$  Moreover, since $||Q(Z)||_{L^\infty}\leq C||Q(Z)||_{H^2},$ with $C$ \textit{real} positive constant, we have that $||Q(Z)||^2_{L^\infty}$ is finite for every $Z\in I.$ In this way we obtain that
\begin{equation}
h(Z)\leq h(0)+\widetilde{C}\int_0^Z\Big[\sqrt{h(Z^\prime)}+h(Z^\prime)\Big]dZ^\prime
\label{gronwallh}
\end{equation}
where  $\widetilde{C}=\max\Bigl\{2\sup_{Z\in I}||W_T(Z)||_{L^2},\sup_{Z\in I}||Q(Z)||^2_{L^\infty}\Bigr\}.$ We should now have that $h(Z)$ is finite for all $Z\in I.$ First we observe that $h(0)=||TW(0)||_{L^2},$ that is finite by hypothesis. By contradiction, we assume that there exists $Z_0\in I$ such that 
$$
\lim_{Z\rightarrow Z_0}h(Z)=\infty;
$$
then there is $\delta >0$ such that $h(Z)>1$ for all $Z\in J=(\delta, Z_0).$ In this way we can write, also because  $h$ is a positive function, that
\begin{eqnarray*}
h(Z_0)&\leq &h(0)+C\int_0^{Z_0}h(Z^\prime)\Big[1+(h(Z^\prime))^{-1/2}\Big]dZ^\prime\\
&\leq&  h(0)+C\int_\delta^{Z_0}h(Z^\prime)\Big[1+(h(Z^\prime))^{-1/2}\Big]dZ^\prime\\
&\leq& h(0)+2C\int_\delta^{Z_0}h(Z^\prime)dZ^\prime
\end{eqnarray*}
and from Gronwall Lemma it follows that $h(Z_0)$ is finite; that is a contradiction. Thus, we obtain that the function $Z\mapsto |T|W(Z,T)$ belongs to $L^2(\RR)$ for every $Z\in I. \Box$
\vspace{.5cm}

\begin{remark}
Let $I=(-L,L)\;(L>0)$ be an  interval of $\RR,$ and $Q(Z,T)$ a $H^2$ solution of the standard NLSE on $I.$ We should claim that $||Q(Z)||_{H^2}$ is finite for every $Z\in \RR.$ Because of Proposition $2.7$ we know that 
$$
||Q(Z)||_{L^2}+||Q_T(Z)||_{L^2}
$$
is finite for every $Z\in\RR.$ Thus, it remains to estimate $||Q_{TT}(Z)||_{L^2}.$

 Let us set $g(Q)=\rho|Q|^2 Q,$ we have that $g\in C(H^2(\RR),L^2(\RR))$ and it is bounded on bounded sets; indeed
$$
||\rho|Q(Z)|^2Q(Z)||_{L^2}\leq |\rho|\; ||Q^2(Z)||_{L^\infty}||Q(Z)||_{L^2}.
$$

Let us set
$$
V(Z)=i\int_0^Z\mathcal{T}(Z-Z^\prime)f(Z^\prime)dZ^\prime.
$$
Since $g(Q)\in L^\infty(I,L^2(\RR))$ and $\frac{d}{dZ}g(Q)\in L^\infty(I,L^2(\RR))$  from [\cite{Cazenave}, Lemma $4.8.5$] we have the following estimate:
\begin{equation*}
||V_{TT}||_{L^\infty(I,L^2)}\leq ||g(Q)||_{L^\infty(I,L^2)}+C||\frac{d}{dZ}g(Q)||_{L^\infty(I,L^2)}+C||g(Q(0))||_{L^2},
\end{equation*}
where $C$ is independent of $I$ and $f.$ If, in addition, $f\in C(I,L^2(\RR)),$ then $V\in C(I,L^2(\RR)).$

Furthermore, from [\cite{Cazenave}, Lemma $4.8.3 $] we have that 
$$
||g(Q)||_{L^\infty(I,L^2(\RR))}\leq L(M)\qquad\mbox{for}\quad ||Q||_{L^\infty(I,H^1(\RR))}\leq M
$$
where $L(M)$ is a coefficient depending on  $M.$

Moreover, from [\cite{Cazenave}, Lemma $4.8.4$] we have that
$$
||\frac{d}{dZ}g(Q)||_{L^\infty(I,L^2)}\leq L(M) ||Q_Z||_{L^\infty(I,L^2)}
$$
where $L(M)$ is a coefficient depending on  $M,$ and $||Q||_{L^\infty(I,H^1(\RR))}\leq M.$

We now notice that $||Q_Z(Z)||_{L^2}=\frac{d}{dZ}||Q(Z)||_{L^2}=0$ since the $L^2$ norm is preserved with respect to the variable $Z.$

Therefore, by setting
$$
Q(Z,T)=\mathcal{T}(Z)Q(0)+i\rho\int_0^Z\mathcal{T}(Z-Z^\prime)|Q(Z^\prime)|^2Q(Z^\prime)dZ^\prime
$$
we have that
$$
||Q_{TT}(Z)||\leq L(M)+|\rho|\;||Q(0)||^2_{L^\infty}||Q(0)||_{L^2}.
$$
Since $||Q(Z)||_{H^1}\leq M$ for every $Z\in \RR $ we have that $||Q(Z)||_{H^2}$ if finite for every $Z\in \RR.$
\end{remark}
\vspace{.5cm}

\begin{remark}
The previous remark allows us to consider the coefficient $\widetilde{C}$ in $(\ref{gronwallh})$ independent of $Z.$ This remarkable consideration will be very useful in the next.
\end{remark}
\vspace{.5cm}

\begin{remark}
Relation $(\ref{gronwallh})$ can be solved with respect the function $h(Z)$ by using a Gronwall argumentation:
let us set $I=[0,L]$ a bounded interval of $\RR $ and 
$$u(Z)\Doteq h(0)+\widetilde{C}\int_0^Z\Big[\sqrt{h(Z^\prime)}+h(Z^\prime)\Big]dZ^\prime.$$
From $(\ref{gronwallh})$ it follows that $h(Z)\leq u(Z)$ for every $Z\in I.$ Furthermore, we can consider the coefficient $\widetilde{C}$ independent on $Z,$ because of Remark $3.4.$ Thus, we can write 
\begin{eqnarray*}
u^\prime(Z)&=&\widetilde{C}\Big[\sqrt{h(Z)}+h(Z)\Big]\\
&\leq&\widetilde{C}\Bigg[\sqrt{h(0)+\widetilde{C}\int_0^Z\Big[\sqrt{h(Z^\prime)}+h(Z^\prime)\Big]dZ^\prime}\\
&+& h(0)+\widetilde{C}\int_0^Z\Big[\sqrt{h(Z^\prime)}+h(Z^\prime)\Big]dZ^\prime\Bigg]
\end{eqnarray*}
Therefore by differentiating the function $\ln[\sqrt{u}+1]$ with respect the variable $Z,$ we obtain that
\begin{eqnarray*}
\frac{d}{dZ}\ln[\sqrt{u(Z)}+1]&=& \frac{u^\prime(Z)}{2[\sqrt{u(Z)}+u(Z)]}\leq \frac{\widetilde{C}}{2}
\end{eqnarray*}
We now integrate with respect to the variable $Z$ and we obtain that
\begin{eqnarray*}
\ln[\sqrt{u(Z)}+1]&\leq&\ln[\sqrt{u(0)}+1]+\int_0^Z\frac{\widetilde{C}}{2}dZ^\prime\qquad \mbox{for every}\quad Z\in I.
\end{eqnarray*}
It then follows that
\begin{eqnarray*}
\sqrt{u(Z)}+1&\leq&[\sqrt{u(0)}+1]e^{\int_0^Z\frac{\widetilde{C}}{2}dZ^\prime},
\end{eqnarray*}
and so being $u(0)=h(0),$ we have that
\begin{eqnarray*}
\sqrt{h(Z)}&\leq&[\sqrt{h(0)}+1]e^{\int_0^Z\frac{\widetilde{C}}{2}dZ^\prime}-1,\qquad \mbox{for every}\quad Z\in I.
\end{eqnarray*}
\end{remark}
\vspace{.7cm}

We are now ready to prove the following result:

\begin{theorem}
Let $I=[0,L]\;(L>0)$ be a bounded interval of $\RR ,$ let $Q_0(T)\in H^2(\RR)$ and $Q(Z,T)$ a $H^2$-solution of
$$
\left\{\begin{array}{l}
iQ_Z+ Q_{TT}+\rho|Q|^2Q=0\\
Q(0,T)=Q_0(T)
\end{array}\right.
$$
on $I.$ If $T^2Q(0,T)\in L^2(\RR)$ and $|T|Q_T(0,T)\in L^2(\RR),$ then the function $Z\mapsto T^2 Q(Z,T)$ belongs to $L^\infty(I,L^2(\RR)).$
\label{teoremadishift}
\end{theorem}
{\bf Proof}\qquad Multiplying formally standard NLSE by $T^4\overline{Q},$ integrating over $\RR $ and taking the imaginary part we have that
$$
\frac 1 2 \frac{d}{dZ}\int_{\RR}|T^2 Q(Z,T)|^2 dT=4\mbox{Im}\int_{\RR}T^2\overline{Q}(Z,T)TQ_T(Z,T) dT
$$
We recall that the previous relation is obtained by formal calculus and it is not a rigorous equation. Now, let us define $ f(Z)=||T^2Q(Z,\cdot)||_{L^2}^2;$ then  from the fundamental theorem of calculus we have that
\begin{eqnarray*}
f(Z)&=&f(0)+\int_0^Z f^\prime(Z^\prime) dZ^\prime\\
&=&f(0)+\int_0^Z 8\mbox{Im}\int_{\RR}T^2\overline{Q}(Z^\prime,T)TQ_T(Z^\prime,T) dT dZ^\prime
\end{eqnarray*}
\\
Since  $TQ_T(0,T)\in L^2(\RR)$ by hypothesis, it follows from proposition $3.2$ that the function $Z\mapsto TQ_T(Z,T)$ belongs to $L^\infty(I,L^2(\RR)).$  From Cauchy-Schwartz inequality it follows that 

\begin{equation}
f(Z)\leq f(0)+8\int_0^Z||TQ_T(Z^\prime)||_{L^2} \sqrt{f(Z^\prime)} dZ^\prime .
\label{stimaf}
\end{equation}

We should now have that $f(Z)$ is finite in $I;$ we first observe that $f(0)=||T^2Q(0)||_{L^2}^2,$ that is finite by hypothesis. Then let us set  $$g(Z)=||TQ_T(Z)||_{L^2} \big(f(Z)\big)^{-1/2}.$$ If $f(Z)$ is not finite  for all $Z\in I,$ there exists $Z_0\in I$ such that $$\lim_{Z\rightarrow Z_0}f(Z)=\infty;$$ for this reason there exists a real number $\delta>0$ such that  one has that $f(Z)>||TQ_T(Z)||^{-2},$ for all $Z\in J=(\delta,Z_0).$ Now, because of $g$ and $f$ are positive functions on $I,$ we have that
\begin{eqnarray*}
f(Z_0)&\leq& ||TQ_0||_{L^2}^2+\int_0^{Z_0} g(Z^\prime)f(Z^\prime) dZ^\prime\\
&\leq&||TQ_0||_{L^2}^2+\int_\delta^{Z_0} g(Z^\prime)f(Z^\prime) dZ^\prime\\
&\leq & ||TQ_0||_{L^2}^2+\int_\delta^{Z_0}f(Z^\prime) dZ^\prime\\
\end{eqnarray*}
where the last inequality holds true because of  $\big(f(Z)\big)^{-1/2}\leq ||TQ_T(Z)||_{L^2},$ for all $Z\in J$ and so $g(Z)\leq 1.$ Therefore we have that
$$
f(Z_0)\leq ||TQ_0||_{L^2}^2+\int_\delta^{Z_0}f(Z^\prime)dZ^\prime
$$
and the contradiction follows from Gronwall Lemma.

We conclude that the function $Z\mapsto T^2 Q(Z,T)$ belongs to $L^\infty(I,L^2(\RR)).\Box$

\begin{remark}
Relation $(\ref{stimaf})$ can be solved with respect to $f(Z):$ let us set $I\ni 0$ a bounded interval of $\RR^+$ and
\begin{equation*}
u(Z)\Doteq f(0)+8\int_0^Z ||TQ_T(Z^\prime)||_{L^2} \sqrt{f(Z^\prime)} dZ^\prime .
\end{equation*}

 By differentiating with respect to the variable $Z$ we have that
\begin{eqnarray*}
u^\prime(Z)&=& 8||TQ_T(Z)||_{L^2} \sqrt{f(Z)}\\
&\leq& 8||TQ_T(Z)||_{L^2} \sqrt{f(0)+8\int_0^Z ||TQ_T(Z^\prime)||_{L^2} \sqrt{f(Z^\prime)} dZ^\prime }
\end{eqnarray*}
where we used the relation $(\ref{stimaf}),$ that is
 $$f(Z)\leq f(0)+8\int_0^Z||TQ_T(Z^\prime)||_{L^2} \sqrt{f(Z^\prime)} dZ^\prime.$$ 
 Therefore by differentiating the function $\sqrt{u(Z)}$ we obtain that
\begin{eqnarray*}
\frac{d}{dZ}\sqrt{u(Z)}&=&\frac{u^\prime(Z)}{2\sqrt{u(Z)}}\\
\\
&\leq& 4||TQ_T(Z)||_{L^2}.
\end{eqnarray*}
We now integrate with respect to the variable $Z$ and obtain that
$$
\sqrt{u(Z)} \leq \sqrt{u(0)}+4\int_0^Z||TQ_T(Z^\prime)||_{L^2}dZ^\prime.
$$
We now observe that $u(0)=f(0)$ and from $(\ref{stimaf})$ and the definition of $u(Z),$ it follows that  $\sqrt{f(Z)}\leq \sqrt{u(Z)}$ for every $Z\in I.$ So, we can conclude that
\begin{equation}
\sqrt{f(Z)}\leq \sqrt{f(0)}+4\int_0^Z||TQ_T(Z^\prime)||_{L^2}dZ^\prime.
\label{gronwallf}
\end{equation}

\end{remark}

\vspace{1cm}

As a consequence of theorem $\ref{teoremadishift}$ we have the following result that will be useful in the next.

\begin{lemma}
Let $I=[0,L]\;(L>0)$ be a bounded interval of $\RR,$ let $Q_0\in H^2,$ and let $Q(Z,T)\in L^\infty(I,H^2(\RR))$ be a  solution of the problem
\begin{equation}
\left\{\begin{array}{l}
iQ_Z(Z,T)+Q_{TT}(Z,T)+\rho|Q(Z,T)|^2Q(Z,T)=0\\
Q(0,T)=Q_0(T)\\
\end{array}\right.
\label{IVPSNLSE}
\end{equation} 
on $I.$ If $T^2Q_0(T)\in L^2(\RR),$ then the function $Z\mapsto |T|Q(Z,T)$ belongs to $L^\infty(I,L^2(\RR)).$ 
\end{lemma}
{\bf Proof}\qquad Since $T^2Q_0(T)\in L^2(\RR),$ because of theorem $3.4.1$ we deduce that the function
$Z\mapsto T^2Q(Z,T)$ belongs to $L^\infty(I,L^2(\RR)).$ Now we have that
$$
\int_{\RR}|T|^2|Q(Z,T)|^2 dT=\int_{\RR}(T^2Q(Z,T))\overline{Q}(Z,T) dT .
$$
We apply the Cauchy-Schwartz inequality and we obtain that
$$
\int_{\RR}|T|^2|Q(Z,T)|^2 dT \leq \Big(\int_{\RR}|(T|^2|Q(Z,T)|)^2 dT\Big)^{1/2}\Big(\int_{\RR}|Q(Z,T)|^2 dT\Big)^{1/2},
$$
for every $Z\in I.$ Thus, because of Theorem $3.4.1$ and  $Q(Z,\cdot)\in H^2(\RR)$ the finiteness of the integrals on the right side of the previous inequality holds. Therefore, the thesis of the Lemma follows. \quad $\Box$

\vspace{.7cm}
We should now make use of Theorem $\ref{teoremadishift}$ and Lemma $\ref{shift}$ to prove the claim that the function $z\mapsto t^2v(z,t)$ belongs to $L^2(\RR)$ for every $z$ in a suitable interval of $\RR^+.$ Because of Theorem $\ref{teoremaditrasformazione}$ we can write
\begin{equation}
v(z,t)=\;e^{i\frac{C_2}{4}t^2-\frac{C_2}{2}z}Q(Z(z),T(z,t)).
\label{trasformata}
\end{equation}
For convenience we write the inverse transformations of $(\ref{trasformazionimodello}):$
\begin{equation}
\left\{\begin{array}{l}
t(Z,T)=\Big(1-2C_2Z\Big)^{-1/2}T\\
\\
z(Z)=-\frac{1}{2C_2}\ln\Big(1-2C_2Z\Big)\\
\\
Q(Z,T)=\;e^{i\frac{C_2}{4}t^2-\frac{C_2}{2}z}v(z(Z),t(Z,T))\\
\end{array}\right.
\label{trasformazioniinversemodello}
\end{equation}
where $Z\in [0,\frac{1}{2C_2}).$
\vspace{.5cm}

\begin{cor}
Let $I=[0,L]\;(L>0)$  be a bounded interval of $\RR,$ let us set $J=Z(I).$ Let  
$
v(z,t)
$
as in Eq. $(\ref{trasformata}),$
with $Q(Z,T)\in C(J,H^2(\RR))$  solution of the standard NLSE $(\ref{modelloSNLSE}).$ If $t^2v(0,\cdot)\in L^2(\RR)$ and $|t|v_t(0,\cdot)\in L^2(\RR)$ then the function $z\mapsto t^2v(z,t)$ belongs to $L^2(\RR)$ for every $z\in I.$
\end{cor}
{\bf Proof}\qquad In order to apply Theorem $\ref{teoremadishift}$ we need $T^2Q(0,T)\in L^2(\RR)$ and $|T|Q_T(0,T)\in L^2(\RR).$ Thus, making use of $(\ref{trasformazioniinversemodello})$ we obtain that
$$
Q(0,T)=e^{-i\frac{C_2}{4}t^2(0,T)}v(0,t(0,T)).
$$
We differentiate this relation  with respect to the variable $T$ and we have that:
\begin{eqnarray*}
Q_T(0,T)&=&e^{-i\frac{C_2}{4}t^2}\big[-i\frac{C_2}{2}tv(0,t)\frac{dt}{dT}+v_t(0,t)\frac{dt}{dT} \big]\\
&=&e^{-i\frac{C_2}{4}t^2}\big[-i\frac{C_2}{2}tv(0,t)+v_t(0,t)\big]
\end{eqnarray*}
where we used  $t(0,T)=T$ and  $\frac{dt}{dT}=1.$
By using the relation $T(0,t)=t$ we obtain that
\begin{eqnarray*}
TQ_T(0,T)&=&te^{-i\frac{C_2}{4}t^2}\big[-i\frac{C_2}{2}tv(0,t)+v_t(0,t)\big]\\
\end{eqnarray*}
We now integrate this relation over $\RR$ with respect the variable $T,$ and by making use of the equation $dT=dt$ we obtain that
\begin{eqnarray*}
\int_{\RR}|TQ_T(0,T)|^2 dT&=&\int_{\RR}\Big|\big[-i\frac{C_2}{2}t^2v(0,t)+tv_t(0,t)\big]\Big|^2 dt\\
&\leq& \Big[\Big(\frac{C_2}{2}\Big)^2\int_{\RR}\Big|t^2v(0,t)\Big|^2 dt+\int_{\RR}\Big|tv_t(0,t)\Big|^2 dt\Big],
\end{eqnarray*}
so by hypothesis we have that $TQ_T(0,T)$ belongs to $L^2(\RR).$

Furthermore, we have that
\begin{eqnarray*}
\int_{\RR}\Big|T^2 Q(0,T)\Big|^2 dT &=& \int_{\RR}\Big|\Big(t\Big)^2 e^{-i\frac{C_2}{4}t^2}v(0,t)\Big|^2 dt\\
&=&\int_{\RR}|t^2v(0,t)|^2 dt,
\end{eqnarray*}
so also $T^2 Q(0,T)$ belongs to $L^2(\RR).$ Therefore, by Theorem $\ref{teoremadishift}$ we have that the function $Z\mapsto T^2Q(Z,T)$ belongs to $L^2(\RR)$ for every $Z\in J.$ Finally,  from Lemma $\ref{shift}$ it follows that
$$
||t^2v(z)||_{L^2}= e^{2C_2 z}||T^2Q(Z)||_{L^2},
$$
thus, we deduce that the function $z\mapsto t^2v(z,t)$ belongs to $L^2(\RR)$ for every $z\in I.\quad \Box$

\vspace{1cm}
As an application of Theorem $\ref{teoremadishift}$ we have the following result 
\begin{pro}
Let $I=[0,L]\;(L>0)$ be a bounded interval of $\RR,$ $Q(Z,T)\in C(I,H^2(\RR))$ be a solution of the problem  $(\ref{IVPSNLSE})$ on $I.$ Let us assume that $ T^2Q(0,T)\in L^2(\RR),\;|T|Q_T(0,T)\in L^2(\RR).$   Then 
$$
 v(z,t)=e^{i\frac{C_2}{4}t^2-\frac{C_2}{2}z}Q(Z,T)
 $$
 belongs to $C(\widetilde{I},H^2(\RR)),$ where $\widetilde{I}=Z^{-1}(I)$ is given by the inverse transform of $ Z(z)=\frac{1}{2 C_2 }(1-e^{-2C_2 z}).$
\end{pro}
{\bf Proof}\qquad We recall that the $H^2$ norm is defined as follows:
$$
||v(z)||_{H^2}^2\Doteq ||v(z)||_{L^2}^2+||v_t(z)||_{L^2}^2+||v_{tt}(z)||_{L^2}^2.
$$
So we begin by evaluating the $L^2$ norm of $v(z,t),$ that is
\begin{eqnarray*}
||v(z)||_{L^2}^2&=&e^{-C_2 z}\int_{\RR}|Q(Z,T)|^2 e^{C_2z} dT\\
&=&\int_{\RR}|Q(Z,T)|^2 dT
\end{eqnarray*}
and so it is finite for all $z\in\widetilde{I}$ because of $Q(Z,\cdot)\in H^2(\RR)$ for every $Z\in I.$ Moreover
\begin{eqnarray*}
\frac{\partial v(z,t)}{\partial t}&=&e^{i\frac{C_2}{4}t^2-\frac{C_2}{2}z}\Big[+i\frac{C_2}{2}tQ(Z,T)+\frac{dT}{dt}Q_T(Z,T)\Big]\\
&=&e^{i\frac{C_2}{4}t^2-\frac{C_2}{2}z}\Big[+i\frac{C_2}{2}e^{C_2z}TQ(Z,T)+e^{-C_2z}Q_T(Z,T)\Big],
\end{eqnarray*}
where we used the inverse transformation $t=e^{-C_2 z}T.$ Thanks to the lemma $3.4.2.$ it follows that the integral $\int_{\RR}|TQ(Z,T)|^2 dT$ is finite; in addition, the integral $\int_{\RR}|Q_T(Z,T)|^2 dT$ is finite because of $Q(Z,\cdot)\in H^2(\RR).$ We can then conclude that the integral
$$
\int_{\RR}|v_t(z,t)|^2 dt
$$
is finite for every $z\in I.$ Furthermore, since $\frac{dT}{dt}=\;e^{-C_2 z}$ and $t= \;e^{C_2 z}T,$ we have that

\begin{eqnarray*}
\frac{\partial^2 v(z,t)}{\partial t^2}&=&e^{i\frac{C_2}{4}t^2-\frac{C_2}{2}z}(i\frac{C_2}{2})t\Big[i\frac{C_2}{2}tQ(Z,T)+e^{-C_2z}Q_T(Z,T)\Big]\\
&&+e^{i\frac{C_2}{4}t^2-\frac{C_2}{2}z}\Big[i\frac{C_2}{2}Q(Z,T)+i\frac{C_2}{2}t\frac{dT}{dt}Q_T(Z,T)\\
&&+e^{-C_2z}Q_{TT}(Z,T)\frac{dT}{dt}\Big]\\
&=&e^{i\frac{C_2}{4}t^2-\frac{C_2}{2}z}(i\frac{C_2}{2})e^{C_2z}T\Big[i\frac{C_2}{2}e^{C_2z}TQ(Z,T)\\
&&+e^{-C_2z}Q_T(Z,T)\Big]+e^{i\frac{C_2}{4}t^2-\frac{C_2}{2}z}\Big[i\frac{C_2}{2}Q(Z,T)\\
&&i\frac{C_2}{2}e^{C_2z}Te^{-C_2z}Q_T(Z,T)+e^{-C_2z}Q_{TT}(Z,T)e^{-C_2z}\\
&=&e^{i\frac{C_2}{4}t^2-\frac{C_2}{2}z}\Big[-\frac{C_2^2}{4}e^{2C_2z}T^2Q(Z,T)\\
&&+i\frac{C_2}{2}TQ_T(Z,T)+i\frac{C_2}{2}Q(Z,T)+i\frac{C_2}{2}TQ_T(Z,T)+e^{-2C_2z}Q_{TT}(Z,T)\Big]\\
&=&e^{i\frac{C_2}{4}t^2-\frac{C_2}{2}z}\Big[-\frac{C_2^2}{4}e^{2C_2z}T^2Q(Z,T)+iC_2 TQ_T(Z,T)\\
&&+e^{-2C_2z}Q_{TT}(Z,T)+i\frac{C_2}{2}Q(Z,T)\Big].\\
\end{eqnarray*}
We now observe that the integral $\int_{\RR}|Q_{TT}(Z,T)|^2 dT$ is finite because of $Q(Z,\cdot)\in H^2(\RR).$ Moreover from theorem $3.4.1$ and proposition $3.2$ it follows that the integrals
$$
\int_{\RR}|T^2Q(Z,T)|^2 dT \qquad \int_{\RR}|TQ_T(Z,T)|^2 dT
$$
are finite. So we easily conclude that the integral
$$
\int_{\RR}|v_{tt}(z,t)|^2 dt
$$
is finite. We deduce that the function $v(z,\cdot)$ belongs to $H^2(\RR)$ for every $z\in \widetilde{I}. \Box$
\vspace{.5cm}

\begin{remark}
Let $I=[0,L]\;(L>0)$ be an interval of $\RR$ and $Z(z)=\frac{1}{2 C_2}(1-e^{-2C_2z})$ for every $z\in I.$ We know that there exists a $H^2$ solution $Q(Z,T)$ of the standard NLSE $(\ref{modelloSNLSE})$ on $Z(I).$ Thus, from Theorem $\ref{teoremaditrasformazione}$ we have a solution of the transformed NLSE given by
$$
v(z,t)=e^{i\frac{C_2}{4}t^2-\frac{C_2}{2}z}Q(Z,T).
$$
Because of Proposition $3.3$ we now know under which assumptions on $Q(Z,T)$ this function is a $H^2$ solution of the transformed NLSE $(\ref{modellointegrabile}).$
\end{remark}

\vspace{1cm}

\vspace{1cm}

Now we can produce the following definition
\begin{definition}
Let $v_0\in H^2(\RR)$ and $I=[0,L]\;(L>0)$ be a bounded interval of $\RR.$ Let $v(z,t)$ be a $H^2$ solution on $I$ of
\begin{equation}
\left\{\begin{array}{l}
iv_z+ v_{tt}+C_1e^{-C_2 z} |v|^2v+\frac{C_2^2}{4}t^2v=0\\
v(0,t)=v_0(t)
\end{array}\right.
\label{IVPTNLSE}
\end{equation}
and $u(z,t)$ a $H^2$ solution on $I$ of
\begin{equation}
\left\{\begin{array}{l}
iu_z+ u_{tt}+C_1e^{-C_2 z} |u|^2u=0\\
u(0,t)=v_0(t)
\end{array}\right.
\label{IVPNLSE}
\end{equation}
We define the distance between $u(z,t)$ and $v(z,t)$ as
\begin{equation}
d(v(z),u(z))=||v(z)-u(z)||_{L^2}
\label{distanza}
\end{equation}
\end{definition}

\vspace{.5cm}
We know that a solution of Eq. $(\ref{IVPTNLSE})$ can be expressed by
\begin{eqnarray*}
v(z)&=&\mathcal{T}(z)v_0+iC_1\int_0^z e^{-C_2 z^\prime}\mathcal{T}(z-z^\prime)|v(z^\prime,\cdot)|^2v(z^\prime,\cdot)dz^\prime\nonumber\\
&+& i\frac{C_2^2}{4}\int_0^z\mathcal{T}(z-z^\prime)t^2v(z^\prime,\cdot)dz^\prime.
\end{eqnarray*}
and similarly we can express a solution of Eq. $(\ref{IVPNLSE})$ as
\begin{eqnarray*}
u(z)&=&\mathcal{T}(z)u_0+iC_1\int_0^z e^{-C_2 z^\prime}\mathcal{T}(z-z^\prime)|u(z^\prime,\cdot)|^2u(z^\prime,\cdot)dz^\prime.\
\end{eqnarray*}

In what follows we state two properties about the unitary semi group operators $(\mathcal{T}(z))_{z\geq 0}.$ We have \cite{Cazenave}
\begin{pro}
For every $\varphi\in L^2(\RR),$ the function $z\rightarrow\mathcal{T}(z)\varphi $ belongs to
$$
L^\infty(\RR,L^2(\RR))\cap C(I,L^2(\RR))
$$
and
$$
||\mathcal{T}(z)\varphi||_{L^2}=||\varphi ||_{L^2}.
$$
\end{pro}

\begin{pro}
Let $I\ni 0$ be an interval of $\RR,$ $J=\overline{I}.$ Let $f\in L^1(I,L^2(\RR)),$ then function 
$$
z\rightarrow\Phi_f(z)=\int_0^z\mathcal{T}(z-z^\prime)f(z^\prime)dz^\prime \qquad \mbox{for}\quad z\in I
$$
belongs to $L^\infty(I,L^2(\RR))\cap C(J,L^2(\RR)).$ Furthermore, there exists a constant $C$ independent of $I,$ such that
$$
||\Phi_f||_{L^\infty(I,L^2)}\leq C||f||_{L^1(I,L^2)}
$$
\end{pro}
\vspace{.5cm}

Let us set $f(z,t)=|v(z,t)|^2v(z,t)-|u(z,t)|^2u(z,t).$  Then we have that
\begin{eqnarray*}
|f(z)|&=&|(|v(z)|^2v(z)-|u(z)|^2u(z))|\\
&\leq& (|v(z)|-|u(z)|)(|u(z)|+|v(z)|)|v(z)|+|u(z)|^2|v(z)-u(z)|\\
&\leq&|v(z)-u(z)|(|u(z)|+|v(z)|)|v(z)|+|u(z)|^2|v(z)-u(z)|\\
&\leq& 3M^2|v(z)-u(z)|
\end{eqnarray*}
with $|u(z)|,|v(z)|\leq M.$ 

Let $I$ be an interval of $\RR $ such that $v(z,t)$ is a $H^2$ solution of the problem $(\ref{IVPTNLSE})$ on $I,$ and $u(z,t)$ is a $H^2$ solution  of the problem $(\ref{IVPNLSE}).$ Let us set $z\in I,$ then we have that
\begin{eqnarray*}
&&d(v(z),u(z))=||v(z)-u(z)||_{L^2}\\
&\leq&\Big|\Big|\int_0^zC_1e^{-C_2 z^\prime}\mathcal{T}(z-z^\prime)\big[|v(z^\prime,\cdot)|^2v(z^\prime,\cdot)-|u(z^\prime,\cdot)|^2u(z^\prime,\cdot)\big]dz^\prime\Big|\Big|_{L^2} \\
&&+\Big|\Big|\int_0^z\frac{C_2^2}{4}\mathcal{T}(z-z^\prime)(\cdot)^2v(z^\prime,\cdot)dz^\prime\Big|\Big|_{L^2}\\
&\leq& C |C_1|\int_0^z||v(z^\prime,\cdot)-u(z^\prime,\cdot)||_{L^2}dz^\prime +\frac{C_2^2}{4}\int_0^z||(\cdot)^2v(z^\prime,\cdot)||_{L^2}dz^\prime\\
\end{eqnarray*}
where $C$ is a positive constant and  proposition $3.5$  was used.

Before to state the main result of this chapter, we let $|I|$ denote the width of a interval $I$ of $\RR.$

\begin{theorem}
Let $I=[0,L]\;(L>0)$ be a bounded interval of $\RR.$ 
Let $v_0\in H^2(\RR)$ such that the functions $t^2v_0(t)$ and $t\frac{\partial}{\partial t}v_0(t)$ belong to $L^2(\RR).$ 
Given $v(z,t)\in L^\infty(I,H^2(\RR))$ a $H^2$-solution of the problem
$$
\left\{\begin{array}{l}
iv_z+ v_{tt}+C_1e^{-C_2 z} |v|^2v+\frac{C_2^2}{4}t^2v=0\\
v(0,t)=v_0
\end{array}\right.
$$

and $u(z,t)\in L^\infty(I,H^2(\RR))$  a $H^2$-solution of the problem
$$
\left\{\begin{array}{l}
iu_z+ u_{tt}+C_1e^{-C_2 z} |u|^2u=0\\
u(0,t)=v_0
\end{array}\right.
$$ 
 For every $\varepsilon >0$ there exists $L(\varepsilon)>0$ such that for every $L\in [0,L(\varepsilon)]$ there exists $\delta >0$ depending on $\varepsilon$ and $L$ such that, if $$||t^2v(0)||_{L^2}<\delta,\qquad||tv_t(0)||_{L^2}<\delta
$$ 
 then
$$||v(z)-u(z)||_{L^2}<\varepsilon $$
for every $z\in I.$
\label{maintheorem}
\end{theorem}

{\bf Proof}\qquad  Because of Theorem $\ref{teoremaditrasformazione}$ we can write
\begin{equation}
v(z,t)=e^{i\frac{C_2}{4}t^2-\frac{C_2}{2}z}Q(Z(z),T(z,t)),
\label{trans}
\end{equation}
where $Q(Z,T)$ is a solution of the standard NLSE $(\ref{modelloSNLSE})$  for every $Z\in \widetilde{I},$ where $\widetilde{I}=Z(I).$

Recalling that  $Z(z)=\frac{1}{2 C_2 }(1-e^{-2C_2 z}),$  the interval $I$ shifts in 
$$
\widetilde{I}=[0,\frac{1}{2 C_2 }(1-e^{2C_2 L})]
$$
and so if $|I|=L$ it follows that $|\widetilde{I}|=\frac{1}{2 C_2 }(1-e^{2C_2 L}).$

Let us set $g(z)=||t^2v(z)||_{L^2}^2$ and $f(Z)=||T^2Q(Z)||_{L^2}^2.$ By Lemma $3.4.1$ we have that 
\begin{equation}
g(z)=e^{4C_2z}f(Z).
\label{norma}
\end{equation}

As pointed out in remark $3.6,$ we have that

\begin{equation}
\sqrt{f(Z)}\leq \sqrt{f(0)}+4\int_0^Z||TQ_T(Z^\prime)||_{L^2}dZ^\prime.
\label{Gronwallf}
\end{equation}

Let us set $h(Z)=||TQ_T(Z)||_{L^2}^2;$  from the remark $3.5$ we have that

\begin{eqnarray}
\sqrt{h(Z)}&\leq&[\sqrt{h(0)}+1]e^{\frac{\widetilde{C}}{2}Z}-1,
\label{Gronwallh}
\end{eqnarray}
where $\widetilde{C}$ is a positive constant independent of $Z,$ as we pointed out in Remark $3.4.$ Let first  observe that $\sqrt{h(0)}=||TQ_T(0)||_{L^2};$ from the inverse transformation $(\ref{trasformazioniinversemodello})$ we write
$$
Q(Z,T)=e^{-i\frac{C_2}{4}t^2(Z,T)+\frac{C_2}{2}z(Z)}v(z(Z),t(Z,T))
$$
and so
$$
Q(0,T)=e^{-i\frac{C_2}{4}t^2(0,T)}v(0,t(0,T)).
$$
Differentiating this equation with respect to the variable $T$ we obtain that:
\begin{eqnarray*}
Q_T(0,T)&=&e^{-i\frac{C_2}{4}t^2}\big[-i\frac{C_2}{2}tv(0,t)\frac{dt}{dT}+v_t(0,t)\frac{dt}{dT} \big]\\
&=&e^{-i\frac{C_2}{4}t^2}\big[-i\frac{C_2}{2}tv(0,t)+v_t(0,t)\big]
\end{eqnarray*}
where we used  $t(0,T)=T$ and  $\frac{dt}{dT}=1.$
By using the relation $T(0,t)=t$ we obtain that
\begin{eqnarray*}
TQ_T(0,T)&=&te^{i\frac{C_2}{4}t^2}\big[i\frac{C_2}{2}tv(0,t)+v_t(0,t)\big]\\
&=&e^{i\frac{C_2}{4}t^2}\big[i\frac{C_2}{2}t^2v(0,t)+tv_t(0,t)\big]
\end{eqnarray*}
We now integrate this relation over $\RR$ with respect the variable $T,$ and by making use of the equation $dT=dt$ we obtain that
\begin{eqnarray*}
||TQ_T(0,T)||_{L^2}&=&\Bigg(\int_{\RR}\Big|\Big[i\frac{C_2}{2}t^2v(0,t)+tv_t(0,t)\Big]\Big|^2dt\Bigg)^{1/2}\\
&=&\Bigg(\int_{\RR}\Big|i\frac{C_2}{2}t^2v(0,t)+tv_t(0,t)\Big|^2 dt\Bigg)^{1/2}\\
&=&||i\frac{C_2}{2}t^2v(0,t)+tv_t(0,t)||_{L^2}\\
&\leq&\frac{C_2}{2}||t^2v(0,t)||_{L^2}+||tv_t(0,t)||_{L^2}\\
&<&\Big(\frac{C_2}{2}+1\Big)\delta.\\
\end{eqnarray*}

Furthermore, we have that
\begin{eqnarray*}
\int_{\RR}|T^2Q(0,T)|^2 dT&=&\int_\RR |t^2 v(0,t)|^2 dt,
\end{eqnarray*}
so we have  that $T^2Q(0,T)\in L^2(\RR),$ because of $t^2v(0,t)\in L^2(\RR)$ by hypothesis. Thus, because of Proposition $3.3,$ without loss of generality, we can assume that $Q(Z,T)$ in Eq. $(\ref{trans})$ belongs to $H^2(\RR)$ for every $Z\in\widetilde{I}.$ Therefore, we can apply Theorem $\ref{teoremadishift}$ to have the function $ T^2Q(Z,T)\in L^2(\RR)$ for every $Z\in\widetilde{I},$ and so to have $t^2 v(z,t)\in L^2(\RR),$ for every $z\in I.$ This claim allow us to consider distance $(\ref{distanza}).$

From $(\ref{Gronwallh})$ it follows that
\begin{equation}
\sqrt{h(Z)}<\Big[\Big(\frac{C_2}{2}+1\Big)\delta+1\Big]e^{\frac{\widetilde{C}}{2}Z}-1.
\label{stimah}
\end{equation}
We notice that the right hand side of this inequality is positive for every $Z\in \widetilde{I}.$
\vspace{.5cm}

By assumption, since $\sqrt{f(0)}=||t^2v(0)||_{L^2},$ we deduce that 
$$
\sqrt{f(0)}<\delta.
$$

Thus,  we have that
\begin{eqnarray}
\sqrt{f(Z)}&\leq &\sqrt{f(0)}+4\int_0^Z\sqrt{h(Z^\prime)}dZ^\prime\nonumber\\
&<&\delta+4\int_0^Z\Big[\Big[\Big(\frac{C_2}{2}+1\Big)\delta+1\Big]e^{\frac{\widetilde{C}}{2}Z^\prime}-1 \Big]dZ^\prime\nonumber\\
&=&\delta+\frac{8}{\widetilde{C}}\Big[\Big(\frac{C_2}{2}+1\Big)\delta+1\Big]\Big(e^{\frac{\widetilde{C}}{2}Z}-1\Big)-4 Z.\\\nonumber
\label{Stima}
\end{eqnarray}

Finally, from $(\ref{norma})$ we can write

\begin{eqnarray*}
\sqrt{g(z)}&= & e^{2C_2z}\sqrt{f(Z)}\nonumber\\
&<&e^{2C_2z}\Bigl\{\delta+\frac{8}{\widetilde{C}}\Big[\Big(\frac{C_2}{2}+1\Big)\delta+1\Big]\Big(e^{\frac{\widetilde{C}}{2}Z}-1\Big)-4Z\Bigr\}\nonumber\\
\end{eqnarray*}
and so
\begin{equation}
\sqrt{g(z)}<e^{2C_2z} \eta(Z(z),\delta)
\label{stimag}
\end{equation}

 where we set

\begin{equation}
\eta(Z,\delta)=\delta+\frac{8}{\widetilde{C}}\Big[\Big(\frac{C_2}{2}+1\Big)\delta+1\Big]\Big(e^{\frac{\widetilde{C}}{2}Z}-1\Big)-4Z
\label{eta}
\end{equation}
 In order to characterize the behavior of $\eta(Z,\delta)$ we consider the first derivative with respect $Z$ and we obtain that
$$
\frac{\partial \eta(Z,\delta)}{\partial Z}=4\Bigl\{\Big[\Big(\frac{C_2}{2}+1\Big)\delta+1\Big]e^{\frac{\widetilde{C}}{2}Z}-1\Bigr\}> 0 \qquad \mbox{for all}\quad Z\geq 0.
$$
Thus $\eta(Z,\delta)$ is increasing on $\widetilde{I}.$ Since $\eta(0,\delta)=\delta>0,$ then the function $\eta(Z,\delta)$ is positive for every $Z\in \widetilde{I}.$ Recalling that $Z(z)$ increasing on $I,$ we can state $\eta(Z(z),\delta)\leq \eta(Z(L),\delta)$ for every $z\in I.$

Therefore, from Eq. $(\ref{stimag})$ we have that 
\begin{equation}
\sqrt{g(z)}<e^{2C_2L} \eta(Z(L),\delta),\quad\mbox{for all}\quad z\in I,
\label{Stimag}
\end{equation}
where we used that $C_2>0$ and so $e^{2C_2z}$ is increasing on $I.$

\vspace{.5cm}

We now recall that
\begin{eqnarray*}
||v(z)-u(z)||_{L^2}&\leq& C |C_1|\int_0^z||v(z^\prime,\cdot)-u(z^\prime,\cdot)||_{L^2}dz^\prime \\
&&+\frac{C_2^2}{4}\int_0^z||(\cdot)^2v(z^\prime,\cdot)||_{L^2}dz^\prime.
\end{eqnarray*}

Recalling that $\sqrt{g(z)}=||(\cdot)^2v(z,\cdot)||_{L^2},$ from $(\ref{Stimag})$ it follows that

\begin{eqnarray}
||v(z)-u(z)||_{L^2}&<& C |C_1|\int_0^z||v(z^\prime,\cdot)-u(z^\prime,\cdot)||_{L^2}dz^\prime\label{stimadistanza} \\
&&+\frac{C_2^2}{4}\int_0^z  e^{2C_2L}\eta(Z(L),\delta)dz^\prime.\nonumber\\\nonumber
\end{eqnarray}
Since $z\in[0,L],$ by $(\ref{stimadistanza})$ we have that

\begin{eqnarray}
||v(z)-u(z)||_{L^2}&<& C |C_1|\int_0^z||v(z^\prime,\cdot)-u(z^\prime,\cdot)||_{L^2}dz^\prime \\
&&+\frac{C_2}{4}\eta(Z(L),\delta)L\;e^{2C_2L}.\nonumber\\\nonumber
\end{eqnarray}
Therefore, by Gronwall Lemma we obtain that
\begin{eqnarray*}
||v(z)-u(z)||_{L^2}&<&\frac{C_2}{4}\eta(Z(L),\delta)L\;e^{2C_2L} e^{\int_0^z C |C_1|dz^\prime}. \\
\end{eqnarray*}
Thus, for every $z\in [0,L]$ we have that
\begin{equation}
||v(z)-u(z)||_{L^2}<\frac{C_2}{4}\eta(Z(L),\delta)L\;e^{2C_2L} e^{C|C_1|L}
\label{Stimadistanza}
\end{equation}
Given $\varepsilon>0,$ we should prove that there exists $\delta>0$ such that
\begin{equation}
\frac{C_2}{4}\eta(Z(L),\delta)L\;e^{2C_2L} e^{C|C_1|L}<\varepsilon.
\label{delta}
\end{equation}

Let us set
$$
G(L,\varepsilon)=\varepsilon\frac{4}{C_2}(1-e^{-C_2L})^{-1} e^{-\frac{C |C_1|}{C_2}(e^{C_2L}-1)},
$$
we notice that $G(L,\varepsilon)>0.$ 

Recalling that
$$
\eta(Z,\delta)=\delta+\frac{8}{\widetilde{C}}\Big[\Big(\frac{C_2}{2}+1\Big)\delta+1\Big]\Big(e^{\frac{\widetilde{C}}{2}Z}-1\Big)-4Z
$$
we also set
$$
H(L)=\frac{8}{\widetilde{C}}\Big(e^{\frac{\widetilde{C}}{2}Z(L)}-1\Big)-4Z(L).
$$

Thus,  we need that $\delta$ fulfils
\begin{equation}
\Bigg[1+\frac{8}{\widetilde{C}}\Big(\frac{|C_2|}{2}+1\Big)\Big(e^{\frac{\widetilde{C}}{2}Z(L)}-1\Big)\Bigg]\delta<G(L,\varepsilon)-H(L).
\end{equation}
We should now have that the function $G(L,\varepsilon)-H(L)$ is positive for any $L>0;$ we notice that the function $G(L,\varepsilon)-H(L)$ is continuous for every $L>0.$ Moreover, we have that
$$
\lim_{L\rightarrow 0}\big(G(L,\varepsilon)-H(L)\big)=+\infty.
$$
Taking the first derivative with respect to $L,$ we have that
\begin{eqnarray*}
G^\prime(L,\varepsilon)-H^\prime(L)&=& -\varepsilon\frac{4}{C_2}\;\frac{2C_2+C|C_1|+1}{L^2}\;e^{-(2C_2+C|C_1|)L}\\
&&+4Z^\prime(L)(1-e^{\frac{\widetilde{C}}{2}Z(L)});
\end{eqnarray*}
since $Z(L)>0,$  we have that $G^\prime(L,\varepsilon)-H^\prime(L)$ is negative for every $L>0.$ Thus, there exists $L(\varepsilon)>0$ such that $G(L,\varepsilon)-H(L)$ is positive for every $L\in[0,L(\varepsilon)].$ Then for
\begin{eqnarray*}
\delta<\Bigg[1+\frac{8}{\widetilde{C}}\Big(\frac{|C_2|}{2}+1\Big)\Big(e^{\frac{\widetilde{C}}{2}Z(L)}-1\Big)\Bigg]^{-1}\Big(G(L,\varepsilon)-H(L)\Big)
\end{eqnarray*}
and $L\in[0,L(\varepsilon)]$ we have that
\begin{equation*}
||v(z)-u(z)||_{L^2}<\varepsilon , \quad
\end{equation*}
for every $z\in [0,L].\quad \Box$

\chapter{Stationary Solutions}
\section{Introduction}
The propagation of optical pulses inside single-mode fiber is governed by the following non linear Schr\"odinger equation
\begin{equation}
i\frac{\partial A(z,t)}{\partial z}=-i\frac{\alpha}{2}A(z,t)+\frac{\beta_2}{2}\frac{\partial^2 A(z,t)}{\partial t^2}-\gamma |A(z,t)|^2A(z,t),
\label{Ing}
\end{equation}
which we profusely described in the previous chapters. Optical pulse can be affected by the amplified spontaneous emission (ASE) noise generated by optical amplifiers. In systems operating in the linear regime, ASE noise is not affected by propagation through the optical fiber and does not change its characteristic after propagation; while, when nonlinear effect are not negligible, signal and noise interaction occur during propagation. For the purpose of investigating the interaction between signal and noise, the signal can be considered a continuous wave (CW), i.e. an electromagnetic wave of constant amplitude, and the noise can be treated as a perturbation of the CW solution of the NLSE.

Let us consider Eq $(\ref{Ing})$ with $\alpha =0,$ that is
\begin{equation}
i\frac{\partial A(z,t)}{\partial z}=\frac{\beta_2}{2}\frac{\partial^2 A(z,t)}{\partial t^2}-\gamma |A(z,t)|^2A(z,t)
\label{standard ing}
\end{equation}
and the propagation of a CW. In the case of CW radiation, the amplitude $A$ is independent of $t$ at the input end of the fiber at $z=0.$ Assuming that $A(z,t)$ remains independent of $t$ during propagation inside the fiber, Eq. $(\ref{standard ing})$ is readily solved to obtain the \textit{steady-state} solution
\begin{equation}
A(z,t)=\sqrt{P_0}\;e^{i\phi_{NL}},
\label{CW}
\end{equation}
where $\phi_{NL}=\gamma P_0z$ is the nonlinear phase shift induced by nonlinear effects. Eq. $\ref{CW}$ implies that CW light should propagate through the fiber unchanged except for the gain of a power-dependent phase shift (and a power reduction in the presence of fiber losses).

We now ask whether the \textit{steady-state} solution $(\ref{CW})$ is stable against small perturbations. To answer this question, we slightly perturb the \textit{steady-state} in such a manner that
\begin{equation}
A=(\sqrt{P_0}+a(z,t))\exp(i\phi_{NL})
\label{PCW}
\end{equation}
and examine the evolution of the complex-valued perturbation $a(z,t)$ using a linear stability analysis. Substituting $(\ref{PCW})$ in Eq. $(\ref{standard ing})$ we have that
\begin{eqnarray*}
ia_ze^{i\phi_{NL}}-(\sqrt{P_0}+a)\gamma P_0e^{i\phi_{NL}}=\frac{\beta_2}{2}a_{tt}e^{i\phi_{NL}}-\gamma|\sqrt{P_0}+a|^2(\sqrt{P_0}+a)e^{i\phi_{NL}},
\end{eqnarray*}
that is
\begin{eqnarray*}
ia_z=\frac{\beta_2}{2}a_{tt}-\gamma P_0(a+\bar{a})-\gamma P_0(a+\bar{a})a-\gamma|a|^2(\sqrt{P_0}+a).
\end{eqnarray*}
By linearizing in $a,$ we obtain that
\begin{equation*}
ia_z=\frac{\beta_2}{2}a_{tt}-\gamma P_0(a+\bar{a}).
\end{equation*}
We now set $a(z,t)=a_1(z,t)+ia_2(z,t),$ with $a_1(z,t)$ and $a_2(z,t)$ \textit{real}-valued functions. Thus, collecting the imaginary part and the real one, we obtain
$$
\left\{\begin{array}{l}
\partial_za_1=\frac{\beta_2}{2}\partial_{tt}a_2\\
\\
\partial_za_2=-\frac{\beta_2}{2}\partial_{tt}a_1+2\gamma P_0a_1
\end{array}\right.
$$
We now take the Fourier transform of functions $a_1$ and $a_2;$ referring to their \textit{real} part as $A_1$ and $A_2$ respectively, we arrive at
$$
\left\{\begin{array}{l}
\partial_z A_1=-\omega^2\frac{\beta_2}{2}A_2\\
\\
\partial_z A_2=\omega^2\frac{\beta_2}{2}A_1+2\gamma P_0A_1
\end{array}\right.
$$
which is solved by
\begin{equation*}
A_1(z)=-\frac{\beta_2\omega^2}{2\kappa}\Big[C_1\sin[kz]-C_2\cos[\kappa z]\Big]
\end{equation*}
\begin{equation*}
A_2(z)=[C_1\cos[kz]+C_2\sin[\kappa z]],
\end{equation*}
where 
\begin{equation}
\kappa=\frac{|\omega\beta_2|}{2}\Big(\omega^2+\mbox{sgn}(\beta_2)\frac{4\gamma P_0}{|\beta_2|}\Big)^{1/2},
\label{k}
\end{equation}
and $C_1,\; C_2$ are \textit{real} coefficients. The same results holds for the \textit{imaginary} part of the Fourier transform of $a_1$ and $a_2.$

We now recall the well-known identities
\begin{equation}
\cos[kz]=\frac{e^{i\kappa z}+e^{-i\kappa z}}{2}\qquad \sin[\kappa z]=\frac{e^{i\kappa z}-e^{-i\kappa z}}{2i}.
\label{cosin}
\end{equation}

From Eq $(\ref{k}),$ we notice that if $\beta_2>0$ then $\kappa$ is real for every $\omega\in \RR ,$ and the steady state is stable against small perturbations. By contrast, if $\beta_2<0$ and $ |\omega|<\Big(\frac{4\gamma P_0}{|\beta_2|}\Big)^{1/2}$ then $\kappa $ becomes imaginary; therefore, from $(\ref{cosin})$ we notice that the perturbation $a(z,t)$ grows exponentially with $z.$ Thus, the solution $(\ref{CW})$ is inherently unstable for $\beta_2<0.$ This instability is referred as \textit{modulation instability} because it leads to a spontaneous temporal modulation of the CW beam.

Thus, in the normal dispersion regime of optical fibers, i.e. when $\beta_2>0$ we refer to Eq. $(\ref{standard ing})$ as the following equation: 
\begin{equation}
iQ_Z+Q_{TT}-|Q|^2Q=0.
\label{NLSEstandard}
\end{equation} 
Therefore, because of (\ref{trasformazionimodello}) we refer to transformed NLSE as
\begin{equation}
iv_z+v_{tt}-e^{-C_2}|v|^2v+\frac{C_2^2}{4}t^2 v=0
\label{NLSEtransformed}
\end{equation}
In this chapter we investigate the stability against small perturbations of CW solutions of the Eq. $(\ref{NLSEstandard}).$ Then, we shall extend the results got in the context of the standard NLSE to the TNLSE $(\ref{NLSEtransformed}).$ At last we investigate the stability against small perturbations of CW solutions of the Eq. $(\ref{Ing}).$

\vspace{.5cm}
The occurrence of \textit{modulation instability} in the anomalous-GVD regime of optical fibers is an indication of a fundamentally different character of Eq. $(\ref{standard ing})$ when $\beta_2<0.$ It turns out that this equation has specific pulse-like solutions that do not change along fiber length, such solutions are known as optical solitons.

\section{Time-Homogeneous Solutions}
We consider the standard NLSE 
\begin{equation}
i Q_Z(Z,T)+Q_{TT}(Z,T)-|Q(Z,T)|^2Q(Z,T)=0
\label{SNLSE}
\end{equation}
and we look for solutions of Eq. $(\ref{SNLSE})$ of the form 
\begin{equation}
\Phi(Z)=\sqrt{P_0}e^{-i P_0 Z},
\label{nonvanish}
\end{equation}
where $P_0$ is a positive \textit{real} coefficient. We notice that these functions do not vanish for $|T|\rightarrow\infty.$ In order to claim the existence of such type of solutions in a suitable functions space, we  should note that the Cauchy problem for the linear problem
$$
\left\{\begin{array}{l}
iQ_Z(Z,T)+Q_{TT}(Z,T)=0\\
\\
Q(0,T)=Q_0(T)
\end{array}\right.
$$
 is ill-posed in the space $C(\RR).$ Indeed, this follows from the fact that the function 
$$
\sqrt{\frac{Z_0}{Z_0-Z}}e^{-i\frac{T^2}{4(Z_0-Z)}}
$$
with an arbitrary $Z_0>0,$ belongs to $C(\RR)$ for each $Z\in[0,T_0)$ and satisfies our linear Schr\"odinger equation, but this function is unbounded in any left half-neighborhood of the point $Z=Z_0.$ 

\subsection{Existence}
Let now introduce the functions spaces where we should look for non vanishing solutions as $|T|\rightarrow\infty$ of Eq. $(\ref{SNLSE}).$  

\begin{definition}
For an arbitrary positive integer $k,$ \quad $X^k$ is the Banach space consisting of functions $Q(T)$ of the argument $T\in \RR ,$ absolutely continuous with their derivatives of order $1,2,\cdots,k-1$ in any finite interval, for each of which the following norm is finite:
$$ |||Q|||_k=\sup_{T\in \RR}|Q(T)|+\sum_{i=1}^k \Big|\Big| \frac{d^iQ}{dT^i}\Big|\Big|_{L^2(\RR)}. $$
\label{spazidiZhidkov}
\end{definition}

In order to define a solution of the problem
\begin{equation}
\left\{\begin{array}{ll}
iQ_Z(Z,T)+Q_{TT}(Z,T)-|Q(Z,T)|^2Q(Z,T)=0&Z\in\RR, \quad T\in\RR \\
Q(0,T)=Q_0(T)
\end{array}\right.
\label{SNLSEIVP}
\end{equation}
in Zhidkov spaces $X^k(\RR),$ we first observe that the Schr\"odinger operator acts in the spaces $L^2(\RR)$ and $H^k(\RR)\; ,k=1,2,3,\ldots, $ as the following integral operator
\begin{equation}
G_Z\Phi=\int_\RR K(T-T^\prime,Z)\Phi(T^\prime)dT^\prime\;,\quad G_0\Phi\Doteq \Phi,
\end{equation}
where $Z\neq 0,$ $K(Z,T)=(4\pi i Z)^{-1/2}\exp\Big[i\frac{ T^2}{4Z}\Big].$ Of course $K(Z,T)$ is the fundamental solution of the operator $L=i\frac{\partial}{\partial Z}+\frac{\partial^2}{\partial T^2}.$ We also set $g(Q)=-|Q|^2 Q.$ In this way the integral formulation of the problem $(\ref{SNLSEIVP})$  can be formally written as follows:
\begin{equation}
Q(Z,T)=G_Z Q_0+\int_0^ZG_{Z-Z^\prime}\;g(Q(Z^\prime,\cdot))dZ^\prime.
\label{INLSE}
\end{equation} 
\vspace{.5cm}

To make rigorous Eq. $(\ref{INLSE})$ in the spaces $X^k(\RR)$ we need the following result
\begin{pro}
For any $k=1,2,3,\ldots$ the operators $G_Z$ considered on $X^k$ satisfy the following properties:
\begin{itemize}
\item[$G1$] for any interval $I\subset\RR $ the family of operator $G_Z:X^k\rightarrow X^k $ is uniformly bounded with respect to $Z\in I;$
\item[$G2$] for any $\Phi\in X^k$ the function $G_Z\Phi:I\rightarrow X^k$ is continuous and $$\lim_{Z\rightarrow 0} G_Z\Phi=\Phi$$  in the sense of the space $X^k.$
\end{itemize}
\label{group}
\end{pro}
{\bf Proof}\qquad [\cite{Zhidkov}, Proposition $I.2.7$].
\vspace{1cm}

Furthermore, because of the next result
\begin{pro}
Let $g(Q)$ be a complex-valued infinitely differentiable function of the complex argument $Q$ and  $I$ be an interval containing zero. Then, a function $Q(Z,T)\in C(I,X^3)\cap C^1(I,X^1)$ satisfies the problem
\begin{equation}
\left\{\begin{array}{ll}
i\frac{\partial }{\partial Z}Q+\frac{\partial^2}{\partial T^2}Q+g(Q)=0&Z\in I, \quad T\in \RR \\
Q(0,T)=Q_0\in X^3
\end{array}\right.
\label{abNLSE}
\end{equation}
if and only if $Q(0,T)=Q_0$ for any $T\in \RR $ and this function is a solution of Eq. $(5.3).$
\label{abstractprop}
\end{pro}
{\bf Proof}\qquad [\cite{Zhidkov}, Proposition $I.2.8$],
\vspace{1cm}

we can accept the following definition:

\begin{definition}
Let $k=1$ or $k=2,$ $g(Q)$ be a complex-valued continuous function of the complex argument $Q$ and $I$ be an interval of $\RR $ containing zero. We call a solution $Q(Z,T)\in C(I,X^k)$ of Eq. $(\ref{INLSE})$ a $X^k$-solution (or generalized solution) of the problem $(\ref{SNLSEIVP}).$
\end{definition}
 Our result on the existence for the standard NLSE in classes of functions non vanishing as $|T|\rightarrow \infty$ is the following.
 
 \begin{theorem}
 Let $k\geq 1$ be an integer. Then, for any $Q_0\in X^k$ there exist $\overline{Z}>0,$ depending only on $|||Q_0|||_k,$ and a unique $X^k$- solution $Q(Z,T)$ of the problem $(\ref{SNLSEIVP})$ defined on the interval $(-\overline{Z},\overline{Z}).$ Furthermore, there exist $Z^\star_1,Z^\star_2 >0$ such that this solution can be continued on the interval $(-Z^\star_1,Z^\star_2)$ and $Z^\star_1=\infty$ (resp. $Z^\star_2=\infty$) if
 $$
 \limsup_{Z\rightarrow{Z^\star_1}^+}|||Q(Z)|||_k<+\infty
 $$
 (resp. if $\limsup_{Z\rightarrow{Z^\star_2}^-}|||Q(Z)|||_k<+\infty$).
 \label{exisitencenonvanish}
 \end{theorem}
 {\bf Proof}\qquad  We now only sketch the proof and, for more details, we refer to \cite{Zhidkov} and \cite{Gallo}. Due to proposition $\ref{abstractprop}, $ Eq. $(\ref{INLSE})$ is equivalent to the problem $(\ref{SNLSEIVP}).$ Therefore, we consider this equation. The Schr\"odinger operator defines a group on $X^k(\RR)$ for every integer $k=1,2,3,\ldots,$ in particular the family of operators $\big(G_Z\big)_{Z\in\RR}$ defines a strongly continuous group on $X^k(\RR).$ Moreover, in \cite{Gallo} it is shown that there exists a constant $C>0$ that depends only on $k,$ such that for every $Q\in X^k(\RR)$ and $Z\in \RR ,$
 \begin{equation}
 |||G_ZQ|||_k\leq C(1+|Z|^{1/4})|||Q|||_k.
 \end{equation}
 Since $g(Q)=-|Q|^2Q,$ it is clear that $g(Q)\in L^\infty,$ for every $Q\in X^k(\RR).$ We now should prove that $ g(Q)\in X^k(\RR).$
 
 Let $\alpha$ be an integer such that $1\leq\alpha<k$ and let us differentiate $\alpha$ times the function $g;$  we obtain a linear combination of terms of type
 $$
 Q^a\;\overline{Q}^b\;\frac{\partial^i Q}{\partial T^i}\;\frac{\partial^j \overline{Q}}{\partial T^j}
 $$ 
 where $a,b,i,j$ are integers such that $1\leq i,j\leq \alpha\leq k.$ By Sobolev immersion and Holder inequality it can be shown easily that $\frac{\partial^\alpha g}{\partial T^\alpha}$ belongs to $L^2(\RR).$
 \vspace{.5cm}
 
  We first show the uniqueness of the solution.  Let $Q_0\in X^1$ and $Q_1(Z,T)$ and $Q_2(Z,T)$ be two solutions of this equation from $C(I,X^1).$  We observe that
  $$
  \rho\Big||Q_1|^2Q_1-|Q_2|^2Q_2\Big|\leq M|Q_1-Q_2|,\qquad \mbox{with}\quad |Q_1|,\;|Q_2|\leq M
  $$ 
  and so, thanks to $(5.5)$ and that $g(Q)\in X^k,$ we arrive at
  $$
  |||Q_1(Z,\cdot)-Q_2(Z,\cdot)|||_k\leq M\int_0^z|||Q_1(Z^\prime,\cdot)-Q_2(Z^\prime,\cdot)|||_k dZ^\prime ;
 $$
 then from the Gronwall Lemma it follows that $Q_1(Z,T)\equiv Q_2(Z,T)$ for all $Z>0.$
 
  Let $l\geq 1$ be integer. For any $R>0$ there exists $\bar{Z}=\bar{Z}(R)>0$ such that, if $|||Q_0|||_l\leq R,$ then the operator $G$ maps the set
 $$
 M_T=\{Q(\cdot)\in C([-\bar{Z},\bar{Z}],X^l)\;:\quad Q(0)=Q_0\in X^l,\quad |||Q(\cdot)|||_l\leq 2|||Q_0|||_l\}
 $$
 into itself. Moreover, for any $R>0,$ there exists a constant $\bar{Z}_1\in (0,\bar{Z}]$ such that the map $G$ is a contraction on $M_{\bar{Z}_1},$ if $|||Q_0|||_l\leq R.$ Since a fixed point of the map $G$ is obviously a solution of the Eq. $(\ref{INLSE}),$ we have that for any $Q_0\in X^l$ there exists $\bar{Z}>0,$ depending only on $|||Q_0|||l,$ such that in the interval $(-\bar{Z},\bar{Z})$ there exists a unique solution $Q(Z,T)\in C((-\bar{Z},\bar{Z}),X^l).$ Since the Eq. $(\ref{SNLSE})$ is autonomous, a result similar to the above one on the existence and uniqueness of a $X^l$ solution takes place if we pose the initial condition not at point $Z=0$ but at an arbitrary point $Z=Z_0\neq 0.$ Therefore, since the length of the interval of the existence is bounded from below by a constant $\bar{Z}>0$ depending only on $|||Q_0|||_l,$ we get the existence of $\bar{Z}_l^- $ and $\bar{Z}_l^+ $ such that our $X^l$ solution can be continued onto the interval $(-\bar{Z}_l^-,\bar{Z}_l^+)$ and
 $$
 \limsup_{Z\rightarrow (\bar{Z}_l^-)^+}|||Q(Z,\cdot)|||_l=+\infty
 $$
 if $\bar{Z}_l^->-\infty.$ (resp. $\limsup_{Z\rightarrow (\bar{Z}_l^+)^-}|||Q(Z,\cdot)|||_l=+\infty )$ if $\bar{Z}_l^+<+\infty$).
 
 Let again $Q_0\in X^l,\;l\geq 2. $ Then obviously for each $k=1,2,\ldots,l$ there exists a unique $X^k$-solution of the problem $(\ref{SNLSEIVP})$ an let $(\bar{Z}_k^-,\bar{Z}_k^+),$ where $\bar{Z}_k^-<0$ and $\bar{Z}_k^+>0,$ be the maximal interval of existence. Then,
 $$-\infty\leq \bar{Z}_1^-\leq\bar{Z}_2^-\ldots\bar{Z}_l^-<0$$
 and
 $$0< \bar{Z}_l^+\leq\ldots\bar{Z}_2^+\leq\bar{Z}_1^+\leq+\infty .$$
 Since it follows from Eq. $(\ref{INLSE})$ that for $k=1,\ldots,l-1,$  $\bar{Z}_{k+1}\in(0,\bar{Z}_k^+)$ and $Z\in [0,\bar{Z}_k+1]$
 $$
 |||Q(Z,\cdot)|||_{k+1}\leq |||Q_0|||_{k}+\widetilde{C}\Big(\max_{Z\in[0,\bar{Z}_k]}|||Q(Z,\cdot)|||_k\int_0^Z|||Q(Z^\prime,\cdot)|||_{k+1}dZ^\prime\Big)
 $$
 and by analogy for $Z<0,$ we have that $ \bar{Z}_1^-=\bar{Z}_2^-=\ldots=\bar{Z}_l^-$ and $ \bar{Z}_1^+=\bar{Z}_2^+=\ldots=\bar{Z}_l^+$. $\Box$
 
 \subsection{Stability}
 In this section, we shall consider a result on the stability of solutions  of the standard NLSE $(\ref{SNLSE})$ non-vanishing as $|T|\rightarrow\infty.$ These solutions are \textit{time-homogeneous} ones $\Phi(Z)=\sqrt{P_0}\;e^{-iP_0 Z},$ where $P_0>0.$
 
 Let $c\in \RR ,\; Q(\cdot)\in X^1$ and $Q(T)\neq 0$ for all $T\in \RR.$ Then, one can easily verify that there exists a unique real-valued function $A(\cdot)\in X^1$ and a real-valued function $\Omega(\cdot),$ absolutely continuous in any finite interval and unique up to adding $2\pi m,\;m=0,\pm 1\pm 2,\ldots $ to it, such that
\begin{equation}
 Q(T)=\big(\sqrt{P_0}+A(T)\big)\;e^{i[c+\Omega(T)]}
 \end{equation}
 and that $\sqrt{P_0}+A(T)=|Q(T)|>0$ for all $T\in \RR ;$ in addition, due to relation
 \begin{equation}
 Q^\prime(T)=[A^\prime(T)+i\big(\sqrt{P_0}+A(T)\big)\Omega^\prime(T)]e^{i[c+\Omega(T)]},
 \end{equation}
 we have $\big(\sqrt{P_0}+A(T)\big)\Omega^\prime(T)\in L^2(\RR).$ In particular, if
 
 $$0<c_1\leq\sqrt{P_0}+A(T) \leq c_2<+\infty$$ for all $T\in \RR ,$ then $\Omega^\prime(\cdot)\in L^2(\RR).$
 
 Conversely, if for a complex-valued function $Q(\cdot)$ there exist real-valued function $A(\cdot)\in X^1(\RR)$ and $\Omega^\prime(\cdot),$ absolutely continuous in any finite interval, such that $\big(\sqrt{P_0}+A(T)\big)\Omega^\prime(T)\in L^2(\RR),\quad \sqrt{P_0}+A(T)>0$ for all $T\in \RR$ and Eq. $(5.7)$ takes place, then $Q(\cdot)\in X^1(\RR)$ and $Q(T)\neq 0$ for all $T\in \RR.$ 
 
 The result on the stability of solutions $\Phi(Z)=\sqrt{P_0}\;e^{-iP_0 Z}$ we consider here is the following. The perturbed solution is written as
  \begin{equation}
 Q(Z,T)=\Big(\sqrt{P_0}+A(Z,T)\Big)\;e^{i[-P_0 Z+\Omega(Z,T)]}
 \label{perturbed}
 \end{equation}
 \begin{theorem}
 Let $\delta_0>0$ be such that $\sup_{T\in \RR}|Q(T)|<\frac{\sqrt{P_0}}{2}$ if $A(\cdot)\in H^1$ and $||A(\cdot)||_{H^1}<\delta_0.$ Then the solution $\Phi(Z)$ is stable for every $Z>0$ in the following sense: for an arbitrary $\varepsilon \in(0,\delta_0)$ there exists $\delta\in (0,\delta_0)$ such that, if $Q_0(\cdot)\in X^1(\RR),$ $|Q_0|-\sqrt{P_0}\in H^1(\RR),\quad ||\;|Q_0|-\sqrt{P_0}||_{H^1}<\delta,\;||\Omega^\prime(\cdot)||_{L^2}<\delta,$ then there exists a $X^1$-solution $Q(Z,T)$ given by $(\ref{perturbed}).$ Moreover, $A(Z,\cdot)\in H^1(\RR)$ and $||A(Z,\cdot)||_{H^1}<\varepsilon$ and $||\Omega^\prime(Z,\cdot)||_{L^2}<\varepsilon.$
 \label{stability}
 \end{theorem}
 {\bf Proof }\qquad Let first $k\geq 1$ be integer and $Q_0(T)\in X^k.$ Then according to theorem $\ref{exisitencenonvanish} ,$ there exists a unique (local) $X^k$-solution of the problem $(\ref{SNLSEIVP}).$ Clearly, for any integer $l=1,2,\cdots,k$ this problem has a unique local $X^l$-solution which, of course, coincides with the $X^k$-solution in all intervals of the existence of both solutions. Let positive $\bar{Z}_1,\bar{Z}_2,\cdots,\bar{Z}_k$ be such that, for any
 $l=1,2,\cdots,k,\quad [0,\bar{Z}_l)$ be the maximal half-interval of the existence of the $X^l$-solution. Clearly, $\bar{Z}_1\geq \bar{Z}_2\geq \cdots \geq \bar{Z}_k >0.$
 
 \begin{lemma}
 $\bar{Z}_1=\bar{Z}_2=\cdots=\bar{Z}_k=Z>0.$
 \label{lemma1stability}
 \end{lemma}
 {\bf Proof}\qquad [\cite{Zhidkov}, Lemma $III.3.2$]
 \vspace{.5cm}
 
 Let $I(Q)=\int_{-\infty}^{+\infty}(|Q|^2-P_0)^2dT.$
 
 \begin{lemma}
 Let the assumptions of theorem $\ref{stability}$ be valid and let $I(Q_0)<\infty$ for some $Q_0\in X^1$ and the corresponding $X^1$-solution $Q(Z,T)$ of the Cauchy problem $(\ref{SNLSEIVP})$ exists for $Z\in [0,Z_0).$ Then $I(Q(Z,T))<\infty$ for all $Z\in [0,Z_0]$ and this function is continuous in $Z.$
 \label{lemma2stability}
 \end{lemma}
 {\bf Proof}\qquad [\cite{Zhidkov}, Lemma $III.3.3$]
 \vspace{.5cm}
 
 \begin{lemma}
 Let the assumptions of theorem $\ref{stability}$ be valid, $Q_0\in X^1$ and the corresponding $X^1$-solution $Q(Z,T)$ of the Cauchy problem $(\ref{SNLSEIVP})$ exists for $Z\in [0,Z_0).$ Let
 $$M(Q)=\int_{-\infty}^{+\infty}\Bigg\{\frac{1}{2} |Q_T|^2-U(|Q|^2)-\frac{P_0}{2} |Q|^2+D \Bigg\}dT,$$
 where $D=\frac{\sqrt{P_0}}{2}P_0+U(P_0)$ and $U(s)=-\frac{s^2}{4}.$ If $I(Q_0)<\infty,$ then the quantity $M(Q(Z,T))$ is determined for all $Z\in [0,Z_0)$ and independent of $Z.$
 \label{lemma3stability} 
 \end{lemma}
  {\bf Proof}\qquad [\cite{Zhidkov}, Lemma $III.3.4$]
  \vspace{.5cm}
  
 Let $Q_0\in X^1$ be such that $I(Q_0)<\infty$ and $Q_0(T)\neq 0$ for all $T\in \RR.$ Let also $[0,Z_0)$ be the maximal half-interval such that the corresponding $X^1$-solution of the problem $(\ref{SNLSEIVP})$ exists for $Z\in[0,Z_0)$ and is non zero for these values of $Z$ and $T\in \RR. $ Then in view of Lemma $\ref{lemma2stability}-\ref{lemma3stability}$ for $Z\in [0,Z_0)$ we can rewrite the functional $M(Q)$ in the following form:
 $$M(Q(Z,T))=M_1+M_2+M_3$$
 where 
 $$M_1(Q(Z,T))=\frac 1 2 \int_{-\infty}^{+\infty}\{A_T^2(Z,T)+2P_0A^2(Z,T)\}dT $$
 $$ M_2(Q(Z,T))=\frac 1 2 \int_{-\infty}^{+\infty}\Omega_T^2(Z,T)(\sqrt{P_0}+A(Z,T))^2 dT $$
 $$M_3(Q(Z,T))=\alpha\Big(\|A(\cdot,Z)\|_{H^1}^2\Big),\quad \lim_{s\rightarrow 0}\frac{\alpha(s)}{s}=0.$$
 Here $Q(Z,T)=(\sqrt{P_0}+A(Z,T))e^{i[-P_0 Z+\Omega(Z,T)]}$ where $\sqrt{P_0}+A(Z,T)>0$ and for any fixed $Z$ as functions of $T$ the function $A(Z,\cdot)$ belongs to $H^1$ and the function $\Omega(Z,T)$ is absolutely continuous in an arbitrary finite interval.
 
 By Lemma $\ref{lemma2stability}$ $||A(Z,\cdot)||_{H^1}$ is a continuous function of $Z\in[0,Z_0).$ Also, there exists $C=\max\Big\{\frac 1 2 ,\frac{P_0}{2}\Big\}$ such that 
 $$M_1(Q)\geq C||A||_{H^1}^2,\quad A\in H^1.$$
 
 Consider an arbitrary $\varepsilon \in (0,\delta_0)$ and let $H(s)=Cs^2+\alpha(s^2).$ Then, there exists $\delta_1 \in (0,\delta_0)$ such that $H(s)\geq \frac{C}{2}$ for any $s\in(0,\delta_1].$ By means of the above-mentioned arguments, there exists $\delta_2\in (0,\delta_1)$ such that
 $$
 H(||A(Z,\cdot)||_{H^1})\leq M_1(Q)<\frac{C}{4}\delta_1
 $$
 if $||A(0,\cdot)||_{H^1}<\delta_2$ and $||\Omega_T^\prime(\cdot,0)<\delta_2.$ Then, using the continuity of $||A(Z,\cdot)||_{H^1},$ we get for such initial data $||A(Z,\cdot)||_{H^1}<\delta_1<\epsilon$ for all $Z\in[0,Z_0).$
 
 Finally, in view of the equality
 $$\frac 1 2 ||\Omega_T^\prime (\sqrt{P_0}+A)||_{L^2}^2=M(Q)-M_1(Q)-M_2(Q)$$
 we obtain $||\Omega_T(Z,\cdot)||_{L^2}<\varepsilon$ if $||A(0,\cdot)||_{H^1}<\delta_2$ and $||\Omega_T(0,\cdot)||_{L^2}<\delta_2$ where $\delta_2>0$ is sufficiently small. Therefore, for these initial data from
 $$Q^\prime(0,T)=\Big[A^\prime(0,T)+i(\sqrt{P_0}+A(0,T))\Omega_T^\prime(0,T)\Big]e^{i\Omega(0,T)}$$
 there exists $C_1>0$ such that $||Q(Z,\cdot)||_{X^1}<C_1$ for all $Z\in [0,Z_0).$ Thus, $Z_0=\infty$ and the theorem is proved. $\Box$
 \vspace{.5cm}
 
 From the Theorem $\ref{stability}$ it immediately follows 
 \begin{cor}
 Let the assumptions of the theorem be valid. Then, the solution $\Phi(Z)$ is also stable in the following sense: for any $\varepsilon \in (0 ,\delta_0)$ and  $d>0$ there exists $\delta\in(0,\delta_0)$ such that if $Q_0(T)\in X^1,$ $|Q_0|-\sqrt{P_0}\in H^1$, $\||Q_0|-\sqrt{P_0}\|_{H^1}<\delta_0$ (consequently $Q_0(T)=(\sqrt{P_0}+A)e^{i\Omega(T)},$with $A=|Q_0|-\sqrt{P_0}$) and for the corresponding $X^1$-solution $Q(Z,T)$ of the problem $(\ref{SNLSEIVP})$ $||A(0,\cdot)||_{H^1}<\delta$ and $||\Omega_T(0,\cdot) ||_{L^2}<\delta,$ then for any $Z>0$ and $T_0\in \RR$ there exists $\Gamma=\Gamma(Z,T_0)$ such that
 $$||Q(Z,T)e^{i\Gamma}-\Phi(Z)||_{H^1(T_0-d,T_0+d)}<\epsilon.$$
 \label{corollariostability}
 \end{cor}
 {\bf Proof}\qquad Let 
 $$
 Q(Z,T)=\Big(\sqrt{P_0}+A(Z,T)\Big)\;e^{i[-P_0 Z+\Omega(Z,T)]}
 $$
 be the perturbed solution given by Theorem $\ref{stability}.$ We now evaluate
 $$||Q(Z,T)e^{i\Gamma}-\Phi(Z)||_{H^1(T_0-d,T_0+d)}.$$
 Thus, we have that
 $$|Q(Z,T)e^{i\Gamma}-e^{-i P_0 Z}|=|(\sqrt{P_0}+A(Z,T))e^{i\Omega(Z,T)+i\Gamma(Z,T_0)}-\sqrt{P_0}|.$$
 We now take the following integral with respect the variable $T:$
 \begin{eqnarray*}
 &&\int_{T_0-d}^{T_0+d}|(\sqrt{P_0}+A(Z,T))e^{i\Omega(Z,T)+i\Gamma(Z,T_0)}-\sqrt{P_0}|^2 dT\\
 &&  \leq \int_{T_0-d}^{T_0+d}|A(Z,T)|^2 dT+\int_{T_0-d}^{T_0+d}|\sqrt{P_0}e^{i\Omega(Z,T)+i\Gamma(Z,T_0)}-\sqrt{P_0}|^2dT\\
 &&=\int_{T_0-d}^{T_0+d}|A(Z,T)|^2 dT+P_0\int_{T_0-d}^{T_0+d}\Big|e^{i\Omega(Z,T)+i\Gamma(Z,T_0)}-1\Big|^2 dT\\
 \end{eqnarray*}
 Moreover by differentiating and then taking the integral with respect the variable $T$ we obtain that
 \begin{eqnarray*}
 &&\int_{T_0-d}^{T_0+d}\Big|A_T(Z,T)+\Omega_T(Z,T)\big(\sqrt{P_0}+A(Z,T)\big)\Big|^2 dT\\
 &&\leq\int_{T_0-d}^{T_0+d}\Big|A_T(Z,T)\Big|^2 dT+\int_{T_0-d}^{T_0+d}\Big|\Omega_T(Z,T)\Big(\sqrt{P_0}+A(Z,T)\Big)\Big|^2 dT .\\
 \end{eqnarray*}
 Thus, mixing the last two relations we have that
 
 \begin{eqnarray*}
 &&\int_{T_0-d}^{T_0+d}|(\sqrt{P_0}+A(Z,T))e^{i\Omega(Z,T)+i\Gamma(Z,T_0)}-\sqrt{P_0}|^2 dT+\\
 &&+\int_{T_0-d}^{T_0+d}\Big|A_T(Z,T)+\Omega_T(Z,T)\big(\sqrt{P_0}+A(Z,T)\big)\Big|^2 dT\\
 &&\leq \int_{T_0-d}^{T_0+d}|A(Z,T)|^2 dT+P_0\int_{T_0-d}^{T_0+d}\Big|e^{i\Omega(Z,T)+i\Gamma(Z,T_0)}-1\Big|^2 dT\\
 &&+\int_{T_0-d}^{T_0+d}\Big|A_T(Z,T)\Big|^2 dT+\int_{T_0-d}^{T_0+d}\Big|\Omega_T(Z,T)\Big(\sqrt{P_0}+A(Z,T)\Big)\Big|^2 dT \\
 &&=||A(Z,\cdot)||_{H^1}^2+\int_{T_0-d}^{T_0+d}\Big|\Omega_T(Z,T)\Big(\sqrt{P_0}+A(Z,T)\Big)\Big|^2 dT\\
 &&+P_0\int_{T_0-d}^{T_0+d}\Big|e^{i\Omega(Z,T)+i\Gamma(Z,T_0)}-1\Big|^2 dT\\
 \end{eqnarray*}
 From theorem $\ref{stability}$ it follows that
 $$
 ||A(Z,\cdot)||_{H^1}<\varepsilon,\quad\mbox{and}\quad \int_{T_0-d}^{T_0+d}\Big|\Omega_T(Z,T)\Big(\sqrt{P_0}+A(Z,T)\Big)\Big|^2 dT<\varepsilon^2 .
 $$
 It remains to evaluate the integral
 $$
 \int_{T_0-d}^{T_0+d}\Big|e^{i\Omega(Z,T)+i\Gamma(Z,T_0)}-1\Big|^2 dT .
 $$
 We claim that there exists $\Gamma(Z,T_0)$ such that it is as small as we wish. In order to show it, we observe that, since $\Omega(Z,\cdot)$ is absolutely continuous for every $Z>0,$ we have that
 \begin{equation}
 \Omega(Z,T)-\Omega(Z,T_0)=\int_{T_0}^T\Omega_T(Z,\tau)d\tau .
 \label{AC}
 \end{equation}
 Then by Holder inequality Eq. $(\ref{AC})$ becomes
 \begin{equation*}
 |\Omega(Z,T)-\Omega(Z,T_0)|\leq||\Omega_T(Z,\cdot)||_{L^2}|T-T_0|^{1/2} ,
 \end{equation*}
 so, recalling that $T\in (T_0-d,T_0+d)$ and that from Theorem $\ref{stability}$ we have that $||\Omega_T(Z,\cdot)||_{L^2}<\varepsilon,$ we obtain that
 \begin{equation}
 |\Omega(Z,T)-\Omega(Z,T_0)|<\sqrt{2d}\;\varepsilon.
 \label{stimaholder}
 \end{equation}
 Therefore if we choose $\Gamma(Z,T_0)=\pi-\Omega(Z,T_0)$ because of $(\ref{stimaholder})$ and  the \textit{Euler identity} $e^{i\pi}+1=0$ we can make the integral
 $$
 \int_{T_0-d}^{T_0+d}\Big|e^{i\Omega(Z,T)+i\Gamma(Z,T_0)}-1\Big|^2 dT 
 $$
 small, and so the thesis of the Corollary easily follows. $\Box$
 \vspace{.5cm}
 
 We would now give an estimate about the coefficient $\delta_0$ that is stated in theorem $\ref{stability}; $ in Chapter $2$ we have claimed that if a function belongs to the Hilbert space $H^2(\RR),$ its $L^\infty$ norm can be controlled by its $H^2(\RR)$ norm. Now we claim that the coefficient that make this control is the real number $1.$ Indeed, let us start with the following statement:
 \begin{pro}
 Let $I\subset\RR$ be an open interval, bounded or not, and let $A(\cdot)\in H^1(I).$ Then there exists a sequence $(A_n)_{n\in \NN}\subset C_c^\infty(\RR)$ such that $(A_n)_{\mid_I}\rightarrow A$ in the sense of $H^1(I).$
 \end{pro}
 {\bf Proof}\qquad [\cite{Brezis}, Theorem VIII.$6$].
 \vspace{.5cm}
 
 We now recall the Young' inequality:
 $$
 ab\leq \frac{1}{p}a^p+ \frac{1}{p^\prime}b^{p^\prime}\qquad \mbox{for all}\quad a\geq 0,\quad b\geq 0,
 $$
 with $\frac{1}{p}+\frac{1}{p^\prime}=1.$
 \vspace{.5cm}
 
 Finally we have:
\begin{pro}
Let  be  $A\in H^1(\RR).$ Then we have
\begin{equation}
||A||_{L^\infty(\RR)}\leq ||A||_{H^1(\RR)}
\end{equation}
\label{stima}
\end{pro}
{\bf Proof}\qquad Let be $v\in C^1_c(\RR)$ and set $G(s)=|s|s.$ The function $w=G(v)$ is in $C_c^1(\RR)$ and $w^\prime =2|v|v^\prime.$ Therefore, for $T\in \RR  $ we have
$$G(v(T))=\int_{-\infty}^T 2|v(x)|v(x)dx.$$
So by using Holder's inequality we obtain
$$|v(T)|^2\leq 2||v||_{L^2}||v^\prime||_{L^2}.$$
From this, by Young's inequality, we deduce that
\begin{equation}
||v||_{L^\infty}\leq ||v||_{H^1}\qquad \mbox{for all}\quad v\in C_c^{\infty}(\RR).
\label{stima2}
\end{equation}
Now, let be $A\in H^1(\RR);$ there exists a sequence $(A_n)_{n\in \NN}\subset C_c^1(\RR)$ such that $A_n\rightarrow A$ in $H^1 (\RR).$ By using $(\ref{stima2})$ we obtain that $(A_n)_{n\in \NN}$ is a Cauchy sequence in $L^\infty,$ so $A_n\rightarrow A$ in $L^\infty$ and we have the thesis. $\Box$
\vspace{.5cm}

\begin{remark}
Returning to theorem $\ref{stability},$ because of Proposition $\ref{stima}$ it is enough to choose 
$$
\delta_0=\frac{\sqrt{P_0}}{2}
$$
to have the implication 
$$\sup_{T\in \RR}|Q(\cdot)|<\frac{\sqrt{P_0}}{2}\quad \mbox{if}\quad ||A(\cdot)||_{H^1}<\delta_0.$$
\label{costante}
\end{remark}

\section{Transformed Time-Homogeneous Solutions}
We should now convert results achieved in the previous section to the TNLSE. Thus, let consider $(\ref{trasformazionimodello})$ by setting $\rho=-1.$ Therefore, the following transformation 

\begin{equation}
\left\{\begin{array}{l}
T(z,t)=e^{-C_2 z}t\\
\\
Z(z)=\frac{1}{2 C_2}(1-e^{-2C_2 z})\\
\\
v(z,t)=e^{i\frac{C_2}{4}t^2-\frac{C_2}{2}z}Q(Z(z),T(z,t))\\
\end{array}\right.
\label{trasformazionimodello1}
\end{equation}
\vspace{.5cm}

allows us to convert the TNLSE
\begin{equation}
iv_z(z,t)+v_{tt}(z,t)-e^{-C_2 z}|v(z,t)|^2v(z,t)+\frac{C_2^2}{4}t^2v(z,t)=0
\label{TNLSE1}
\end{equation}
into the standard NLSE
\begin{equation}
iQ_Z(Z,T)+Q_{TT}(Z,T)- |Q(Z,T)|^2Q(Z,T)=0.
\label{SNLSE2}
\end{equation}

The function $\Phi(Z)=\sqrt{P_0}\;e^{-iP_0Z}$ is solution of $(\ref{SNLSE2})$ for every $Z>0.$ Let consider $Z\in[0,\frac{1}{2C_2});$ then from $(\ref{trasformazionimodello1})$ it follows that
\begin{equation}
v(z,t)=e^{i\frac{C_2}{4}t^2-\frac{C_2}{2}z}\Phi(Z(z))
\label{nonvanishtrasformata}
\end{equation}
solves Eq. $(\ref{TNLSE1})$ for every $z\geq 0.$

\vspace{.5cm}

Let consider the perturbed solution of $(\ref{SNLSE2})$ given by Theorem $\ref{stability},$ that is
\begin{equation}
Q(Z,T)=\Big(\sqrt{P_0}+A(Z,T)\Big)\;e^{i\big[- P_0 Z+\Omega(Z,T)\big]},
\label{perturbedsolution}
\end{equation}
with $Z\in[0,\frac{1}{2C_2}).$ From $(\ref{trasformazionimodello1})$ we have that
\begin{equation}
w(z,t)=\;e^{i\frac{C_2}{4}t^2-\frac{C_2}{2}z}Q(Z(z),T(z,t)),
\label{wdiperturbazione}
\end{equation}
 solves Eq. $(\ref{TNLSE1})$ for every $z\geq 0.$ 
 
 Let $t_1,t_1\in \RR,$ such that $t_1<t_2$ and  $I=(t_1,t_2)$ be a bounded interval of $\RR ;$ we have easily that $v(z,\cdot)\in H^1(I),$ for every $z\geq 0.$ Furthermore, because of Theorem $\ref{stability}$ it follows straightforwardly that also $w(z,\cdot)\in H^1(I)$ for every $z\geq 0.$ Indeed, differentiating Eq. $(\ref{wdiperturbazione})$ with respect to the variable $t$ we have that
 $$
 w_t(z,t)=\Big[\Big(i\frac{C_2}{2}t\Big)Q(Z(z),T(z,t))+e^{-C_2 z}Q_T(Z(z),T(z,t))\Big]
 $$
 and taking the square integral over $I$ with respect to the variable $t$ we arrive at
\begin{eqnarray*}
 \int_I|w_t(z,t)|^2 dt&\leq& e^{-C_2 z}\int_I\Big(\frac{C_2}{2}t\Big)^2|Q(Z(z),T(z,t))|^2 dt\\
 &&+e^{-3C_2 z}\int_I|Q_T(Z(z),T(z,t))|^2 dt
 \end{eqnarray*}
 that is obviously finite for every $z>0.$
\vspace{.5cm}

We are now ready to state the following result as consequence of Corollary $\ref{corollariostability}:$
\begin{theorem}
Let $\tau>0,$ $\varepsilon\in (0,\frac{\sqrt{P_0}}{2}) $ and $$Q(Z,T)=\Big(\sqrt{P_0}+A(Z,T)\Big)\;e^{i\big[- P_0 Z+\Omega(Z,T)\big]},$$ with $||A(Z)||_{H^1(\RR)}\leq \varepsilon,\quad ||\Omega_T(Z)||_{L^2(\RR)}<\varepsilon,$ be a solution of $(\ref{SNLSE2})$ for every $Z\in[0,\frac{1}{2C_2})$ given by Theorem $\ref{stability}.$ Let us consider
$$
v(z,t)=\;e^{i\frac{C_2}{4}t^2-\frac{C_2}{2}z}\Big(\sqrt{P_0}\;e^{-i P_0 Z(z)}\Big)
$$
and
$$
w(z,t)=\;e^{i\frac{C_2}{4}t^2-\frac{C_2}{2}z}\;Q(Z(z),T(z,t)).
$$

Then, for every $z\geq 0$ and $t_0\in \RR,$ there exists $\Gamma=\Gamma(z,t_0)$ such that
\begin{equation}
||w(z)e^{i\Gamma}-v(z)||_{H^1(t_0-\tau,t_0+\tau)}<\varepsilon\; H(z,\tau),
\label{stimaTNLSEnonvanish}
\end{equation}
where $H$ is a smooth function depending on $\tau,\;z.$
\end{theorem}
{\bf Proof}\qquad In order to evaluate $||w(z)e^{i\Gamma}-v(z)||_{H^1(t_0-\tau,t_0+\tau)}$ we take the first derivative of $w(z,t)$ with respect the variable $t:$
\begin{eqnarray*}
w_t(z,t)&=&\Big[\big(i\frac{C_2}{2}t\big)Q(Z(z),T(z,t))+Q_T(Z(z),T(z,t))\frac{dT}{dt}\Big]e^{i\frac{C_2}{4}t^2-\frac{C_2}{2}z}\\
&=&\Big[\big(i\frac{C_2}{2}t\big)Q(Z(z),T(z,t))+Q_T(Z(z),T(z,t))e^{-C_2 z}\Big]e^{i\frac{C_2}{4}t^2-\frac{C_2}{2}z}\\
\end{eqnarray*}
we now take the first derivative of $v(z,t)$ with respect the variable $t:$
\begin{eqnarray*}
v_t(z,t)&=&\Big[\big(i\frac{C_2}{2}t\big)\Big(\sqrt{P_0}\;e^{-i P_0 Z(z)}\Big)\Big]e^{i\frac{C_2}{4}t^2-\frac{C_2}{2}z}.
\end{eqnarray*}

Because of $(\ref{trasformazionimodello1})$ we have that $z=z(Z)$ and $t_0=t_0(Z,T).$ So, let us set $\Gamma$ as in Corollary $\ref{corollariostability}.$ Thus, we have that
\begin{eqnarray*}
&&\int_{t_0-\tau}^{t_0+\tau}|w_t(z,t)e^{i\Gamma}-v_t(z,t)|^2 dt\\
&&=\;e^{-C_2 z}\int_{t_0-\tau}^{t_0+\tau}\Big|(i\frac{C_2}{2}t\big)Q(Z(z),T(z,t))e^{i\Gamma}+e^{-C_2 z}Q_T(Z(z),T(z,t))e^{i\Gamma}\\
&&-(i\frac{C_2}{2}t\big)\Big(\sqrt{P_0}\;e^{-i P_0 Z(z)}\Big)\Big|^2 dt\\
&&=\;e^{-C_2 z} \int_{\widetilde{I}}\Big|i\frac{C_2}{2}\;e^{C_2 z}T\Big(Q(Z,T)e^{i\Gamma}-\sqrt{P_0}\;e^{-i P_0 Z}\Big)+e^{-C_2 z}Q_T(Z,T)\Big|^2 e^{C_2 z} dT\\
\\
&&= \int_{\widetilde{I}}\Big|i\frac{C_2}{2}\;e^{C_2 z}T\Big(Q(Z,T)e^{i\Gamma}-\sqrt{P_0}\;e^{-i P_0 Z}\Big)+e^{-C_2 z}Q_T(Z,T)\Big|^2 dT,
\end{eqnarray*}
where $\widetilde{I}=\Big(e^{-C_2 z}(t_0-\tau),e^{-C_2 z}(t_0+\tau)\Big).$ Thus, we have that
\begin{eqnarray*}
||w_t(z)e^{i\Gamma}-v_t(z)||_{L^2(I)}&\leq& \frac{C_2}{2}\;e^{C_2 z}\Bigg(\int_{\widetilde{I}}\Big|T\Big(Q(Z,T)e^{i\Gamma}-\sqrt{P_0}\;e^{-i P_0 Z}\Big)\Big|^2 dT\Bigg)^{1/2}\\
&&+e^{-C_2 z}\Bigg(\int_{\widetilde{I}}\Big|Q_T(Z,T)\Big|^2 dT\Bigg)^{1/2}.
\end{eqnarray*}
Taking into consideration the first integral of the previous inequality, by Holder inequality and Theorem $\ref{stability}$ we arrive at
\begin{eqnarray*}
&&\int_{\widetilde{I}}\Big|T\Big(Q(Z,T)e^{i\Gamma}-\sqrt{P_0}\;e^{-i P_0 Z}\Big)\Big|^2 dT\\
&&\leq \sup_{T\in\widetilde{I}}|T|^2\;||Q(Z)\;e^{i\Gamma}-\sqrt{P_0}\;e^{-iP_0 Z}||^2_{L^2(\widetilde{I})}\\
&&<4\tau^2\; e^{-2C_2 z}\varepsilon^2.
\end{eqnarray*}
Furthermore, from Theorem $\ref{stability}$ it easily follows that
$$
\int_{\widetilde{I}}\Big|Q_T(Z,T)\Big|^2 dT\leq \varepsilon^2.
$$

Moreover, we have that
\begin{eqnarray*}
\int_{t_0-\tau}^{t_0+\tau}|w(z,t)e^{i\Gamma}-v(z,t)|^2 dt&=&e^{-C_2 z}\int_{t_0-\tau}^{t_0+\tau}|Q(Z,T)e^{i\Gamma}-\sqrt{P_0}e^{i\rho P_0Z}|^2 e^{C_2 z} dT\\
&=&\int_{t_0-\tau}^{t_0+\tau}|Q(Z,T)e^{i\Gamma}-\sqrt{P_0}e^{i\rho P_0Z}|^2 dT\\
&<&\varepsilon^2
\end{eqnarray*}
Summing up, we have that
\begin{eqnarray*}
&&||w(z,t)e^{i\Gamma }-v(z,t)||_{H^1(t_0-\tau,t_0+\tau)}\leq\\
&&\leq \frac{C_2}{2}\;e^{C_2 z}\Big(4\tau^2\;e^{-2C_2 z}\varepsilon^2+\varepsilon^2\Big)^{1/2}+\varepsilon
\end{eqnarray*}
Finally, setting
$$
H(z,\tau)=\Big[\frac{C_2}{2}\;e^{C_2 z}\Big(4\tau^2\;e^{-2C_2 z}+1\Big)^{1/2}+1\Big]
$$
we have the thesis of the Theorem, that is  $$||w(z,t)e^{i\Gamma }-v(z,t)||_{H^1(t_0-\tau,t_0+\tau)}\leq H(z,\tau)\;\varepsilon\quad \Box$$

\section{Optical pulse into absorption regime}
In \cite{Forestieri} the propagation of light in optical fibers is governed by the equation
\begin{equation}
iu_z(z,t)+\frac{\beta_2}{2}u_{tt}(z,t)-\gamma e^{-\alpha z}|u(z,t)|^2u(z,t)=0,
\label{alfaing}
\end{equation}
that we obtain from $(\ref{Ing})$ by setting the transformation $$A(z,t)\Doteq \exp[(-\alpha z/2)u(z,t)],$$
and then by reversing the $z$- axis.

In \cite{Forestieri} is given special attention to the propagation of a stochastic process, such as an optical signal affected by the amplified spontaneous emission (ASE) noise generated by optical amplifiers. For the purpose of investigating the interaction between signal and noise, the signal is considered a continuous wave (CW), and the noise is treated as a perturbation of the CW solution of the Eq. $(\ref{alfaing}).$ In order to have stability against small perturbation of CW solution, we assume $\beta_2 >0,$ whereas $\alpha$ and $\gamma$ are always positive \textit{real} coefficients.
\\
Taking a non-rigorous analysis  based on the approach described in \cite{Forestieri} and on the results carried out in section $4.2.2,$ we investigate small perturbations against CW solution of  Eq.  $(\ref{alfaing}).$

Let consider the CW solution as
\begin{equation}
\Psi(z)=\sqrt{P_0}\;e^{-i\phi_{NL}(z)},
\label{CWsolution}
\end{equation}
where $P_0$ is a positive \textit{real} coefficient and 
\begin{equation}
\phi_{NL}(z)=\int_0^z\gamma P_0\exp[-\alpha \zeta]d\zeta 
\label{time-independent nonlinear phase rotation}
\end{equation}
is the deterministic time-independent nonlinear phase rotation of the noise-free solution. We write the perturbed solution of Eq. $(\ref{alfaing})$ as
\begin{equation}
u(z,t)=\sqrt{P_0}\;[1+c(z,t)]\;e^{-i\theta(z,t)}
\label{perturbedCW}
\end{equation}
where $c(z,t)$ is a \textit{complex}-valued function and $\theta(z,t)$ is a \textit{real}-valued function. Substituting Eq. $(\ref{perturbedCW})$ into Eq. $(\ref{alfaing})$ we obtain the complex differential equation
\begin{eqnarray}
&&c_z(z,t)-i(1+c(z,t))\theta_z(z,t)=\nonumber\\
&&i\frac{\beta_2}{2}\Big[c_{tt}(z,t)-i(1+c(z,t))\theta_{tt}(z,t)-2i c_t(z,t)\theta_t(z,t)-(1+c(z,t))\theta_t^2(z,t)\Big]\nonumber\\
&&-i\gamma P_0\; e^{-\alpha z}|1+c(z,t)|^2(1+c(z,t)).
\label{complexdifferentialequation}
\end{eqnarray}
If we write $c(z,t)=a(z,t)+i\;b(z,t)$ with $a(z,t)$ and $b(z,t)$ \textit{real}-valued function, Eq. $(\ref{perturbedCW})$ has one additional degree of freedom; as a consequence, in $(\ref{complexdifferentialequation}),$ we have two equations for three unknown quantities $a,\;b$ and $\theta,$ which makes the system undetermined. Hence, we have the freedom to arbitrarily add an additional equation to relate these three variables. In \cite{Forestieri} it is found an equation for $\theta$ in order to minimize the impact of the terms that will be neglected in the linearization process. Since the term $\theta_z(z,t)$ on the left-hand side of $(\ref{complexdifferentialequation})$ is multiplied by the factor $i(1+c(z,t)),$ we may equate it to the fourth and the fifth term on the right-hand side of $(\ref{complexdifferentialequation})$ and eliminate the common factor, obtaining

\begin{equation}
\theta_z(z,t)=\gamma P_0\exp(-\alpha z)|1+c(z,t)|^2+\frac{\beta_2}{2}\theta_t^2(z,t).
\label{forestieri}
\end{equation}
\\
We now consider the following assumptions: let  $\phi(z,t)$ be a \textit{real}-valued function such that $||\phi_t(z,\cdot)||_{L^2}$ is finite for every $z>0.$ Thus, we set

\begin{equation}
\theta(z,t)=\phi_{NL}(z)+\phi(z,t),
\label{faseperturbata}
\end{equation}

where $\phi_{NL}(z)$ is given by $(\ref{time-independent nonlinear phase rotation}).$ Furthermore, Theorem $\ref{stability}$ and Remark $\ref{costante}$ suggest to assume

\begin{equation}
\int_{\RR}\Big(\phi_t(z,t)\Big)^2 dt \quad \in \quad \Big(0,\frac{\sqrt{P_0}}{2}\Big) \quad \mbox{for all}\quad z>0
\label{Ipotesi1}
\end{equation}
and
\begin{equation}
\int_{\RR}|c(z,t)|^2 dt \quad \in \quad \Big(0,\frac{\sqrt{P_0}}{2}\Big)\quad \mbox{for all}\quad z>0,
\label{Ipotesi2}
\end{equation}
where $\Big(0,\frac{\sqrt{P_0}}{2}\Big)$ is a bounded interval of $\RR.$ If we assume that the loss $\alpha$ in optical fiber affects both the functions $\phi(z,t)$ and $c(z,t),$ we have that a possible function that fulfils $(\ref{Ipotesi1})$ should be

\begin{equation}
\phi_t^2(z,t)=e^{-\alpha z}\; e^{-\frac{4\pi t^2}{P_0}}.
\label{Gaussiana1}
\end{equation}
Indeed, at $z=0$ we have that
$$
\int_\RR \Big(\phi_t(0,t)\Big)^2 dt=\int_\RR e^{-\frac{4\pi t^2}{P_0}} dt = \frac{\sqrt{P_0}}{2},
$$
so, it is clear that the function $e^{-\alpha z}\; e^{-\frac{4\pi t^2}{P_0}}$ satisfies $(\ref{Ipotesi1}).$ Similarly, we set
\begin{equation}
|c(z,t)|^2 =e^{-\alpha z}\; e^{-\frac{4\pi t^2}{P_0}}.
\label{Gaussiana2}
\end{equation}
In view of all these assumptions, Eq. $(\ref{forestieri})$ becomes
\begin{equation*}
\gamma P_0 e^{-\alpha z}+\phi_z=\gamma P_0 e^{-\alpha z}(1+2a+|c|^2)+\frac{\beta_2}{2}(\phi_t)^2,
\end{equation*}

that because of $(\ref{Gaussiana1})$ and $(\ref{Gaussiana2})$ turns into

\begin{equation}
\phi_z(z,t)=2\gamma P_0\; e^{-\alpha z}\; a(z,t)+\gamma P_0\; e^{-2\alpha z}\; e^{-\frac{4\pi t^2}{P_0}}+\frac{\beta_2}{2}\;e^{-\alpha z}\; e^{-\frac{4\pi t^2}{P_0}}.
\label{Forestieri2}
\end{equation}

The other two equations of the system  are obtained by dividing the remaining terms of $(\ref{complexdifferentialequation})$ into real and imaginary parts. Writing explicitly $c(z,t)=a(z,t)+ib(z,t),$ the system of three real equations is

\begin{equation}
\left\{\begin{array}{l}
a_z(z,t)=\frac{\beta_2}{2}\Big(-b_{tt}(z,t)+\theta_{tt}(z,t)+a(z,t)\theta_{tt}(z,t)+2a_t(z,t)\theta_t(z,t)\Big)\\
\\
b_z=\frac{\beta_2}{2}\Big(a_{tt}(z,t)+b(z,t)\theta_{tt}(z,t)+2b_t(z,t)\theta_t(z,t)\Big)\\
\\
\phi_z(z,t)=2\gamma P_0\; e^{-\alpha z}\; a(z,t)+\gamma P_0\; e^{-2\alpha z}\; e^{-\frac{4\pi t^2}{P_0}}+\frac{\beta_2}{2}\;e^{-\alpha z}\; e^{-\frac{4\pi t^2}{P_0}}
\end{array}\right.
\label{system}
\end{equation}

Recalling $(\ref{faseperturbata})$ and $(\ref{Gaussiana1})$ we obtain that
\begin{equation}
\theta_{tt}(z,t)=\phi_{tt}(z,t)=-\frac{\pi}{P_0}\;t\; e^{-\frac{\alpha}{2}z}\;e^{-\frac{2\pi}{P_0}t^2}.
\label{derivatasecondadellafase}
\end{equation}

We now compute the Fourier transforms of the functions $e^{-\frac{4\pi t^2}{P_0}}$ and $t\;e^{-\frac{2\pi}{P_0}t^2}$ with respect the variable $t.$ It is well-known that
$$
\mathcal{F}\Big(e^{-\kappa t^2}\Big)(\omega)=\sqrt{\frac{\pi}{\kappa}}\;e^{-\frac{\omega^2}{4\kappa}}\quad \mbox{and}\quad \mathcal{F}\Big(te^{-\kappa t^2}\Big)(\omega)=-\frac{\omega}{2 \kappa}\sqrt{\frac{\pi}{\kappa}}\;e^{-\frac{\omega^2}{4\kappa}}.
$$
Thus, we obtain that

$$
\mathcal{F}\Big(e^{-\alpha z}\;e^{-\frac{4\pi t^2}{P_0}}\Big)=\frac{\sqrt{P_0}}{2}\;e^{-\frac{P_0\omega^2}{16\pi}}
$$
and
$$
\mathcal{F}\Big(-\frac{\pi}{P_0}\;e^{-\frac{\alpha z}{2}}t\;e^{-\frac{2\pi t^2}{P_0}}\Big)=\sqrt{\frac{P_0}{2}}\;e^{-\frac{\alpha z}{2}}\omega\;e^{-\frac{\omega^2P_0}{8\pi}}.
$$
Finally, by neglecting the quadratic terms in the first and the second equations of the system $(\ref{system})$ and denoting the Fourier transform by uppercase letters of the corresponding quantities in lowercase, we obtain the following system:
\begin{equation}
\left\{\begin{array}{l}
A_z(z,\omega)=\frac{\beta_2}{2}\omega^2B(z,\omega)+\sqrt{\frac{P_0}{2}}e^{-\frac{\alpha z}{2}}\omega\;e^{-\frac{P_0\omega^2}{8\pi}}\\
\\
B_z(z,\omega)=-\frac{\beta_2}{2}\omega^2 A(z,\omega)\\
\\
\Phi_z(z,\omega)=2\gamma P_0\;e^{-\alpha z}A(z,\omega)+\gamma P_0\frac{\sqrt{P_0}}{2}\;e^{-2\alpha z}\;e^{-\frac{P_0\omega^2}{16\pi}}+\frac{\beta_2}{2}\frac{\sqrt{P_0}}{2}\;e^{-\alpha z}\;e^{\frac{P_0\omega^2}{16\pi}}
\end{array}\right.
\label{Fouriersystem}
\end{equation}

Let us first solve the first two equations of system $(\ref{Fouriersystem}).$ Since $B_z=-\frac{\beta_2}{2}\omega^2 A,$ we have that
\begin{equation}
B(z)=B(0)+\int_0^z\frac{\beta_2}{2}\omega^2 A(z^\prime)dz^\prime.
\label{Bintermedia}
\end{equation}
Substituting Eq. $(\ref{Bintermedia})$ in the first equation of the system $(\ref{Fouriersystem})$ we obtain that
$$
A_z=\frac{\beta_2}{2}\omega^2\Big(B(0)+\int_0^z\frac{\beta_2}{2}\omega^2 A(z^\prime)dz^\prime\Big)+\sqrt{\frac{P_0}{2}}\omega\;e^{-\frac{P_0\omega^2}{8\pi}}\;e^{-\frac{\alpha z}{2}}.
$$
Differentiating with respect the variable $z$ we have that
$$
A_{zz}=\Big(\frac{\beta_2}{2}\omega^2\Big)^2 A-\frac{\alpha}{2}\sqrt{\frac{P_0}{2}}\;\omega\;e^{-\frac{\omega^2 P_0}{8\pi}}\;e^{-\frac{\alpha z}{2}},
$$
whose solution is easily found to be
\begin{equation}
A(z,\omega)=\frac{2\alpha \sqrt{P_0/2}\;\omega\;e^{-\frac{\omega^2 P_0}{8\pi}}}{(\beta_2\omega^2)^2-\alpha^2}\;e^{-\frac{\alpha z}{2}}+c_1\;e^{-\frac{\beta_2\omega^2}{2}z}+c_2\;e^{\frac{\beta_2\omega^2}{2}z},
\label{A}
\end{equation}
where $c_1$ and $c_2$ are arbitrary \textit{real} coefficients. Substituting Eq. $(\ref{A})$ into Eq. $(\ref{Bintermedia})$ we obtain that

\begin{equation}
B(z,\omega)=-\frac{\beta_2}{2}\omega^2\Big[-4\frac{\sqrt{P_0/2}\;\omega\;e^{-\frac{\omega^2 P_0}{8\pi}}}{(\beta_2\omega^2)^2-\alpha^2}\;e^{-\frac{\alpha z}{2}}-\frac{2c_1}{\beta_2\omega^2}\;e^{-\frac{\beta_2\omega^2}{2}z}+\frac{2c_2}{\beta_2\omega^2}\;e^{\frac{\beta_2\omega^2}{2}z}\Big]+\kappa_1,
\label{B}
\end{equation}
where $\kappa_1$ is a \textit{real} coefficient. Finally, substituting Eq. $(\ref{A})$ into the third equation of the system $(\ref{Fouriersystem}),$ and integrating with respect the variable $z$ we have that
\begin{eqnarray*}
\Phi(z)&=&2\gamma P_0\Big[-\frac{4}{3}\frac{\sqrt{P_0/2}\;\omega\;e^{-\frac{\omega^2 P_0}{8\pi}}}{(\beta_2\omega^2)^2-\alpha^2}\;e^{-\frac{3\alpha z}{2}}-\frac{2c_1\;e^{-\frac{\beta_2\omega^2}{2}z-\alpha z}}{\beta_2\omega^2+2\alpha}+\frac{2c_2\;e^{\frac{\beta_2\omega^2}{2}z-\alpha z}}{\beta_2\omega^2-2\alpha}\Big]\\
\\
&-& \frac{\gamma P_0}{2\alpha}\frac{\sqrt{P_0}}{2}\;e^{-2\alpha z}\;e^{-\frac{P_0\omega^2}{16\pi}}-\frac{\beta_2}{2\alpha}\frac{\sqrt{P_0}}{2}\;e^{-\alpha z}\;e^{-\frac{P_0\omega^2}{16\pi}}.\\
\end{eqnarray*}
Thus, the stability against small perturbation is provided with $\omega\neq \pm \sqrt{\frac{\alpha}{\beta_2}}$ and $\omega\neq \pm \sqrt{\frac{2\alpha}{\beta_2}}.$

\section{Soliton-like solutions}
\textit{Modulation instability} occurs in the regime of optical fibers when in Eq. $(\ref{Ing})$ $\beta_2<0.$ In this case, it turns out that Eq. $(\ref{Ing})$ has specific solutions that do not change along fiber length. If the fiber loss are ignored, i.e. $\alpha=0,$ Eq. $(\ref{Ing})$ becomes Eq. $(\ref{standard ing}).$  Because of $\beta_2<0,$ we can refer to the Eq. $(\ref{standard ing})$ as the dimensionless model
\begin{equation}
iQ_Z(Z,T)+Q_{TT}(Z,T)+|Q(Z,T)|^2Q(Z,T)=0,
\label{standardfocusing}
\end{equation}
where $Z,T\in \RR.$ We look for solutions of Eq. $(\ref{standardfocusing})$ as 
\begin{equation}
Q(Z,T)=\Phi(T)\;e^{i\theta Z},
\label{solitarywawes}
\end{equation}
where the profile $\Phi(T)$ is a \textit{real}-valued positive function independent of the variable $Z,$ and $\theta$ is a \textit{real} positive coefficient. The solutions of these kinds will be called the \textit{solitary waves} if $\Phi$ is a bounded function possessing limits as $T\rightarrow\pm \infty.$

Substitution of Eq. $(\ref{solitarywawes})$ into Eq. $(\ref{standardfocusing})$ leads to consider the following equation:
\begin{equation}
\Phi^{\prime\prime}(T)-\theta \Phi(T)+\Phi(T)^3=0.
\label{equationestazionaria}
\end{equation}
Adding the condition: $\Phi(+\infty)=\Phi(-\infty)=0$ we solves Eq. $(\ref{equationestazionaria})$ making a qualitative analysis. First of all, we note that if a solution of this equation is bounded on a half interval $[0,b),$ it can be continued onto a right half-neighborhood of the point $b.$ Indeed, if a solution $\Phi$ of $(\ref{equationestazionaria})$ is bounded, then it follows from this equation that the second derivative $\Phi^{\prime \prime }(\cdot)$ is bounded, too, therefore the first derivative $\Phi^{\prime}(\cdot)$ of this solution is also bounded. Setting $\Phi_0=\Phi(0)+\int_0^b\Phi^\prime(T)dT$ and $\Phi_0^\prime=\Phi^\prime(0)+\int_0^b\Phi^{\prime \prime}(T)dT$ and considering the Cauchy problem for equation $(\ref{equationestazionaria})$ with the initial data $\Phi(b)=\Phi_0,\quad \Phi^\prime(b)=\Phi_0^\prime,$ we immediately get our statement.

Let $f_1(\Phi)=\Phi^3-\theta \Phi$ and $F_1(\Phi)=\int_0^\Phi f_1(p)dp=\frac{\Phi^2}{2}\Big( \frac{\Phi^2}{2} -\theta\Big).$ Then the equality
\begin{equation} 
\{ \frac 1 2 (\Phi^\prime)^2+F_1(\Phi) \}^\prime=0
\label{conservazioneenergia}
\end{equation}
follows from $(\ref{equationestazionaria}).$ Let be $b=\sqrt{2\;\theta}>0$ such that $F_1(b)=0,\quad f_1(b)>0$ and $F_1(\Phi)<0$ if $\Phi \in (0,b).$ Let us prove that equation $(\ref{equationestazionaria})$ has a solution satisfying the conditions $\Phi(\infty)=0.$ Indeed we take an arbitrary point $T_0\in \RR$ and the following initial data for equation $(\ref{equationestazionaria}):$
$$
 \Phi(T_0)=b,\qquad \Phi^\prime(T_0)=0.
$$
therefore, by $(\ref{equationestazionaria})$ we have that $\Phi^{\prime \prime}(T_0)<0,$ thus $\Phi^\prime(T)<0$ in a right half-neighborhood of the point $T_0.$ There cannot exist a point $T>T_0$ such that $\Phi(T)=0$ because otherwise the function $E=\frac 1 2 [\Phi^\prime(T)]^2+F_1(\Phi(T))$ must be positive at the point $T$ and, hence, non-equal to itself at $T_0$ ($\Phi^\prime(T)\neq 0$ for this point $T$ by the uniqueness theorem because $\Phi \equiv 0$ is a solution of the equation). Furthermore, $\Phi^\prime(T)\neq 0$ for any $T>T_0$ such that $\Phi(T)\in (0,b)$ because otherwise there is no conservation of the energy $E.$ Hence $\Phi(T)\in (0,b)$ for all $T>T_0.$ The above facts easily imply, in particular, that the solution $\Phi$ is global, i.e. it can be continued onto the whole half-line $T>T_0.$ Also, in view of the above arguments, the graphs of the function $\Phi$ has a horizontal asymptote as $T\rightarrow +\infty. $ A simple corollary of the above considerations is that $\Phi(+\infty)=0.$ By analogy, $\Phi(-\infty)=0,\quad \Phi(T)\in (0,b)$ for all $T$ and $\Phi^\prime(T)>0$ from the left of the point $T_0.$
\\
\\
Let us prove that the above-constructed solution $\Phi=\Phi(\theta,T)$ satisfying equation $(\ref{equationestazionaria})$ and the conditions $\Phi(\theta,\pm \infty)=0,\quad \Phi(\theta,T)>0$ for $T\in \RR$ and $\frac{\partial \Phi(\theta,T_0)}{\partial T}=0$ is continuously differentiable as a function of the argument $\theta$ for an arbitrary $T_0$(in view of the invariance of the equation with respect to translations in $T,$ one cannot state the differentiability with respect to $\theta$ of an arbitrary family of solutions depending on the parameter $\theta$). It follows from that the parameter $b=\sqrt{2\;\theta}$ is locally  continuously differentiable as a function of $\theta \in (\theta_0-\delta,\theta_0+\delta)$ for some $\delta>0,$ and $\theta_0>0. $ This is true from the implicit function theorem because $b$ is a solution of the algebraic equation $F_1(\theta,b)=0,$ where $\frac{\partial}{\partial r}F_1(\theta,r)\mid_{r=b}\neq 0,$ and, thus, the continuous differentiability of the function $\phi(\theta,T)$ with respect to $\theta$ is proved.

We shall call the corresponding \textit{solitary waves} of the above-constructed function $\Phi(T)$ as \textit{soliton-like} solutions. Moreover, we refer to $\Phi(T)$ as \textit{ground state}. Defining $$R(T)=\frac{\Phi(\theta^{-1}T)}{\sqrt{\theta}}$$ we have that
 
 \begin{equation}
 R^{\prime\prime}-R+R^3=0.
 \label{equationGS}
 \end{equation}
 
 There exists a unique positive solution of Eq. $(\ref{equationGS})$  of the form
 
 \begin{equation}
 R(T)=\frac{\sqrt{2}}{\cosh(T)}
 \label{GS}
 \end{equation}
 that satisfies the zero boundary conditions at infinity. We refer to the function $R(T)$ as the fundamental \emph{ground state}. Thus, for every fixed $\theta >0$ the unique solution $\Phi$ of Eq. $(\ref{equationestazionaria})$  satisfying the conditions $\Phi(\theta,\pm \infty)=0,\quad \Phi(\theta,T)>0$ for $T\in \RR$ and $\frac{\partial \Phi(\theta,0)}{\partial T}=0$ can be expressed as
 \begin{equation}
 \Phi(\theta,T)=\frac{\sqrt{2\;\theta}}{\cosh(\theta \;T)}\;.
 \label{gs}
 \end{equation}
 Taking the first derivative with respect to the variable $T$ and recalling the well-known relations between hyperbolic functions and exponential, we obtain that
 \begin{eqnarray*}
 \Phi^\prime(\theta,T)&=&-\theta\sqrt{2\;\theta}\;\;\frac{\tanh(\theta\;T)}{\cosh(\theta\;T)}\\
 \\
 &=&-\frac{2\theta\sqrt{2\;\theta}\;e^{\theta T}(e^{2\theta T}-1)}{(e^{2\theta T}+1)^2}.\\
 \end{eqnarray*}
 Therefore, we have that
 \begin{eqnarray*}
 |\Phi(T)|+|\Phi(T)|^\prime&=&\frac{\sqrt{2\;\theta}}{\cosh(\theta \;T)}+\theta\sqrt{2\;\theta}\;\;\frac{\tanh(\theta\;T)}{\cosh(\theta\;T)}\\
 \\
 &=&\frac{2\sqrt{2\;\theta}\;e^{\theta T}}{(e^{2\theta T}+1)^2}\big[(1+\theta)e^{2\theta T}+(1-\theta)\big].\\
 \end{eqnarray*} 
 Thus, we easily have that
 
 \begin{equation}
|\Phi(T)|+|\Phi(T)|^\prime\leq A_1\;e^{-A_2|T|}
\label{stimags}
 \end{equation}
 for some positive \textit{real} constants $A_1$ and $A_2.$ Then, we obtain that the \textit{ground state} $\Phi(\cdot)\in H^1(\RR).$ 
 
 \subsection{Transformed soliton-like solutions}
 Let us recall the transformations $(\ref{trasformazionimodello})$ where we set $\rho=1.$ As consequence we have that $C_1=1.$ Thus, we obtain that
 
 \begin{equation}
 \left\{\begin{array}{l}
 T(z,t)=e^{-C_2 z}t\\
 \\
 Z(z)=(1-e^{-2C_2 z})\\
 \\
 v(z,t)=\;e^{i\frac{C_2}{4}t^2-\frac{C_2}{2}z}Q(Z(z),T(z,t))\\
 \end{array}\right.
 \label{trasformazionimodellofocusing}
 \end{equation}
 \\
 \\
where $Q(Z,T)$ and $v(z,t)$ are solutions of
\begin{equation}
iQ_Z(Z,T)+Q_{TT}(Z,T)+|Q(Z,T)|^2Q(Z,T)=0
\label{SNLSEfocusing}
\end{equation}
and
\begin{equation}
iv_z(z,t)+v_{tt}(z,t)+e^{-C_2 z}|v(z,t)|^2v(z,t)+\frac{C_2^2}{4}t^2v(z,t)=0
\label{TNLSEfocusing}
\end{equation}
 respectively. In the previous section we pointed out the existence of soliton- like solution of Eq. $(\ref{SNLSEfocusing}),$ such as
 \begin{equation}
 Q(Z,T)=\frac{\sqrt{2\;\theta}}{\cosh(\theta T)}\;e^{i\theta Z},
 \end{equation}
 with $\theta$ \textit{real} positive coefficient. Because of Theorem $(\ref{teoremaditrasformazione})$ we have that the function
 \begin{equation}
 v(z,t)=\;e^{i\frac{C_2}{4}t^2-\frac{C_2}{2}z}\frac{\sqrt{2\;\theta}}{\cosh(\theta T(z,t))}\;e^{i\theta Z(z)}
 \end{equation}
 solves Eq. $(\ref{TNLSEfocusing}).$
 \vspace{.5cm}

 \begin{pro}
 Let $I$ be an interval of $\RR ^+;$ then the function $v(z,t)$ belongs to $C(I,H^1(\RR)).$
 \end{pro}
 {\bf Proof}\qquad Without loss of generality, we set $I=[0,L).$  We have that
 $$
 ||v(z)||_{H^1}^2\Doteq ||v(z)||_{L^2}^2+||v_t(z)||^2_{L^2}.
 $$
 Let us set
 $$
 \Phi(T)=\frac{\sqrt{2\;\theta}}{\cosh(\theta T)}.
 $$
 Taking the first derivative with respect to the variable $t$ we have that
 \begin{eqnarray*}
 \frac{\partial v(z)}{\partial t}&=&\Big[i\frac{C_2}{2}t\Phi(T(z,t))+\Phi^\prime(T(z,t))\frac{dT(z,t)}{dt}\Big]\;e^{i(\frac{C_2}{2}t^2+\theta Z(z))-\frac{C_2}{2}z}\\
 \\
 &=&\Big[i\frac{C_2}{2}e^{C_2 z}T\Phi(T(z,t))+\Phi^\prime(T(z,t))e^{-C_2 z}\Big]\;e^{i(\frac{C_2}{2}t^2+\theta Z(z))-\frac{C_2}{2}z}.\\
 \end{eqnarray*}

 Therefore, we obtain that
 \begin{eqnarray*}
 |v_t(z)|\leq \frac{C_2}{2}\;e^{\frac{C_2}{2}z}|T|\;|\Phi(T)|+e^{\frac{-3C_2}{2}z}|\Phi^\prime(T)|.
 \end{eqnarray*}
 
 We now make use of $(\ref{stimags})$ to obtain that
 
 \begin{eqnarray*}
 ||v_t(z)||_{L^2}^2
 &\leq& \Big(\frac{C_2}{2}\Big)^2\;e^{C_2 z}\int_\RR  |T|^2|\Phi(T)|^2 e^{C_2 z}dT+e^{-3C_2 z}\int_\RR|\Phi^\prime(T)|^2e^{C_2 z}dT\\
 &\leq&\Big(\frac{C_2}{2}\Big)^2e^{2C_2 z}A_1\int_\RR  (T)^2e^{-A_2|T|}dT+e^{-2C_2 z}A_1\int_\RR  e^{-A_2|T|}dT,
 \end{eqnarray*}
 that is finite for every $z\in I.$
 
 Moreover we have that
 \begin{eqnarray*}
 ||v(z)||_{L^2}^2&=&e^{-C_2 z}\int_\RR  |\Phi(T)|^2e^{C_2 z}dT\\
 &=&\int_\RR  |\Phi(T)|^2 dT\leq A_1\int_\RR e^{-A_2|T|}dT
 \end{eqnarray*}
 that is finite for every $z\in I.$ Therefore, summing up the last two inequalities the thesis of the Proposition easily follows.$\Box$
 \vspace{1cm}
 
 We should now have the existence of such solutions also for Eq. $(\ref{Ing})$ when $\beta_2<0.$ By reversing the $z$-axis (i.e. we take the change of variable $\bar{z}=-z$) and defining $A(z,t)=\exp(-\alpha z)U(z,t)$ we have that Eq. $(\ref{Ing})$ becomes
 \begin{equation}
 iU_z(z,t)+\frac{\beta_2}{2} U_{tt}(z,t)-\gamma e^{-\alpha z} |U(z,t)|^2 U(z,t)=0
 \label{Uing}
 \end{equation}
where we continue to refer to the variable $\bar{z}$ as $z.$ By means of scaling- transformation, as showed into the introduction of Chapter $2,$ because of $\beta_2<0$ we refer to Eq. $(\ref{Uing})$ as the following dimensionless equation:
\begin{equation}
iu_z(z,t)+u_{tt}(z,t)+e^{-C_2 z}|u(z,t)|^2u(z,t)=0,
\label{ING}
\end{equation}
where the fiber loss is taken into consideration by the \textit{real} positive constant $C_2.$ Thus, we will achieve our goal by means of a straightforwardly application of Theorem $\ref{maintheorem}.$

Let us consider the function
\begin{equation}
v(z,t)=e^{i\frac{C_2}{4}t^2-\frac{C_2}{2}z}\Phi(T)\;e^{i\theta Z},
\label{v}
\end{equation}
where $\Phi(T)=\frac{\sqrt{2\theta}}{\cosh(\theta T)}$ and $T(z,t)=e^{-C_2 z}t,\quad Z(z)=\frac{1}{2 C_2}(1-e^{-C_2 z}).$ We know that $v(z,t)$ is a solution of Eq. $(\ref{TNLSEfocusing}).$ In order to apply Theorem $\ref{maintheorem}$ we compute the first derivative with respect to $t$ of Eq. $(\ref{v})$ 
$$
v_t(z,t)=\Big(i\frac{C_2}{2}t\frac{\sqrt{2\theta}}{\cosh(\theta \;e^{-C_2 z}t)}-\frac{\theta\sqrt{2\theta}e^{-C_2 z}\tanh(\theta \;e^{-C_2 z}t)}{\cosh(\theta \;e^{-C_2 z}t)}\Big)e^{i\frac{C_2}{4}t^2-\frac{C_2}{2}z+i\theta Z(z)}
$$

and the following integrals:
\begin{eqnarray}
\int_\RR |t^2 v(z,t)|^2 dt &=& 2\theta e^{-C_2 z}\int_\RR\frac{t^4}{\cosh^2(\theta \;e^{-C_2 z}t)}dt
\label{integraletallaquartav}\\\nonumber
\\
\int_\RR |tv_t(z,t)|^2 dt &=& e^{-C_2 z}\int t^2\Big|i\frac{C_2}{2}\frac{t\sqrt{2\theta}}{\cosh(\theta \;e^{-C_2 z}t)}-\frac{\theta\sqrt{2\theta}e^{-C_2 z}\tanh(\theta \;e^{-C_2 z}t)}{\cosh(\theta \;e^{-C_2 z}t)}\Big|^2 dt\nonumber\\
\nonumber\\
&\leq&2\theta\Big(\frac{C_2}{2}\Big)^2e^{-C_2 z}\int_\RR \frac{t^4}{\cosh^2(\theta \;e^{-C_2 z}t)}dt\nonumber\\
\nonumber\\
&&+2\theta^3 e^{-3C_2 z}\int_\RR \frac{t^2\tanh^2(\theta \;e^{-C_2 z}t)}{\cosh^2(\theta \;e^{-C_2 z}t)}dt .\label{integralederivatadiv}
\end{eqnarray}
We notice that the integrals on the right-hand side of the Eqs. $(\ref{integraletallaquartav}),(\ref{integralederivatadiv})$ are finite. Thus, by choosing a suitable $\theta$ we can consider the quantities $||(\cdot)^2v(z,\cdot)||_{L^2}$ and $||\;|\cdot|v_t(z,\cdot)||_{L^2}$ small as we wish for every $z$ in a suitable interval of $\RR^+.$ Therefore, let us set $\varepsilon>0;$ then, by means of the Theorem $\ref{maintheorem}$ we deduce  for an arbitrary solution $u(z,t)$ of Eq. $(\ref{ING})$ that
\begin{equation}
||v(z)-u(z)||_{L^2}<\varepsilon,
\end{equation}
for every $z$ in a suitable interval $I\subset\RR^+$ containing $0$ and depending on $\varepsilon.$ 

Finally, we can conclude that we approximate the profile of the  function $u(z,t)$ as
\begin{equation}
e^{-\frac{C_2}{2}z}\frac{\sqrt{2\theta}}{\cosh(\theta \;e^{-C_2 z}t)}
\label{solitonsolution}
\end{equation}
for almost all $t\in \RR $ and for every $z$ in a suitable interval of $\RR^+.$ Here, $\theta$ is taken to be enough to hold hypothesis of the Theorem $\ref{maintheorem}$ true.

\subsection{Orbital Stability}
Let consider Eq. $(\ref{standard ing})$ with $\beta_2<0.$ In this case, we pointed out that special solutions exist which conserve their shape along the optical fiber. Referring to Eq. $(\ref{standard ing})$ as Eq. $(\ref{SNLSEfocusing})$  we called such solutions as \textit{soliton} and presented them as
$$
Q(Z,T)=\Phi(\theta,T)\;e^{i\theta Z},
$$
where $\Phi(\theta,T)$ is referred as \textit{ground state}. Let consider the fundamental \textit{ground state} $R(T),$ which is solution of the equation:
\begin{equation}
R^{\prime\prime}(T)-R(T)+R^3(T)=0.
\label{fundamentalgs}
\end{equation}
We shall now discuss about a type of stability of the fundamental \textit{ground state}, which turns out from the symmetries of the standard NLSE $(\ref{SNLSEfocusing}).$
\vspace{.5cm}

NLSE $(\ref{SNLSEfocusing})$  has phase and translation symmetries, i.e. if $Q(Z,T)$ solves NLSE then $e^{i\Gamma}Q(Z,T+T_0)$ solves NLSE for any $T_0\in \RR $ and $\Gamma\in [0,2\pi).$ By orbital stability \cite{Weinstein} means stability modulo these symmetries. An \emph{orbit} of a function $Q(Z,T)$ is defined as
\begin{equation}
\mathcal{O}_Q\Doteq\{Q(Z,T+T_0)e^{i\Gamma}\big| (T_0,\Gamma)\in \RR \times [0,2\pi) \}
\label{orbita}
\end{equation}
The deviation of the solution $\Psi(Z,T)$ of Eq. $(\ref{SNLSEfocusing})$ from $\mathcal{O}_Q$ is measured by the following metric 
\begin{eqnarray}
\Big[d_{\theta}(\Psi(Z),\mathcal{O}_Q)\Big]^2&\Doteq&\inf_{T_0\in\RR,\;\Gamma\in[0,2\pi)}\Bigl\{||\Psi_T(Z,\cdot+T_0)e^{i\Gamma}-Q_T||^2_{L^2}\nonumber\\
&&+\theta||\Psi(Z,\cdot+T_0)e^{i\Gamma}-Q||^2_{L^2}\Bigr\}.\label{distanzaorbitale}
\end{eqnarray}
Since the minimum in $(\ref{distanzaorbitale})$ is attained, this defines functions $T_0(Z)$ and $\Gamma(Z).$ In \cite{Weinstein1} a system of coupled non linear differential equations (\textit{modulation equations}) are derived for $T_0(Z)$ and $\Gamma(Z).$ It is proved that if $T_0(Z)$ and $\Gamma(Z)$ evolve according to the modulation equations, then the 'modulated' \textit{ground state} is linearly stable. Because of this fact the proof of the following result given by Weinstein is based on Lyapunov analysis.

\begin{theorem}
Let $Q(Z,T)$ be the unique solution of the standard NLSE with initial data $Q_0\in H^1(\RR).$ Then the ground state is orbitally stable, i.e., for any $\varepsilon>0,$ there is a $\delta(\varepsilon),$ such that if
\begin{equation}
d_\theta(Q_0,\mathcal{O}_R)<\delta(\varepsilon),
\label{datoiniziale}
\end{equation}
then for all $Z>0$
\begin{equation}
d_\theta(Q(Z),\mathcal{O}_R)<\varepsilon
\label{datofinale}
\end{equation}
\label{teoremadistabilitàorbitale}
\end{theorem}
\vspace{.5cm}

We only sketch the main steps of the proof given by \cite{Weinstein}.

\emph{Step $1$: Lyapunov functional.}\qquad We consider the Lyapunov functional

\begin{equation}
E[Q]\Doteq \int_{\RR}\Big[\frac 1 2 |Q_T(Z,T)|^2-\frac{1}{4}|Q(Z,T)|^4+\theta|Q(Z,T)|^2\Big]dT,
\label{funzionale}
\end{equation}
and we write the solution
\begin{equation}
Q(Z,T+T_0)e^{i\Gamma}\Doteq R(T)+W,
\label{gsperturbato}
\end{equation}
where $W(Z,T)=U(Z,T)+iV(Z,T)$ with $U,\;V$ \textit{real}-valued functions. The quantity $\mathcal{H}=\int_{\RR}\Big[\frac 1 2 |Q_T(Z,T)|^2-\frac{1}{4}|Q(Z,T)|^4 dT$ and $\mathcal{N}=\int_\RR|Q(Z,T)|^2\Big]dT$ are respectively the Hamiltonian and the square integral functionals, which are conserved for standard NLSE.

Then,
\begin{eqnarray}
\Delta E&\Doteq& E[Q_0(\cdot)]-E[R(\cdot)]\nonumber\\
&=&E[Q(Z,\cdot)]-E[R(\cdot)]\qquad\mbox{by conservation of E}\nonumber\\
&=&E[Q(Z,\cdot+T_0)e^{i\Gamma}]-E[R(\cdot)]\qquad \mbox{by scale invariance}\label{Delta}\\
&=& E[R+W]-E[R]\qquad \mbox{by (\ref{gsperturbato})}\nonumber\\
&\geq&(L_+U,U)+(L_-V,V)-K_1||W||_{H^1}^{2+\eta}-K_2||W||_{H^1}^6,\quad \mbox{with}\quad \eta>0.\nonumber
\end{eqnarray}
Here,
\begin{equation}
L_+=-\frac{\partial^2}{\partial T^2}+1-3R^2\qquad L_-=-\frac{\partial^2}{\partial T^2}+1-R^2
\label{linearizzato}
\end{equation}
are, respectively, the real and the imaginary parts of the linearized NLSE operator about the ground state and $K_1,\;K_2$ \textit{real} positive constants. The inequality in $(\ref{Delta})$ is arrived at as follows. First Taylor expand the previous line about $R.$ The first variation of $E$ at $R$ vanishes because $R$ solves Eq $(\ref{fundamentalgs}).$ the second variation is the quadratic functional in $U$ and $V.$ The remaining terms $o(|W|^3)$ can be estimated from below by an interpolation estimate of Nirenberg and Gagliardo (see for example \cite{Friedman}).

\emph{Step $2:$ Spectral properties of $L_+$ and $L_-$}\qquad If $T_0=T_0(Z),$ and $\Gamma=\Gamma(Z)$ are chosen to minimize $d_\theta(Q(Z),\mathcal{O}_R),$ then 
\begin{equation}
\int R^2(T)R^\prime(T)U(Z,T)dT=0,
\label{vincolo1}
\end{equation}
\begin{equation}
\int R^3(T)V(Z,T)dT=0.
\label{vincolo2}
\end{equation}
To estimate the first two terms in the inequality $(\ref{Delta}),$ one has to consider the spectral properties of the linear operators $L_+$ and $L_-.$ Since $L_-R=0,$ $R>0,$ which is non degenerate. Therefore, $L_-$ can be rewritten as
$$
L_-=-\frac{1}{R}\frac{\partial}{\partial T}\Big(R^2\frac{\partial}{\partial T}\Big(\frac{1}{R}\cdot\Big)\Big),
$$
and so 
$$(L_- V,V)=\int \Big|\frac{\partial }{\partial T}\Big(\frac{V}{R}\Big)\Big|^2R^2 dT\geq 0.$$
 Therefore, $L_-$ is a non negative operator.
 Also, the infimum of $(L_-V,V)/(V,V)$ under the constrain $(\ref{vincolo2})$ can be zero only if $V=R,$ but this contradicts $(\ref{vincolo2}).$ Thus
\begin{equation}
(L_-V,V)>c||V||_{L^2}^2,
\end{equation}
and also,
\begin{equation}
(L_-V,V)>c_1||V||_{H^1}^2.
\label{lowerbound}
\end{equation}
The analysis of the term $(L_+U,U)$ is slightly more delicate. Let first assume that the perturbed solution has the same $L^2$ norm as the ground state. Setting $U=U_{\parallel} +U_\perp,$ where $U_{\parallel}=(U,R)R$ and $U_\perp=U-U_\parallel,$ condition $||Q(Z)||_{L^2}=||R||_{L^2},$ implies that
\begin{equation}
(U,R)=-\frac{1}{2}((U,U)+(V,V))
\label{prodottoscalare}
\end{equation}
Now,
$$
(L_+U,U)=(L_+ U_\parallel,U_\parallel)+2(L_+U_\parallel,U_\perp)+(L_+U_\perp,U_\perp).
$$
Since $L_+R^\prime=0$ and $(R^\prime,R)=0,$ we have that
$$
\inf_{(f,R)}(L_+ f,f)=0,
$$
and the infimum is attained at $f=KR^\prime,\;K\in\RR. $ But this violates $(\ref{vincolo1});$ thus,
\begin{eqnarray*}
(L_+U_\perp,U_\perp)&\geq& K_1(U_\perp,U_\perp)\\
&=&K_1\big[(U,U)-\frac{1}{4}\big[(U,U)+(V,V)\big]^2\big],\quad K_1>0
\end{eqnarray*}
Also, by $(\ref{prodottoscalare}),$
$$
(L_+ U_\parallel,U_\parallel)=\frac{1}{4}(L_+ R,R)\big[(U,U)+(V,V)\big]^2,
$$
where $(L_+R,R)\leq 0.$

Finally one has that
$$
(L_+U_\perp,U_\parallel)\geq K_2||W||_{L^2}||W_T||_{L^2}.
$$
Summing up one can find that
\begin{equation}
(L_+U,U)\geq C_1||U||_{H^1}^2-C_2||W_T||^2_{L^2}||W||_{L^2}^2-C_3||W||_{L^2}^2.
\label{lowerbound1}
\end{equation}
To obtain $(\ref{lowerbound1})$ for a more general perturbation one can rescales the ground state $R(T)$ as $\widetilde{R}(T)=\theta R(\theta T)$ such that $\widetilde{R}$ has the same $L^2$- norm as $Q,$ and $||\widetilde{R}-R||_{H^1}<\varepsilon.$

\emph{Step $3$: A lower bound for $\Delta E$}\qquad Thanks to $(\ref{lowerbound})$ and $(\ref{lowerbound1})$ we have that 
\begin{equation}
(L_+U,U)+(L_-V,V)\geq K_3||W||^2_{H^1}-K_4||W||_{H^1}^3-K_5||W||_{H^1}^4.
\end{equation}
Furthermore, since
\begin{equation}
\sqrt{\min(\theta,1)}||W(Z)||_{H^1}\leq d_\theta(Q(Z),\mathcal{O}_R)\leq\sqrt{\max(\theta,1)}||W(Z)||_{H^1}
\end{equation}
it follows that 

\begin{equation}
\Delta E\geq g\big[d_\theta(Q(Z),\mathcal{O}_R)\big],
\label{DeltaE}
\end{equation}
where
\begin{equation}
g(Z)=cZ^2(1-aZ^\eta-bZ^4)\qquad \mbox{with}\quad a,b,c,\eta >0.
\end{equation}
The essential properties of $g$ are that $g(0)=0$ and $g(Z)>0$ for $0<Z\ll 1.$

\emph{Step $4:$ Stability result}\qquad 
The stability result can be derived from $(\ref{DeltaE})$ as follows. Let $\varepsilon>0$ be sufficiently small. Then, by continuity of $E$ in $H^1,$ there is a $\delta(\varepsilon)>0$ such that is
\begin{equation}
d_\theta(Q_0,\mathcal{O}_R)<\delta(\varepsilon),
\end{equation}
then
\begin{equation}
\Delta E(0)<g(\varepsilon).
\end{equation}
Since $\Delta E$ is constant in $Z,$ $(\ref{DeltaE})$ implies that $g\big[d_\theta(Q(Z),\mathcal{O}_R)\big]<g(\delta(\varepsilon))$ for all $Z>0.$ Therefore, since $d_\theta(Q(Z),\mathcal{O}_R)$ is a continuous function of $Z,$
\begin{equation}
d_\theta(Q(Z),\mathcal{O}_R)<\varepsilon \qquad \mbox{for all}\quad Z>0,
\end{equation}
i.e. the ground state is stable. $\Box$
\vspace{1cm}

We should now convert Theorem $\ref{teoremadistabilitàorbitale}$ into the context of the TNLSE. Let $Q(Z,T)$ be a solution of Eq. $(\ref{SNLSEfocusing}).$ Because of translations and phase shift symmetries of the standard NLSE , we have that the function $Q(Z,T+T_0)e^{i\Gamma}$ is still a solution of Eq. $(\ref{SNLSEfocusing}),$ for some $T_0\in \RR $ and $\Gamma\in[0,2\pi).$ Thus, from $(\ref{trasformazionimodellofocusing})$ we can write
$$
w(z,t)=e^{i\frac{C_2}{4}t^2-\frac{C_2}{2}z}Q(Z(z),T(z,t)+T_0)e^{i\Gamma},
$$
which is, because of Theorem $\ref{teoremaditrasformazione},$ a solution of Eq. $(\ref{TNLSEfocusing}).$ With this in mind we can produce the following definition:
\begin{definition}
Let 
 $v(z,t)$ be a solution of the Eq. $(\ref{TNLSEfocusing}),$ such that
  $$v(z,t)=e^{i\frac{C_2}{4}t^2-\frac{C_2}{2}z}Q(Z(z),T(z,t))$$ with $Q(Z,T)$ a solution of Eq. $(\ref{SNLSEfocusing}).$ Then we define the orbit of $v(z,t)$ as 
\begin{equation}
\mathcal{G}_v\Doteq \Bigl\{v(z,t,T_0,\Gamma)=e^{i\frac{C_2}{4}t^2-\frac{C_2}{2}z}Q(Z(z),T(z,t)+T_0)e^{i\Gamma}\; |\; T_0\in \RR, \;\Gamma\in [0,2\pi)\Bigr\}.
\label{orbitaTNLSE}
\end{equation}
\label{definizioneorbitaTNLSE}
\end{definition}

\vspace{.5cm}
Reversing the first equation of transformations $(\ref{trasformazionimodellofocusing})$ with respect to the variable $t$ and the third one to have $Q(Z,T)$ as function depending on $v(z,t)$ we obtain that
\begin{equation}
 \left\{\begin{array}{l}
 t(z,T)=e^{C_2 z}T\\
 \\
 Z(z)=\frac{1}{2C_2}(1-e^{-2C_2 z})\\
 \\
 Q(Z(z),T)=\;e^{-i\frac{C_2}{4}t^2+\frac{C_2}{2}z}v(z,t(z,T))\\
 \end{array}\right.
 \label{trasformazioniinversemodellofocusing}
 \end{equation}
 \vspace{.3cm}
 
Let us set
\begin{equation}
v(z,t)=e^{i\frac{C_2}{4}t^2-\frac{C_2}{2}z}\;R(T(z,t))\;e^{i Z(z)}
\label{GStrasformato}
\end{equation}
where $R(T)$ is referred as the fundamental \textit{ground state}. Let consider
\begin{equation}
w(z,t)=e^{i\frac{C_2}{4}t^2-\frac{C_2}{2}z}Q(Z(z),T(z,t)
\label{w}
\end{equation}
and
\begin{equation}
w(z,t,T_0,\Gamma)=e^{i\frac{C_2}{4}t^2-\frac{C_2}{2}z}Q(Z(z),T(z,t)+T_0)e^{i\Gamma}.
\label{orbitatrasformata}
\end{equation}

We now take into consideration distance $(\ref{distanzaorbitale}).$ By means of $(\ref{trasformazioniinversemodellofocusing})$ we compute
\begin{eqnarray}
&&\int_\RR |Q_T(Z,T+T_0)\;e^{i\Gamma}-R^\prime(T)\;e^{iZ}|^2 dT\label{distanzadellederivate}\\
&&=\int_\RR \Big|(w_t(z,t,T_0,\Gamma)-v_t(z,t))+i\frac{C_2}{2}t(v(z,t)-w(z,t,T_0,\Gamma))\Big|^2\Big(\frac{dt}{dT}\Big)^2 dt\nonumber\\
&&=e^{2C_2 z}\int_\RR \Big|(w_t(z,t,T_0,\Gamma)-v_t(z,t))+i\frac{C_2}{2}t(v(z,t)-w(z,t,T_0,\Gamma))\Big|^2 dt\nonumber
\end{eqnarray}
and
\begin{equation}
\int_\RR |Q(Z,T+T_0)\;e^{i\Gamma}-R(T)\;e^{iZ}|^2 dT=\int_\RR |w(z,t,T_0,\Gamma)-v(z,t)|^2 dt .
\label{distanzadellefunzioni}
\end{equation}
Let us define
\begin{eqnarray}
&&\widetilde{d}_\theta (w(z,T_0,\Gamma)-v(z))\nonumber\\
&&=\inf_{T_0\in \RR,\Gamma\in[0,2\pi)}\Big[e^{2C_2 z}\int_\RR \Big|(w_t(z,t,T_0,\Gamma)-v_t(z,t))+i\frac{C_2}{2}t(v(z,t)-w(z,t,T_0,\Gamma))\Big|^2 dt\nonumber\\
&&+\theta\int_\RR |w(z,t,T_0,\Gamma)-v(z,t)|^2 dt\Big]\label{distanzatrasformata}
\end{eqnarray}
Therefore, we can produce the following result by means of straightforwardly application of Theorem $\ref{teoremadistabilitàorbitale}.$
\begin{cor}
Let $v(z,t)$ as in $(\ref{GStrasformato})$ and $$w(z,t)=e^{i\frac{C_2}{4}t^2-\frac{\alpha}{2}z} Q(Z(z),T(z,t)),$$ where $Q(Z,T)$ satisfies the assumptions of the Theorem $\ref{teoremadistabilitàorbitale}.$ Then, for any $\varepsilon>0$ there is a $\delta(\varepsilon)$ such that if
\begin{equation}
\widetilde{d}_\theta(w(0),\mathcal{G}_v)<\delta(\varepsilon)
\label{trasformatodiddatoiniziale}
\end{equation}
then for all $z\geq 0$
\begin{equation}
\widetilde{d}_\theta(w(z),\mathcal{G}_v)<\varepsilon.
\label{trasformatodid}
\end{equation}

\end{cor}
{\bf Proof.}\qquad Because of Eqs. $(\ref{distanzadellederivate}),\;(\ref{distanzadellefunzioni})$ we have that
\begin{equation}
\widetilde{d}_\theta(w(z),v(z))0=d_\theta(Q(z),R).
\label{1}
\end{equation}
Thus, from Theorem $\ref{teoremadistabilitàorbitale},$ restricted to the interval $Z\in[0,\frac{1}{2C_2}),$ we have that for any $\varepsilon>0$ there exists a $\delta(\varepsilon)>0$ such that $\widetilde{d}_\theta(w(0),v(0))<\delta(\varepsilon)$ implies that $d_\theta(w(z),v(z))<\varepsilon,$ for every $z$ such that $Z(z)\in[0,\frac{1}{2C_2})\quad \Box$

\end{document}